%% file: main.tex
\documentclass[
 amsmath,amssymb,
 reprint,
 superscriptaddress,aps,
 nofootinbib,prd%
]{revtex4-2}

\usepackage{rotating} % packages to rotate tables

\usepackage{graphicx}
\usepackage{booktabs}
\usepackage{amsmath}
\usepackage{arydshln}
\usepackage{latexsym}
\usepackage{mathrsfs}
\usepackage{mathtools,slashed}
\usepackage{soul}
\usepackage{amssymb}
\usepackage{multirow}                   % for tabling
\usepackage{pifont}                     % for tick and cross marks
\usepackage{dcolumn}% Align table columns on decimal point
\usepackage{bm}% bold math
\usepackage{hyperref}% add hypertext capabilities

\usepackage[T1]{fontenc}
\usepackage[dvipsnames,table]{xcolor}   % colours
\usepackage{youngtab}
\usepackage{comment}
\usepackage[normalem]{ulem}

\definecolor{red}{rgb}{0.6,.0706,.1373}
\definecolor{blue}{rgb}{0,0.396,0.741}
\definecolor{PadovaRed}{rgb}{0.60546875,0,0.078125}

\colorlet{mylinkcolor}{violet}
\colorlet{mycitecolor}{violet}
\colorlet{myurlcolor}{violet}

\hypersetup{
  linkcolor  = PadovaRed,
  citecolor  = PadovaRed,
  urlcolor   = PadovaRed,
  colorlinks = true
}
\pretocmd{\tableofcontents}{\hypersetup{linkcolor=black}}{}{}
\apptocmd{\tableofcontents}{\hypersetup{linkcolor=red}}{}{}

\newcommand{\cL}{\mathcal{L}}
\newcommand{\cO}{\mathcal{O}}
\newcommand{\cR}{\mathcal{R}}

\newcommand{\cQ}{{\mathcal Q}}

\newcommand{\cI}{{\mathcal I}}
\newcommand{\cC}{{\mathcal C}}

\newcommand{\cT}{{\mathcal T}}

\newcommand{\gev}{\mathrm{GeV}}
\newcommand{\tev}{\mathrm{TeV}}
\newcommand{\hermc}{\text{h.c.}}
\newcommand{\eminus}{\vcenter{\hbox{\scalebox{0.6}[1]{$ - $}}}}	%Narrow minus signed (for e.g. negative exponents)
\newcommand{\rep}[1]{\mathbf{#1}}
\newcommand{\repbar}[1]{\overline{\mathbf{#1}}}
\newcommand{\sscript}[1]{{\scriptscriptstyle \mathrm{#1}}}

\newcommand{\U}{\mathrm{U}}
\newcommand{\SU}{\mathrm{SU}}
\newcommand{\ud}[2]{\phantom{}^{#1}\phantom{}_{#2}}
 
\newcommand{\sig}{\sigma}
\newcommand{\lzm}{\left(}
\newcommand{\dzm}{\right)}

\newcommand{\lzv}{\left\{}
\newcommand{\dzv}{\right\}}

\begin{document}

\title{Implications of Flavor Symmetries for Baryon Number Violation}

\author{Arnau Bas i Beneito}%
 \email{arnau.bas@ific.uv.es}
 \affiliation{%
  Institut de F\'isica Corpuscular (IFIC), Consejo Superior de Investigaciones
Cient\'ificas (CSIC) and Universitat de Val\`encia (UV), 46980 Val\`encia, Spain
}%

\author{Ajdin Palavri\'c}%
\email{ajdin.palavric@ific.uv.es}
\affiliation{%
  Institut de F\'isica Corpuscular (IFIC), Consejo Superior de Investigaciones
Cient\'ificas (CSIC) and Universitat de Val\`encia (UV), 46980 Val\`encia, Spain
}%
\affiliation{%
Department of Physics, University of Basel, Klingelbergstrasse 82, CH-4056 Basel, Switzerland
}

\author{Andrea Sainaghi}%
\email{andrea.sainaghi@phd.unipd.it}
\affiliation{%
Istituto Nazionale di Fisica Nucleare, Sezione di Padova, 35131 Padova, Italy
}%
\affiliation{%
Dipartimento di Fisica e Astronomia ”Galileo Galilei”, Universit\`a di Padova, 35131 Padova, Italy}%
\affiliation{%
Physik-Institut, Universit\"at Z\"urich, CH-8057 Z\"urich, Switzerland}%

% \date{\today}

\begin{abstract}
In the Standard Model, baryon number is an accidental symmetry, whose violation would constitute unambiguous evidence of new physics, with proton decay providing its most prominent experimental signature. At the same time, the peculiar structure of flavor can serve as a guiding principle for exploring possible new-physics effects. In this work, we present a systematic classification of dimension-six baryon-number-violating (BNV) SMEFT operators across several flavor-symmetry assumptions and analyze the resulting phenomenology. Interestingly, in certain flavor scenarios the non-trivial interplay with tiny neutrino masses leads to proton-decay constraints compatible with BNV scales in the multi-TeV range. Finally, we complement the EFT analysis by identifying one-particle UV completions of the BNV operators, revealing scenarios in which the leading-order EFT description may not fully account for their underlying dynamics.
\end{abstract}

\maketitle

%%%%%%%%%%%%%%%%%%%%%%%%%%%%%%%%%%%%%%%%%%%%%%%%%%%%
\section{Introduction}
%%%%%%%%%%%%%%%%%%%%%%%%%%%%%%%%%%%%%%%%%%%%%%%%%%%%

One of the most striking features of the Standard Model (SM) is the intricate pattern of fermion masses and mixings that defines its flavor structure. In the absence of Yukawa interactions, the SM would exhibit an extensive global flavor symmetry allowing independent unitary rotations among the three generations of quarks and leptons. The introduction of the Yukawa couplings explicitly breaks this symmetry, giving rise to the observed hierarchies and to the principal sources of flavor and CP violation within the SM. Framing the flavor structure in terms of this approximate global symmetry clarifies the origin of the observed hierarchies and provides a systematic guide for constructing extensions of the SM consistent with flavor constraints.

The Minimal Flavor Violation (MFV) framework~\cite{DAmbrosio:2002vsn,Chivukula:1987py}, originally developed in the quark sector, provides a symmetry-based strategy for addressing the flavor problem associated with TeV-scale new physics. MFV posits that the only sources of flavor and CP violation are the Yukawa couplings, which explicitly break the SM flavor symmetry. By promoting these Yukawa matrices to spurionic fields with definite transformation properties, one ensures that any higher-dimensional operator or new-physics interaction respects the SM flavor alignment. Consequently, new-physics effects acquire CKM-like suppression factors through structures such as $Y_u Y_u^\dagger$ and $Y_d Y_d^\dagger$, thereby mitigating potentially dangerous FCNCs and CP-violating contributions and ensuring compatibility with precision measurements in kaon, $B$-meson, and $D$-meson systems.

Building on this quark-sector construction, the MFV framework was later generalized to the lepton sector~\cite{Cirigliano:2005ck,Davidson:2006bd,Alonso:2011jd,Gavela:2009cd} to accommodate neutrino mass generation and to analyze charged-lepton flavor violation (cLFV). In this context, processes such as $\mu \to e\gamma$ and $\mu \to 3e$ remain controlled by the same spurionic structures that dictate flavor breaking in the SM, ensuring the characteristic MFV suppression. The framework naturally interfaces with seesaw scenarios~\cite{Minkowski:1977sc,Mohapatra:1980yp,Foot:1988aq}, where lepton-number violation (LNV) introduces additional breaking of the flavor symmetry in the neutrino sector.

While MFV was initially developed to address the flavor problem in the quark and lepton sectors, the same organizing principle has been applied in a variety of broader contexts. A notable application is found in leptoquark frameworks, where flavor covariance of the new interactions can be ensured either by introducing appropriate spurion fields or by allowing the leptoquarks to transform nontrivially under the flavor symmetry~\cite{Davidson:2010uu}. Beyond the leptoquark models, MFV has also been explored in scenarios featuring baryon-number violation (BNV). The embedding of MFV within supersymmetric models has been shown to suppress proton-decay operators to levels consistent with experimental bounds, thereby removing the need for imposing $R$-parity~\cite{Nikolidakis:2007fc}. The potential correlation between LNV and proton stability was later investigated, suggesting that the smallness of neutrino masses and the longevity of the proton may originate from a common symmetry-breaking pattern, potentially allowing the BNV scale to lie near the TeV regime~\cite{Smith:2011rp}. This possibility was subsequently revisited within a broader effective field theory (EFT) framework~\cite{Helset:2019eyc}.

Focusing more directly on the BNV sector, proton decay plays a central role among low-energy probes of physics beyond the SM~\cite{Weinberg:1979sa,Wilczek:1979hc,Abbott:1980zj,Weinberg:1980bf}. Its observation would provide unambiguous evidence for BNV interactions, while the continued absence of a signal places exceptionally stringent limits on such dynamics. More concretely, current searches by Super-Kamiokande (Super-K) constrain the proton lifetime to the level of $10^{32}-10^{33}$ years for many decay channels~\cite{Super-Kamiokande:2020wjk,Super-Kamiokande:2013rwg,Super-Kamiokande:2022egr}, with the most sensitive modes, such as $p \to \pi^0 e^+$, already setting lower limits exceeding $10^{34}$ years. Future experiments, including Hyper-Kamiokande (Hyper-K)~\cite{Hyper-Kamiokande:2018ofw} and DUNE~\cite{DUNE:2020ypp}, are expected to substantially improve proton-decay sensitivities, with projected reach up to lifetimes of $\cO(10^{35})$ years in selected channels after several years of data taking. Interpreted within the Standard Model EFT (SMEFT) framework~\cite{Grzadkowski:2010es,Buchmuller:1985jz}, these bounds translate into extremely high scales for the dimension-six BNV operators, reaching up to $\Lambda_\sscript{B}\sim\cO(10^{16}\,\gev)$ in the anarchic limit~\cite{Beneito:2023xbk}. Additional constraints arise from direct searches for BNV processes at colliders involving heavy quarks~\cite{CMS:2024dzv,BaBar:2011yks,LHCb:2022wro}, but these are negligible compared to the limits set by proton stability. Proton-decay limits therefore dominate the phenomenology of all SMEFT BNV operators, including those contributing only at loop level or via multi-body final states~\cite{Hou:2005iu,Beneke:2024hox,Crivellin:2023ter,Gisbert:2024sjw,Heeck:2024jei,Dong:2011rh,Marciano:1994bg,ThomasArun:2025dav,Kumar:2025aek}. This pronounced suppression of BNV interactions involving light-family fermions thus provides a compelling motivation to treat the BNV sector within a flavor-oriented framework, in line with the MFV considerations introduced above.

In this paper, we investigate the implications of embedding BNV and LNV within well-defined flavor frameworks. Our primary analysis is carried out in an \textit{extended} MFV scenario, in which both Yukawa couplings and neutrino masses serve as the spurions breaking the underlying symmetry. In addition, we explore several alternative flavor symmetries to compare their predictions with the ones obtained assuming MFV, particularly those in which new physics couples preferentially to third-generation fermions~\cite{Barbieri:2011ci,Isidori:2012ts,Barbieri:2012uh,Allwicher:2023shc}. Concentrating on the four dimension-six BNV operators of the SMEFT, we construct their flavor-invariant combinations and identify the least suppressed structures. This allows for a quantitative comparison between the characteristic scales governing BNV and LNV processes. To complement the EFT analysis, we also examine ultraviolet (UV) completions that generate these operators at tree level, allowing us to assess the corresponding phenomenological implications within and beyond the EFT description.

The remainder of this work is organized as follows. In Sec.~\ref{sec:MFV_overview}, we provide an overview of the extended MFV framework, outlining the relevant spurion structures, with particular emphasis on the inclusion of neutrino mass effects. In Sec.~\ref{sec:EFT_analysis_BNV_LFV}, we perform the EFT analysis by constructing the flavor invariants associated with the four dimension-six BNV SMEFT operators and identifying the least suppressed combinations, followed by a discussion of their phenomenological implications. Sec.~\ref{sec:UV_compl} is devoted to the UV completions that can generate these operators, establishing connections between the low-energy EFT and possible high-scale dynamics. In Sec.~\ref{sec:MFV_alter}, we explore alternative flavor frameworks built on smaller symmetry groups and discuss their implications. Finally, Sec.~\ref{sec:conc} summarizes our findings and presents the conclusions. Additional technical details and derivations are collected in the appendices.

%%%%%%%%%%%%%%%%%%%%%%%%%%%%%%%%%%%%%%%%%%%%%%%%%%%%
\section{Overview of the Extended MFV}
\label{sec:MFV_overview}
%%%%%%%%%%%%%%%%%%%%%%%%%%%%%%%%%%%%%%%%%%%%%%%%%%%%

The gauge interactions of the SM are invariant under a large global flavor symmetry in the limit where all Yukawa couplings vanish. This symmetry corresponds to independent unitary transformations acting on each type of fermion field, leading to the maximal flavor group
\begin{equation}
    G_\sscript{F}\equiv\U(3)^5=\U(3)_q\times\U(3)_u\times\U(3)_d\times\U(3)_\ell\times\U(3)_e\,,
\end{equation}
where the five $\U(3)$ factors correspond to independent unitary rotations of the left-handed quark doublets ($q$), right-handed up-type quarks ($u$), right-handed down-type quarks ($d$), left-handed lepton doublets ($\ell$), and right-handed charged leptons ($e$), respectively. The transformation properties of the SM fermions under $G_\sscript{F}$ are given by
\begin{equation}
    \begin{alignedat}{4}
        q&\sim(\rep{3},\rep1,\rep1,\rep1,\rep1)\equiv\rep3_q\,,
        &~~\quad
        \ell&\sim(\rep1,\rep1,\rep1,\rep3,\rep1)\equiv\rep3_\ell\,,
        \\
        u&\sim(\rep1,\rep3,\rep1,\rep1,\rep1)\equiv\rep3_u\,,
        &~~\quad 
        e&\sim(\rep1,\rep1,\rep1,\rep1,\rep3)\equiv\rep3_e\,.\\
        d&\sim(\rep{1},\rep1,\rep3,\rep1,\rep1)\equiv\rep3_d\,,
    \end{alignedat}
\end{equation}
The only interaction terms in the Standard Model Lagrangian that explicitly break $G_\sscript{F}$ are the Yukawa couplings
\begin{equation}\label{eq:Yuk_Lagr}
    -\cL_{\sscript{Y}}=\bar q_p [Y_d]\ud{p}{r} d^r H+\bar q_p [Y_u]\ud{p}{r}u^r\widetilde H+\bar\ell_p [Y_e]\ud{p}{r}e^rH+\hermc\,,
\end{equation}
where $H$ denotes the Higgs doublet and $\widetilde H=i\sigma_2H^*$. Throughout this work, we retain explicit flavor indices and use a covariant index notation, making $G_\sscript{F}$ transformations manifest and streamlining the construction of invariants.

The MFV framework~\cite{DAmbrosio:2002vsn} formalizes the idea that the flavor-breaking structure of the SM persists even in the presence of new physics. This is achieved by promoting the Yukawa matrices to spurion fields that transform under $G_\sscript{F}$ in such a way that the full Lagrangian, including higher-dimensional operators, remains formally invariant. The transformation properties of the spurions are defined such that the Yukawa interaction terms become formally invariant under $G_\sscript{F}$:
\begin{equation}\label{eq:Yuk_spurions}
    \begin{alignedat}{2}
        Y_d&\sim(\rep3,\rep1,\repbar3,\rep1,\rep1)\equiv(\rep3_q,\repbar3_d)\,,
        \\
        Y_u&\sim(\rep3,\repbar3,\rep1,\rep1,\rep1)\equiv(\rep3_q,\repbar3_u)\,,
        \\
        Y_e&\sim(\rep1,\rep1,\rep1,\rep3,\repbar3)\equiv(\rep3_\ell,\repbar3_e)\,.
    \end{alignedat}
\end{equation}
The MFV hypothesis asserts that these three spurions, in absence of neutrino masses, govern all flavor-violating interactions, both in the SM and in any effective extension, i.e. any higher-dimensional operator that violates flavor must do so through insertions of $Y_d$, $Y_u$, or $Y_e$, arranged to preserve invariance under $G_\sscript{F}$. This reasoning applies naturally to the SMEFT framework, where MFV provides a symmetry-based principle that constrains the flavor structure of higher-dimensional operators. It ensures that all such operators respect the same flavor-breaking pattern as the SM itself, thus providing a controlled and phenomenologically consistent extension of the low-energy theory~\cite{Greljo:2022cah,Greljo:2023adz,Greljo:2023bdy}.

While the MFV framework successfully reproduces the flavor-breaking pattern of the SM in the limit of massless neutrinos, its minimal implementation fails to accommodate the observed lepton-flavor violation (LFV) implied by neutrino oscillations~\cite{Esteban:2024eli}. In the absence of right-handed neutrinos or Majorana mass terms, the SM lepton sector contains a single flavor-breaking spurion, the charged-lepton Yukawa matrix $Y_e$, which breaks the global $\U(3)_\ell \times \U(3)_e$ symmetry. Introducing neutrino masses necessarily requires new spurions in order to formally restore the flavor symmetry.

Focusing on the EFT framework, we consider neutrino masses generated by the dimension-five Weinberg operator~\cite{Weinberg:1979sa}\footnote{The charge-conjugated field $\psi^c$ is defined as $\psi^c\equiv C\bar\psi^\intercal$, where $C$ is the charge-conjugation matrix. In two-component (Weyl) notation, the charge-conjugated fields are defined as $\psi_L^c\equiv -i\sigma_2\psi_L$ and $\psi_R^c=i\sigma_2\psi_R$.}
\begin{equation}\label{eq:Weinberg_op}
    [\cO_{\sscript{W}}]_{pr}=(\bar\ell_p\widetilde H)(\widetilde H^\intercal \ell_r^c)\,,
\end{equation}
where $\cO_{\sscript{W}}$ violates total lepton number by two units and induces a Majorana mass term for the left-handed neutrinos after Electroweak Symmetry Breaking (EWSB). In addition, with lepton doublets transforming as $\ell\sim\rep3_\ell$ under the MFV assumption, the Weinberg operator explicitly breaks the flavor symmetry. To restore invariance under the flavor symmetry, the coefficient of the operator must be promoted to a spurion field that transforms appropriately under $\U(3)_\ell$. Since the operator in Eq.~\eqref{eq:Weinberg_op} involves the tensor product of the form
\begin{equation}
    \bar\ell\,\ell^c\sim\repbar3_\ell\otimes\repbar3_\ell=\repbar6_\ell\oplus\rep3_\ell\,,
\end{equation}
the Weinberg operator selects out the symmetric part of this decomposition, and thus requires the introduction of a symmetric spurion $\Upsilon_\nu$ transforming as 
\begin{equation}
   \Upsilon_\nu\sim(\rep1,\rep1,\rep1,\rep6,\rep1)\equiv\rep6_\ell\,.
\end{equation}
This additional spurion encodes the flavor-breaking effects associated with Majorana neutrino masses and represents the minimal extension of the lepton flavor sector beyond the charged lepton Yukawa coupling within the MFV framework~\cite{Cirigliano:2005ck}.

The flavor-invariant interaction term can then be written as
\begin{equation}
    \cL_{\sscript{SMEFT}}\supset -\frac{1}{2\Lambda_\sscript{L}}[\Upsilon_\nu]^{pr}(\bar\ell_p\widetilde H)(\widetilde H^\intercal \ell_r^c)+\hermc\,,
\end{equation}
where $\Lambda_\sscript{L}$ denotes the LNV scale. Upon EWSB, this interaction term gives rise to a Majorana mass term for the left-handed neutrinos
\begin{equation}\label{eq:neutrino_mass}
    \cL_\nu\supset -\frac{1}{2}[m_\nu]^{pr}(\bar\nu_{p}\nu^c_r)+\hermc\,,
    \quad
    [m_\nu]^{pr}=\frac{v^2}{2\Lambda_\sscript{L}}[\Upsilon_\nu]^{pr}\,,
\end{equation}
where the Higgs vacuum expectation value is taken to be $\langle H\rangle=\frac{1}{\sqrt2}[0~~v]^\intercal$. In this \textit{extended} MFV setup, characterized by $Y_e$ and $\Upsilon_\nu$ spurions, one may exploit the $\U(3)_\ell \times \U(3)_e$ flavor symmetry to bring the charged lepton Yukawa matrix into a convenient form. Specifically, by performing unitary transformations
\begin{equation}\label{eq:lep_rotation}
    \ell\to V_\ell\,\ell\,,
    \quad 
    e\to V_e\,e\,,
    \quad
    V_{\ell,e}\in\U(3)_{\ell,e}\,,
\end{equation}
the matrix $Y_e$ can be brought to the diagonal form
\begin{equation}
    Y_e\to \widehat Y_e=V_\ell^\dag Y_e V_e\,,
    \quad
    \widehat Y_e=\sqrt2\,\frac{\widehat m_e}{v}\,,
\end{equation}
which fixes the $V_\ell$ and $V_e$ rotations. After EWSB, the neutrino mass matrix $m_\nu$ in Eq.~\eqref{eq:neutrino_mass} can be diagonalized by a single unitary matrix $U_\nu$:
\begin{equation}
    m_\nu\to\widehat m_\nu= U_\nu^\dag m_\nu U_\nu^*\,,
\end{equation}
which can be used to express $\Upsilon_\nu$ as
\begin{equation}\label{eq:Ynu_via_PMNS}
    \Upsilon_\nu\to 
    \frac{2\Lambda_{\sscript{L}}}{v^2}U_\nu\widehat m_\nu U_\nu^\intercal\,.
\end{equation}
In this basis, where the charged lepton Yukawa matrix $Y_e$ is diagonal, the unitary matrix $U_\nu$ that diagonalizes the neutrino mass matrix coincides with the leptonic mixing matrix, i.e. we directly identify $U_\nu = U_{\sscript{PMNS}}$, which governs neutrino oscillations through its appearance in the charged-current interactions.\footnote{If we chose not to use the $\U(3)_\ell \times \U(3)_e$ flavor rotations to diagonalize the charged lepton Yukawa matrix $Y_e$, then the leptonic mixing matrix appearing in the charged current would take slightly modified form given by $U{\sscript{PMNS}} = V_\ell^\dagger U_\nu$.} We adopt the standard parameterization of the PMNS matrix, which can be found in Ref.~\cite{Esteban:2024eli}. This relation allows us to express the spurion $\Upsilon_\nu$ directly in terms of the physical neutrino masses and mixing angles. Eq.~\eqref{eq:Ynu_via_PMNS} can further be expressed in its component form as
\begin{equation}\label{eq:Ynu_components}
    [\Upsilon_\nu]_{ij}= \frac{2\Lambda_{\sscript{L}}}{v^2}\sum_k\, [U_\nu]_{ik}[U_\nu]_{jk}\,\widehat m_{\nu,k}\,,
\end{equation}
which can be used to analyze the parametrization of $\Upsilon_\nu$ depending on the mass ordering. In addition, the Majorana phases in the PMNS matrix are neglected, as they do not influence the operator structures relevant for our analysis. Given the unknown absolute neutrino mass scale, we derive analytical approximations for $\Upsilon_\nu$ in the normal (NO) and inverted (IO) ordering by expanding around the lightest neutrino mass, employing the known mass-squared splittings and mixing parameters. As input, we adopt the central values from the most recent global fit to neutrino-oscillation data~\cite{Esteban:2024eli}.

%%%%%%%%%%%%%%%%%%%%%%%%%%%%%%%%%%%%%%%%%%%%%%%%%%%%
\vspace{0.2cm}
\noindent
\textbf{Normal ordering (NO).} In the case of normal ordering, where $\widehat m_{\nu,1} < \widehat m_{\nu,2} < \widehat m_{\nu,3}$, the experimentally measured parameters that can be used are the mass-squared splittings
\begin{equation}
    \Delta m_{21}^2 = \widehat m_{\nu,2}^2 - \widehat m_{\nu,1}^2\,, \qquad \Delta m_{32}^2 = \widehat m_{\nu,3}^2 - \widehat m_{\nu,2}^2\,.
\end{equation}
Assuming $\widehat m_{\nu,1} \ll \widehat m_{\nu,2} \ll \widehat m_{\nu,3}$, we expand the neutrino masses in terms of the heavier eigenvalues using
\begin{equation}
    \begin{alignedat}{2}
        \widehat m_{\nu,3} &\approx \sqrt{\Delta m_{32}^2 + \Delta m_{21}^2} \approx \sqrt{\Delta m_{32}^2} + \frac{\Delta m_{21}^2}{2\sqrt{\Delta m_{32}^2}} + \cdots\,,
        \\
         \widehat m_{\nu,2} &\approx \sqrt{\Delta m_{21}^2}\,,
    \end{alignedat}
\end{equation}
which, once plugged into Eq.~\eqref{eq:Ynu_components}, yields
\begin{equation}
    \begin{alignedat}{2}
        [\Upsilon_\nu]_{ij}&\approx  \frac{2\Lambda_{\sscript{L}}}{v^2}\bigg[\delta_{ij}\sqrt{\Delta m_{21}^2}-[U_\nu]_{i1}[U_\nu]_{j1}\sqrt{\Delta m_{21}^2}
        \\&~~~~~~~~~+[U_\nu]_{i3}[U_\nu]_{j3}\bigg( \sqrt{\Delta m_{32}^2}-\sqrt{\Delta m_{21}^2} \bigg)
        \\&~~~~~~~~~+\cO\lzm{\Delta m_{21}^2}/{\sqrt{\Delta m_{32}^2}}\dzm\bigg]\,.
    \end{alignedat}
\end{equation}
%%%%%%%%%%%%%%%%%%%%%%%%%%%%%%%%%%%%%%%%%%%%%%%%%%%%

%%%%%%%%%%%%%%%%%%%%%%%%%%%%%%%%%%%%%%%%%%%%%%%%%%%%
\vspace{0.2cm}
\noindent
\textbf{Inverted ordering (IO).} For the inverted mass ordering, where $m_{\nu,3} < m_{\nu,1} < m_{\nu,2}$, it is convenient to express the mass-squared 
splittings as
\begin{equation}
    \begin{alignedat}{2}
        \Delta m_{23}^2 &= \widehat m_{\nu,2}^2 - \widehat m_{\nu,3}^2\,,
        \\
        \Delta m_{13}^2 &= \widehat m_{\nu,1}^2 - \widehat m_{\nu,3}^2 = \Delta m_{23}^2 - \Delta m_{21}^2\,.
    \end{alignedat}
\end{equation}
Assuming $\widehat m_{\nu,3} \ll \widehat m_{\nu,1}, \widehat m_{\nu,2}$, we expand around $\widehat m_{\nu,3}$, and, once again, rewrite the heavier masses in terms of $\sqrt{\Delta m_{13}^2}$ and $\sqrt{\Delta m_{23}^2}$. The resulting parametrization for $\Upsilon_\nu$ is then given as
\begin{equation}
    \begin{alignedat}{2}
    [\Upsilon_\nu]_{ij}&\approx \frac{2\Lambda_{\sscript{L}}}{v^2}\bigg[[U_\nu]_{i1}[U_\nu]_{j1}\sqrt{\Delta m^2_{13}}
    \\
    ~~~~~~~~~~~~~&+[U_\nu]_{i2}[U_\nu]_{j2}\sqrt{\Delta m^2_{23}}+\mathcal{O}\lzm{\widehat m_{\nu,3}}\dzm\bigg] \,.
    \end{alignedat}
\end{equation}
%%%%%%%%%%%%%%%%%%%%%%%%%%%%%%%%%%%%%%%%%%%%%%%%%%%%

%%%%%%%%%%%%%%%%%%%%%%%%%%%%%%%%%%%%%%%%%%%%%%%%%%%%
% \vspace{-0.4cm}
% \vspace{-0.0cm}
\section{EFT Analysis and Phenomenology}
\label{sec:EFT_analysis_BNV_LFV}
%%%%%%%%%%%%%%%%%%%%%%%%%%%%%%%%%%%%%%%%%%%%%%%%%%%%
Building on the framework introduced in Sec.~\ref{sec:MFV_overview}, we now analyze the flavor-invariant structures of the dimension-six BNV SMEFT operators. Our goal is to systematically construct the minimal spurion insertions required to render these operators invariant under the full flavor symmetry group and to identify the dominant phenomenological channels associated with each structure. Throughout this analysis, we assume a hierarchical separation of scales, $\Lambda_\sscript{L}\gtrsim  \Lambda_\sscript{F} \gg \Lambda_\sscript{B}$, where $\Lambda_\sscript{F}$ characterizes flavor-symmetry breaking. This hierarchy allows us to treat the spurions as background fields that can be consistently inserted when constructing BNV operators, whose UV dynamics occur at the scale $\Lambda_\sscript{B}$. This analysis establishes a framework for quantitatively linking the effective scales of BNV and LNV within the extended MFV scenario. For completeness, in App.~\ref{app:cLFV_ex_MFV} we also apply this formalism to the relevant cLFV transitions, illustrating their structure and suppression patterns within the same framework.

%%%%%%%%%%%%%%%%%%%%%%%%%%%%%%%%%%%%%%%%%%%%%%%%%%%%
\subsection{Flavor-Invariant BNV Structures}
\label{sec:flavor_invariant_BNV_ops}
%%%%%%%%%%%%%%%%%%%%%%%%%%%%%%%%%%%%%%%%%%%%%%%%%%%%

%%%%%%%%%%%%%%%%%%%%%%%%%%%%%%%%%%%%%%%%%%%%%
\begin{table*}[t]
    \input{tables/Dim6_BNV_SMEFT_Spurion_Exp}
    \caption{Overview of the dimension-six BNV SMEFT operators and the flavor-invariant structures in the extended MFV framework. The first column lists the relevant operator structures, while the second gives their definitions in terms of SM fermion fields. The third column displays the minimal spurion insertions required to render each operator invariant under the full flavor symmetry group $G_\sscript{F}$, expressed in terms of the SM Yukawa matrices and the additional $\Upsilon_\nu$ spurion. The final column reports the overall suppression order in spurions, highlighting the flavor-breaking structure induced by the extended MFV hypothesis. The constructions account for Fierz identities, permutation symmetries, and charge-conjugation conventions, with the goal of identifying the least suppressed, non-vanishing flavor-invariant structures relevant for BNV processes. In the operator definitions, Greek indices $\alpha,\beta,\gamma$ label $\SU(3)_\sscript{C}$, Latin dotted indices $\dot a,\dot b,\dot d,\dot e$ denote $\SU(2)_\sscript{L}$, and $p,r,s,t$ represent flavor indices. In the spurion expansions, all indices are understood to refer to flavor. Symmetry relations, arising from the Fierz identities and the permutation symmetries, can be found in Ref.~\cite{Alonso:2014zka}.}
    \label{tab:dim6_BNV_SMEFT_Spurions}
\end{table*}
%%%%%%%%%%%%%%%%%%%%%%%%%%%%%%%%%%%%%%%%%%%%%

%%%%%%%%%%%%%%%%%%%%%%%%%%%%%%%%%%%%%%%%%%%%%
\begin{table}[t]
    \input{tables/Dim6_LEFT_matching}
    \caption{Dimension-six $\Delta B=\Delta L=1$ operators in the LEFT and their corresponding tree-level matching to the SMEFT basis. The matching coefficients are presented without specifying a particular flavor basis; the basis choice and transformation to the up/down mass basis are discussed in the main text.}
    \label{tab:SMEFT_LEFT_matching}
\end{table}
%%%%%%%%%%%%%%%%%%%%%%%%%%%%%%%%%%%%%%%%%%%%%

The non-observation of proton decay places extremely stringent lower limits on the scale suppressing BNV operators, typically requiring $\Lambda_\sscript{B} \gtrsim 10^{16}\,\gev$ for generic, unsuppressed flavor structures~\cite{Beneito:2023xbk}. However, when the flavor symmetry is imposed, the structure of these operators becomes highly constrained: all sources of flavor breaking must arise through spurion insertions built from the known fermion mass matrices and mixing parameters. This naturally raises the question of whether such flavor-induced suppression can relax the lower bound on $\Lambda_\sscript{B}$, potentially bringing BNV phenomena even within the reach of current or future experiments.

To quantify the impact of flavor symmetries on BNV processes, we proceed with the analysis of the flavor invariants, with particular emphasis placed on the role of $\Upsilon_\nu$, which opens up the novel possibilities for constructing invariant structures. For each operator, we identify the transformation properties of the fermion bilinears under the flavor group $G_\sscript{F}$ and construct the minimal set of spurion insertions required to restore flavor invariance. In this construction, it is convenient to define the following composite spurionic tensors in the lepton sector
\begin{equation}
    \begin{alignedat}{2} \label{eq:leptonflavourdef}
        [\mathcal T_\ell]_{s} &\equiv  \varepsilon_{p r s}[Y_e Y_e^\dag\Upsilon_\nu]^{rp} \sim \repbar3_\ell\,,
        \\
        [\mathcal T_e]_{t} &\equiv  \varepsilon_{p r s}[Y_e Y_e^\dag\Upsilon_\nu]^{rp}[Y_e]\ud{s}{t}\sim \repbar3_e\,.
    \end{alignedat}
\end{equation}
Additional group-theoretical aspects of the flavor-invariant construction and the associated numerical values are provided in App.~\ref{app:GT_invariants}. The resulting flavor-invariant structures are summarized in Table~\ref{tab:dim6_BNV_SMEFT_Spurions}, which lists the four independent  dimension-six BNV SMEFT operators together with their minimal spurion insertions under the extended MFV framework. As a notational convention, the SMEFT Wilson coefficients are denoted by $\cC$.

Several symmetry properties of the operators directly determine their phenomenological relevance. For example, at the lowest non-vanishing order in the spurion expansion, $[\mathcal{C}_{duue}]_{prst}$ is antisymmetric under $r \leftrightarrow s$, which prevents this operator from mediating tree-level two-body proton decay~\cite{IBeneito:2025nby,Dorsner:2012nq}. Likewise, the nominally leading spurion structure $[\mathcal{C}_{qque}]_{prst} \sim \varepsilon_{prx} [Y_u]\ud{x}{s}[\cT_e]_t$ vanishes since $[\mathcal{O}_{qque}]^{prst}$ is symmetric under $p \leftrightarrow r$ and, as a result, the first non-vanishing flavor-invariant structure arises only at higher order in the spurion expansion. 

Another particularly illustrative example is provided by the operator $\cO_{qqq\ell}$, whose flavor structure is subject to nontrivial permutation relations among its quark indices. As shown in Ref.~\cite{Alonso:2014zka}, the corresponding Wilson coefficient decomposes into independent components with fully antisymmetric (A), fully symmetric (S), and mixed (M) flavor structures. In the spurion expansion, these components contribute at different orders, and in our EFT analysis we retain only the leading flavor invariants, namely those requiring the minimal number of spurion insertions. The fully antisymmetric component yields an $\cO(1)$ invariant without any spurion insertions, while the next-to-leading contribution arises from an insertion of the form $Y_u Y_u^\dagger$, leading to structures of the type $\cC_{prst}\sim \varepsilon_{prx} [Y_u Y_u^\dagger]\ud{x}{s} [\cT_\ell]_t$~\cite{Smith:2011rp}. These contributions, however, must satisfy the intrinsic flavor-operator identity~\cite{Abbott:1980zj}
\begin{equation} \label{eq:flavorrelqqql}
    [\cO_{qqq\ell}]^{prst}+[\cO_{qqq\ell}]^{rpst}=[\cO_{qqq\ell}]^{sprt}+[\cO_{qqq\ell}]^{srpt} \, ,
\end{equation}
which enforces a definite symmetry of the Wilson coefficient. Imposing this condition yields
\begin{equation}
    \cC_{prst} \sim\lzv \varepsilon_{prx} [Y_u Y_u^\dagger]\ud{x}{s} 
    - (p\leftrightarrow s) - (r\leftrightarrow s) \dzv [\mathcal T_\ell]_t \, ,
\end{equation}
which, like the spurion-free contribution, is fully antisymmetric under all index permutations. Thus, the leading invariants selected in our analysis correspond to well-defined symmetry components of the coefficient, and, as we show in the phenomenological section, their numerical impact is parametrically comparable to that of the nominal $\cO(1)$ term.

Given these structural features, it is worth highlighting that previous work, most notably Ref.~\cite{Smith:2011rp}, provided an important first exploration of MFV assumption applied to BNV operators offering valuable insight into how flavor symmetries shape the corresponding interactions. Since it was carried out prior to the development of the modern SMEFT framework, improved determinations of hadronic matrix elements, and the current precision in neutrino data, our present analysis is able to revisit and extend those results within a fully consistent spurion decomposition. This allows us to make the symmetry structure of dimension-six BNV operators explicit and to identify the complete set of non-vanishing flavor-invariant contributions.

%%%%%%%%%%%%%%%%%%%%%%%%%%%%%%%%%%%%%%%%%%%%%%%%%%%%
\subsection{Phenomenology of BNV Operators}
\label{sec:BNV_dim6_pheno}
%%%%%%%%%%%%%%%%%%%%%%%%%%%%%%%%%%%%%%%%%%%%%%%%%%%%

%%%%%%%%%%%%%%%%%%%%%%%%%%%%%%%%%%%%%%%%%%%%%
\subsubsection{Methodology}
%%%%%%%%%%%%%%%%%%%%%%%%%%%%%%%%%%%%%%%%%%%%%

With the complete set of minimal spurion structures identified, we now turn to the phenomenological analysis of the BNV operators. We first perform the tree-level SMEFT-LEFT matching for the relevant dimension-six BNV operators. At the electroweak scale, with the heavy degrees of freedom integrated out, SMEFT operators collected in Table~\ref{tab:dim6_BNV_SMEFT_Spurions} match onto a corresponding set of four-fermion operators in the low-energy effective theory (LEFT). The resulting LEFT Lagrangian contains the leading $\Delta B=\Delta L=1$ structures mediating proton decay~\cite{Nath:2006ut,Jenkins:2017jig}. The explicit operator definitions, together with their corresponding tree-level matching relations, are collected in Table~\ref{tab:SMEFT_LEFT_matching}. As a notational convention, the LEFT Wilson coefficients are denoted by $L$.

In addition, the explicit form of the SMEFT-LEFT matching relations depends on the choice of flavor basis, which determines how the fermion fields are rotated to their mass eigenstates after electroweak symmetry breaking. In the quark sector, one may work either in the up or down basis, with the two choices being related by CKM rotations of the left-handed quark fields:
\begin{equation}
\label{eq:up_down_basis}
    \mathrm{up:}~~d_{L,p}\to V_{pr}\,d_{L,r}\,,
    \qquad
    \mathrm{down:}~~u_{L,p}\to V^\dag_{pr}\,u_{L,r}\,,
\end{equation}
where $V$ denotes the CKM matrix. Consequently, the CKM matrix elements enter the flavor contractions of the BNV operators differently depending on the chosen basis. As input for quark masses and CKM parameters, we use the most recent PDG values~\cite{ParticleDataGroup:2024cfk}. In the remainder of the phenomenological analysis, we consider both arrangements to illustrate how the resulting CKM insertions affect the phenomenological predictions, in particular the relative magnitudes of the decay widths associated with different BNV channels.

Moving forward, in order to perform the computation of nucleon decay rates, the effective four-fermion interactions derived in the LEFT must be further matched onto hadronic degrees of freedom at low energies, where quarks and gluons are replaced by baryons and mesons. This matching is performed within the framework of baryon chiral perturbation theory (B$\chi$PT), which provides a systematic low-energy expansion of QCD~\cite{Claudson:1981gh,Beneito:2023xbk}. The quark-level BNV operators are thus translated into hadronic operators involving nucleon and meson fields, parameterized by a small number of low-energy constants that encode non-perturbative QCD dynamics, which are determined from lattice QCD calculations or phenomenological fits. In particular, for the nucleon decay channels relevant to our analysis, parity conservation of the strong interactions and isospin symmetry imply that only two independent nucleon matrix elements contribute.\footnote{Alternatively, Refs.~\cite{Yoo:2021gql,Aoki:2017puj} determine proton-decay matrix elements directly from lattice QCD, bypassing B$\chi$PT.}

For the derivation of the proton-decay bounds, which constitute the main results of this work, we employ the experimental lower limits on the relevant two-body proton-decay channels as reported in Table~1 of Ref.~\cite{Beneito:2023xbk}. While our numerical analysis covers all accessible two-body nucleon decay channels into pseudoscalar mesons and anti-lepton states, the dominant experimental constraints stem from the $p\to K^+\nu$, $p\to \pi^0 \ell_r^+$, and $p\to K^0\ell^+_r$ modes. When evaluated within the extended MFV framework, these channels provide the strongest sensitivity to the BNV operators considered in this work. For these modes, we present analytical expressions for the partial decay widths in terms of the LEFT WCs, providing a transparent connection between the underlying flavor structures and the observable nucleon decay rates~\cite{Nath:2006ut,Beneito:2023xbk}:
\begin{widetext}
    \begin{equation}
    \label{eq:decay_widths_3ch}
        \begin{alignedat}{2}
            &\Gamma(p\to K^+\nu_r)=\frac{(m_p^2-m_K^2)^2}{32\pi f_\pi^2 m_p^3}
            \Biggl|\alpha [L^{S,RL}_{dud}]_{112r}-\beta [L^{S,LL}_{udd}]_{112r}+\frac{m_p}{2m_\Sigma}(D-F)\lzm \alpha [L^{S,RL}_{dud}]_{211r}-\beta[L^{S,LL}_{udd}]_{121r} \dzm
            \\&
            ~~~~~~~~~~~~~
            ~~~~~~~~~~~~~~~~~~~~~~~~~~
            +\frac{m_p}{6m_\Lambda}(D+3F)\lzm \alpha[L^{S,RL}_{dud}]_{211r}-\beta[L^{S,LL}_{udd}]_{121r}+2\alpha[L^{S,RL}_{dud}]_{112r}-2\beta[L^{S,LL}_{udd}]_{112r} \dzm\Biggr|^2
            \,,
            \\[0.4cm]
            &\Gamma(p\to \pi^0\ell^+_r)=\frac{(m_p^2-m_\pi^2)^2}{32\pi f_\pi^2 m_p^3}\frac{(1+D+F)^2}{2}\Biggl\{ \Biggl|\alpha[L^{S,RL}_{duu}]_{111r}+\beta[L^{S,LL}_{duu}]_{111r}\Biggr|^2+\Biggl|\alpha[L^{S,LR}_{duu}]_{111r}+\beta[L^{S,RR}_{duu}]_{111r}\Biggr|^2 \Biggr\}
            \,,
            \\[0.4cm]
            &\Gamma(p\to K^0\ell^+_r)=\frac{(m_p^2-m_K^2)^2}{32\pi f_\pi^2 m_p^3}
            \Biggl\{
            \Biggl|\beta[L^{S,LL}_{duu}]_{211r}-\alpha[L^{S,RL}_{duu}]_{211r}+\frac{m_p}{m_\Sigma}(D-F)\left(\beta[L^{S,LL}_{duu}]_{211r}+\alpha[L^{S,RL}_{duu}]_{211r}\right)\Biggl|^2
            \\
            &~~~~~~~~~~~~~~~~~~~~~~~~~~~~~~~~~~~~~~
            + \Biggl|\beta[L^{S,RR}_{duu}]_{211r}-\alpha[L^{S,LR}_{duu}]_{211r}+\frac{m_p}{m_\Sigma}(D-F)\left(\beta[L^{S,RR}_{duu}]_{211r}+\alpha[L^{S,LR}_{duu}]_{211r}\right)\Biggl|^2\Biggl\}\,.
        \end{alignedat}
    \end{equation}
\end{widetext}
In Eq.~\eqref{eq:decay_widths_3ch}, $m_X$ denotes the mass of the hadron or meson, and $f_\pi=139$ MeV is the pion decay constant. The parameters $\alpha$ and $\beta$ are real low-energy constants that parameterize the two independent nucleon matrix elements arising from the three-quark operators. Their numerical values, extracted from lattice QCD, are taken as $\alpha=-0.01257(111)\,\gev^3$ and $\beta=0.01269(107)\,\gev^3$~\cite{Yoo:2021gql,Aoki:2017puj}. The coefficients $D$ and $F$ denote the axial couplings of the baryon octet, which appear in the B$\chi$PT Lagrangian. These constants can be obtained either phenomenologically, with $D=0.80(1)$ and $F=0.47(1)$~\cite{Aoki:2008ku}, or from lattice computations, which yield $D=0.730(11)$ and $F=0.447^{(6)}_{(7)}$~\cite{Bali:2022qja}. Also note that for neutrino final states, the decay width implicitly sums over the three neutrino flavors ($r=e,\mu,\tau$), which are experimentally indistinguishable.

\begin{figure}[t]
    \centering
    \includegraphics[width = 0.9\linewidth]{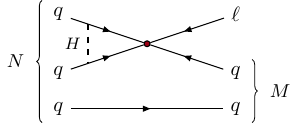}
    \caption{Representative one-loop diagram contributing to two-body proton decay, $N \to M \ell$, where $N$ denotes a nucleon decaying into a pseudoscalar meson $M$ and a lepton $\ell$. The red blob indicates a BNV vertex involving fermions of heavier generations, generated via the spurion expansion of a generic dimension-six BNV operator. The dashed line corresponds to the exchange of a SM Higgs boson, which mediates the flavor-changing transition to light-generation fermions.}
    \label{fig:1loopprotondecay}
\end{figure}

As an important complementary step to the matching framework outlined above, our analysis incorporates the full renormalization-group (RG) evolution of the BNV operators down to hadronic scales. This evolution is carried out using \texttt{DsixTools}~\cite{Celis:2017hod,Fuentes-Martin:2020zaz}, which includes the one-loop anomalous dimensions of the BNV operators~\cite{Alonso:2014zka} and consistently implements the SMEFT-LEFT matching conditions. Radiative effects play a crucial role, as operator mixing at one loop generates the LEFT structures that mediate proton decay even when absent at tree level; a representative contribution is displayed in Fig.~\ref{fig:1loopprotondecay}. In cases where two-loop effects are required to connect high-scale operators to low-energy observables~\cite{Banik:2025wpi}, sequential one-loop running provides an efficient approximation, evolving the coefficients down to the $2\,\gev$ scale relevant for hadronic matrix elements. This mechanism is especially relevant for certain flavor structures appearing in the reduced flavor symmetries described in Sec.~\ref{sec:MFV_alter}. A representative example is the operator $[\cO_{duue}]^{3333}$, for which two SM Higgs exchanges are necessary to generate LEFT Wilson Coefficients entering nucleon decay rates in Eq.~\eqref{eq:decay_widths_3ch}. As highlighted in Ref.~\cite{Gisbert:2024sjw}, these contributions scale as $\left(\frac{1}{16\pi^2}\log\frac{\Lambda^2}{m_\sscript{EW}^2} \right)^2$.

%%%%%%%%%%%%%%%%%%%%%%%%%%%%%%%%%%%%%%%%%%%%%
%%%%%%%%%%%%%%%%%%%%%%%%%%%%%%%%%%%%%%%%%%%%%

%%%%%%%%%%%%%%%%%%%%%%%%%%%%%%%%%%%%%%%%%%%%%
\subsubsection{Results}
%%%%%%%%%%%%%%%%%%%%%%%%%%%%%%%%%%%%%%%%%%%%%

%%%%%%%%%%%%%%%%%%%%%%%%%%%%%%%%%%%%%%%%%%%%%
\begin{table*}[t]
    \input{tables/Dim6_BNV_table_bounds}
    \caption{Summary of the leading proton-decay constraints for the dimension-six BNV operators and their minimally spurion-suppressed flavor completions under the extended MFV hypothesis. The first column lists the effective operators, while the second displays the corresponding spurion expansion. The third (fifth) column reports the bounds on the ratio $\cR_{\sscript{BL}}\equiv\Lambda_\sscript B^2 / \Lambda_\sscript L$ in the up (down) basis, denoted as $\cR_{\sscript{BL}}^{\sscript{up}}$ ($\cR_{\sscript{BL}}^{\sscript{down}}$), while the adjacent columns indicate the decay channel that provides the most stringent limit in each case. We note that the RG effects between $\Lambda_{\sscript B}$ and $m_p$ induce a mild logarithmic dependence, which slightly affects the bound on $\cR_{\sscript{BL}}$. For $\cO_{duq\ell}$ (up basis), the $p \to \pi^0\mu^{+}$ limit is accompanied by $p \to K^+\nu$, as shown in Fig.~\ref{fig:ratios}. For $\cO_{qque}$ (down basis), the constraint from $p \to \pi^0\mu^{+}$ is accompanied by $p \to K^{0}\mu^{+}$, which exhibits comparable sensitivity. Limits quoted in the table are obtained using $\Upsilon_\nu$ evaluated for the NO spectrum.}
    \label{tab:dim6bnvtablelim}
\end{table*}
%%%%%%%%%%%%%%%%%%%%%%%%%%%%%%%%%%%%%%%%%%%%%

The main results of the numerical analysis are summarized in Table~\ref{tab:dim6bnvtablelim}. The corresponding bounds are expressed in terms of the effective ratio $\cR_\sscript{BL}=\Lambda_\sscript{B}^2/\Lambda_\sscript{L}$, which connects the two UV scales. All results in Table~\ref{tab:dim6bnvtablelim} are quoted for the NO spectrum. At the precision relevant for this analysis, the bounds on the ratio $\cR_\sscript{BL}$ in the IO case can be obtained by applying the inverse of the rescaling associated with the leptonic flavor invariants in NO and IO, as given in Eqs.~\eqref{eq:numvalTl} and~\eqref{eq:numvalTe} of App.~\ref{app:GT_invariants}. Consequently, for operators involving $\mathcal{T}_\ell$ ($\mathcal{T}_e$), the IO bound on $\mathcal{R}_\sscript{BL}$ is scaled by a factor of $0.8$ ($1.7$) relative to the NO one. For each operator we report the strongest bound obtained in both the up and down flavor bases, corresponding to the most constraining decay channel in each case. 

\begin{figure*}[t!]
\centering
\begin{tabular}{cc}
\includegraphics[width=80mm]{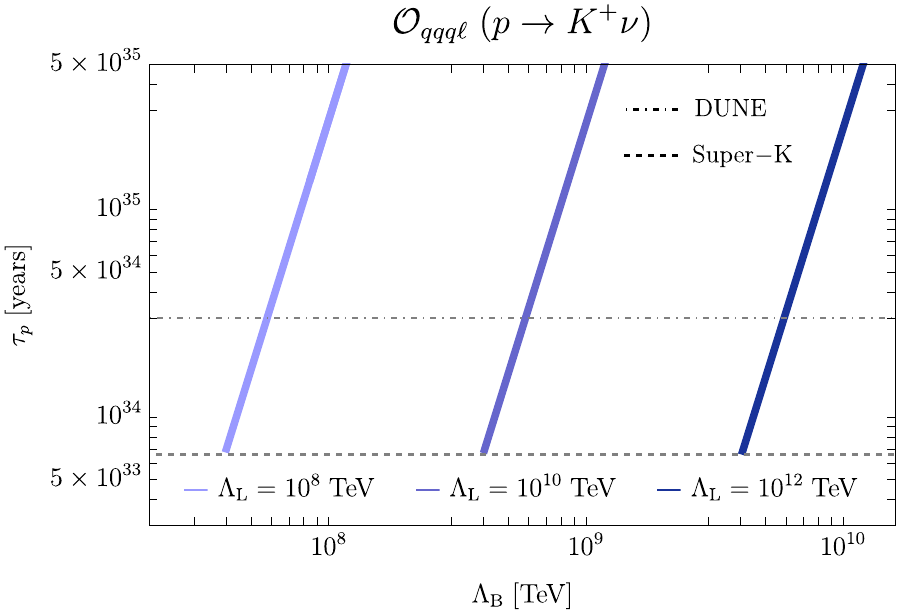} 
% &\qquad\quad
&\hspace{+0.5cm}
\includegraphics[width=79mm]{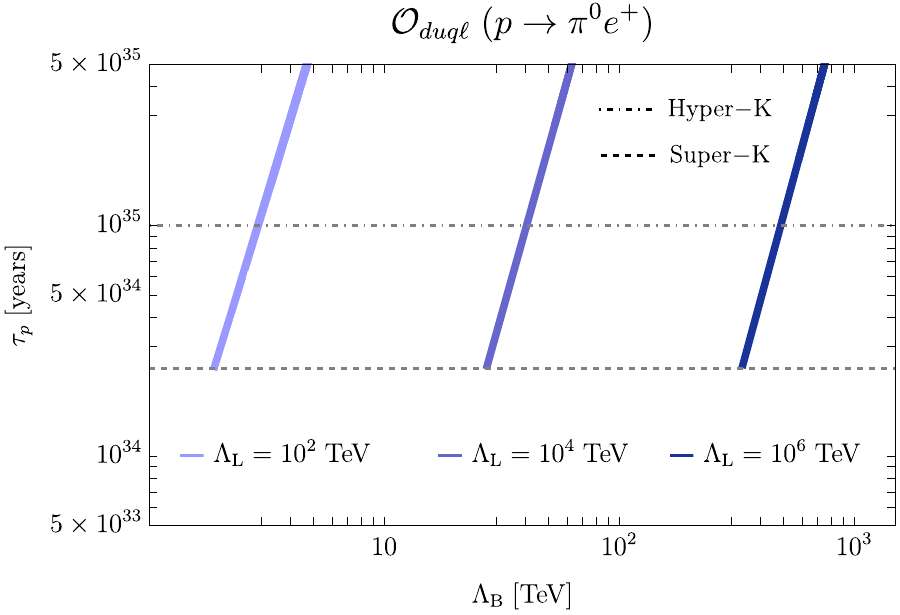} 
\\[5pt]
\includegraphics[width=79mm]{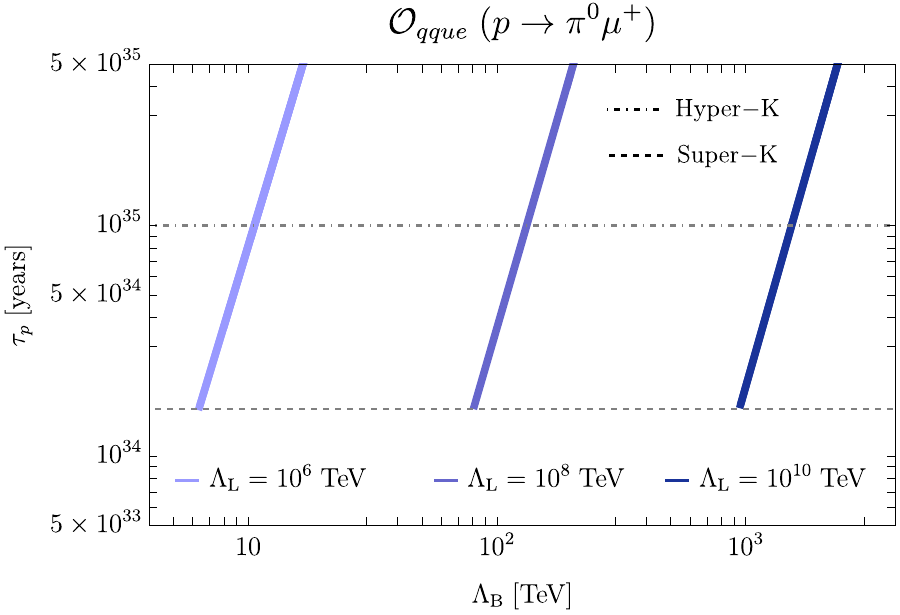}
% &\qquad\quad
&\hspace{+0.5cm}
\includegraphics[width=80mm]{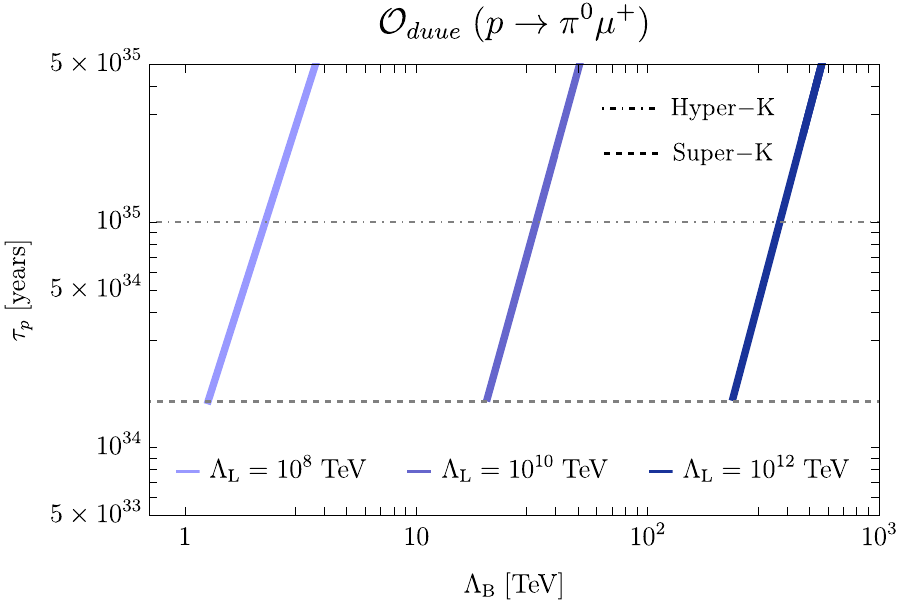}
\end{tabular}
\caption{Predicted proton lifetimes $\tau_p$ from the leading spurion expansion of baryon-number violating operators $\mathcal{O}_{qqq\ell}$, $\mathcal{O}_{duq\ell}$, $\mathcal{O}_{qque}$, and $\mathcal{O}_{duue}$ from Table~\ref{tab:dim6_BNV_SMEFT_Spurions}. The dashed and dot-dashed curves represent current (Super-K) and projected (Hyper-K, DUNE) experimental sensitivities, respectively. For each operator, we vary $\Lambda_{\sscript L}$ across three benchmark values to map the allowed $\Lambda_{\sscript B}$ parameter space. The limiting decay channel for each case follows Table~\ref{tab:dim6bnvtablelim}, with all lifetimes computed in the up-quark basis for concreteness.}
\label{fig:BNV_plots_proj}
\end{figure*}

Additionally, the overall impact of the extended MFV framework on the allowed BNV parameter space in the context of present and future experiments is illustrated in Fig.~\ref{fig:BNV_plots_proj}. For each of the four dimension-six BNV operators, we use the leading spurion structures identified above and compute the proton lifetime $\tau_p$ as a function of the BNV scale $\Lambda_\sscript{B}$, while fixing $\Lambda_\sscript{L}$ to several benchmark values displayed in the figure obeying the consistency relation $\Lambda_\sscript{B} \leq \Lambda_\sscript{L}$. The diagonal bands show the predicted lifetimes for fixed values of $\Lambda_\sscript{L}$, while the horizontal lines indicate the current limits from Super-K and the projected sensitivities of Hyper-K~\cite{Hyper-Kamiokande:2018ofw} for the $p \to \pi^{0}\ell^{+}$ channel and of DUNE~\cite{DUNE:2020ypp} for the $p \to K^{+}\nu$ mode. The intersections of these curves determine the minimal values of $\Lambda_\sscript{B}$ compatible with a given choice of $\Lambda_\sscript{L}$.

These results reveal a substantial spread in the BNV scales associated with different operators. For three of the four operators, in particular $\cO_{duq\ell}$, $\cO_{qque}$, and $\cO_{duue}$, whose leading contribution to proton decay is induced radiatively at one-loop level (see the following subsection for details), proton-decay limits are compatible with BNV scales in the multi-TeV range, provided the LNV scale $\Lambda_\sscript{L}$ lies around $\cO(10^{3}\,\tev)$, $\cO(10^{6}\,\tev)$, and $\cO(10^{9}\,\tev)$, respectively. In contrast, $\cO_{qqq\ell}$ remains exceptionally constrained, requiring $\Lambda_\sscript{B}\gtrsim 10^{7}$ TeV. Although the values of $\Lambda_\sscript{L}$ inferred in this way are considerably lower than the naive expectation $\Lambda_\sscript{L}\sim\cO(10^{12}\,\tev)$ suggested by a minimal type-I seesaw origin of $\Upsilon_\nu$~\cite{Minkowski:1977sc,Cirigliano:2005ck}, such hierarchies arise naturally in alternative neutrino-mass constructions. Examples include type-II seesaw models~\cite{Magg:1980ut,Cheng:1980qt} and inverse-seesaw scenarios~\cite{Wyler:1982dd}, where the smallness of neutrino masses is governed by a technically natural source of LNV~\cite{tHooft:1980xss}, allowing the corresponding scale to be significantly lower than the naive expectation without introducing substantial fine-tuning.

\begin{figure}[t]
    \centering
    \includegraphics[width = 1.0\linewidth]{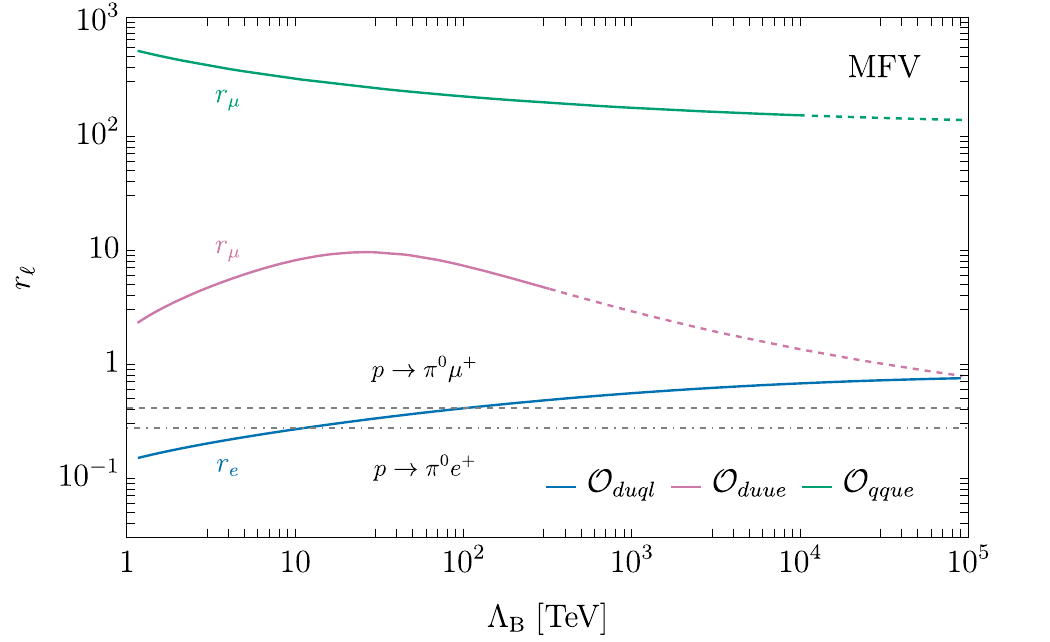}
    \caption{Ratio of the proton-decay channels $p \to \pi^0 \ell^+$ and $p \to K^+ \nu$ defined in Eq.~\eqref{eq:ratio_r_def} as a function of $\Lambda_{\sscript B}$ for the three spurion expansions relevant in the MFV framework (see Table~\ref{tab:dim6_BNV_SMEFT_Spurions}).
    The predictions for $\mathcal{O}_{duue}$ and $\mathcal{O}_{qque}$ correspond to the muonic channel, while those for $\mathcal{O}_{duq\ell}$ correspond to the electron mode $p \to \pi^{0} e^{+}$. The horizontal dashed (dot--dashed) gray lines indicate the experimental values from Super-K for the electronic (muonic) channel; operator curves lying above (below) these lines are therefore more tightly constrained by $p \to \pi^{0}\ell^{+}$ ($p \to K^{+}\nu$), in agreement with Table~\ref{tab:dim6bnvtablelim}. For $\mathcal{O}_{duue}$ and $\mathcal{O}_{qque}$, solid (dashed) segments denote the regions where $\Lambda_{\sscript{L}}$ is below (above) $10^{12}\,\mathrm{TeV}$, with the latter being the maximal scale compatible with a weakly-coupled regime of a UV seesaw origin of $\Upsilon_\nu$. The dashed portions remain phenomenologically viable, but observable proton decay in this regime would require $\Lambda_{\sscript{L}} >10^{12} \,\mathrm{TeV}$. All decay widths are computed in the up-quark basis; see the main text for details.}
    \label{fig:ratios}
\end{figure}

Lastly, to further quantify the impact of the different flavor structures listed in Table~\ref{tab:dim6bnvtablelim}, it is instructive to examine the dominant decay widths associated with the \textit{golden channels} $p \to \pi^0 \ell_r^+$ and $p \to K^+ \nu$, whose lifetime limits are expected to improve about an order of magnitude in upcoming experiments~\cite{Hyper-Kamiokande:2018ofw,DUNE:2020ypp}. A convenient way to display their relative importance is through the predicted ratio
\begin{equation}\label{eq:ratio_r_def}
r_\ell \equiv \frac{\Gamma(p \to \pi^0 \ell^+)}{\Gamma(p \to K^+ \nu)}\,,
\qquad
\ell\in\{e,\mu\}\,,
\end{equation}
expressed as a function of $\Lambda_{\sscript B}$. This ratio renders the renormalization-group dependence particularly transparent and eliminates the explicit dependence on $\Lambda_{\sscript L}$. In contrast, the projections shown in Fig.~\ref{fig:BNV_plots_proj} necessarily depend on the choice of $\Lambda_{\sscript L}$ in order to quantify experimental sensitivities. Fig.~\ref{fig:ratios} shows the ratio $r_\ell$ for the leading spurion structures of $\cO_{duq\ell}$, $\cO_{duue}$, and $\cO_{qque}$. We omit $\cO_{qqq\ell}$, whose ratio is strongly suppressed ($\sim10^{\eminus4}$) owing to the fact that the neutrino mode is generated at tree level, whereas the charged-lepton channel is generated only radiatively. As discussed in Sec.~\ref{sec:flavor_invariant_BNV_ops}, constructing MFV flavor invariants for the BNV operators necessarily involves inserting the leptonic triplet spurions $\cT_e$ or $\cT_\ell$. Numerically, the dominant entry of $\cT_\ell$ is always the component associated with $e$ or $\nu_e$, whereas for $\cT_e$ the second component provides the leading contribution for both NO and IO (see App.~\ref{app:GT_invariants}). Consequently, operators built with $\cT_e$ predominantly yield muonic final states. For this reason, in Fig.~\ref{fig:ratios} we take $\ell=\mu$ for $\cO_{duue}$ and $\cO_{qque}$, while for $\cO_{duq\ell}$ we set $\ell=e$. Within this MFV framework, observing a proton-decay mode with a $\mu^+$ in the final state, rather than an $e^+$ or a neutrino, would naturally single out $\cO_{duue}$ or $\cO_{qque}$ as the underlying operators. Conversely, if BNV were mainly induced by $\cO_{duq\ell}$ with $\Lambda_\sscript{B}$ near the multi-TeV regime, one would anticipate similar decay rates for $p \to \pi^0 e^+$ and $p \to K^+ \nu$.

%%%%%%%%%%%%%%%%%%%%%%%%%%%%%%%%%%%%%%%%
%%%%%%%%%%%%%%%%%%%%%%%%%%%%%%%%%%%%%%%%

%%%%%%%%%%%%%%%%%%%%%%%%%%%%%%%%%%%%%%%%%%%%%
%%%%%%%%%%%%%%%%%%%%%%%%%%%%%%%%%%%%%%%%%%%%%

%%%%%%%%%%%%%%%%%%%%%%%%%%%%%%%%%%%%%%%%%%%%%
\subsubsection{Operator-by-Operator Discussion}
%%%%%%%%%%%%%%%%%%%%%%%%%%%%%%%%%%%%%%%%%%%%%

In what follows, we briefly discuss the phenomenological implications of the individual BNV operators and the features of their dominant flavor structures.

%%%%%%%%%%%%%%%%%%%%%%%%%%%%%%
\vspace{0.2cm}
\noindent
%%%%%%%%%%%%%%%%%%%%%%%%%%%%%%
{$\bm{\cO_{qqq\ell}.}$} Within the spurion counting, the leading flavor structures are $\cO(Y_e^2\Upsilon_\nu)$ and, at the next-to-leading order, $\cO(Y_u^2 Y_e^2\Upsilon_\nu)$. Although the latter carries an extra $Y_u^2$ insertion, the dominance of the top Yukawa ($y_t\simeq 1$) renders the numerical suppression modest, so both contributions are of comparable size. These structures contribute already at tree level to the proton decay, with the most constraining channel being $p\to K^+\nu_r$. Since the $\cO(Y_e^2\Upsilon_\nu)$ invariant does not involve quark-Yukawa insertions at the SMEFT level, no distinction between the up and down bases emerges until the subsequent matching onto the LEFT. Nonetheless, the resulting LEFT Wilson coefficients turn out to be identical in the two bases. In the up basis these coefficients can be expressed as
\begin{equation}
    [L^{S,LL}_{udd}]_{112r}=-[L^{S,LL}_{udd}]_{121r}=\frac{3[\mathcal T_\ell]_r}{\Lambda_\sscript{B}^2}\lzm V_{td}V_{cs}-V_{ts}V_{cd} \dzm\,,
\end{equation}
whereas in the down basis they are given by
\begin{equation}
    [L^{S,LL}_{udd}]_{112r}=-[L^{S,LL}_{udd}]_{121r}=- \frac{3[\mathcal T_\ell]_r}{\Lambda_\sscript{B}^2} V_{ub}^*\,.
\end{equation}
This equality follows from a general identity for $n\times n$ matrices. For any matrix $\mathbf{V}$, the adjugate satisfies $\mathbf{V}\operatorname{adj}(\mathbf{V}) = \det(\mathbf{V})\,\mathbf{I}$. When $\mathbf{V}$ is unitary, this reduces to $\operatorname{adj}(\mathbf{V}) = \mathbf{V}^\dagger$. Evaluating the $(1,3)$ component then yields the relation
\begin{equation}
    V_{ts}V_{cd} -V_{td}V_{cs} = V_{ub}^* \,,
\end{equation}
which precisely equates the CKM combinations appearing in the two expressions above.

For the $\cO(Y_u^2 Y_e^2\Upsilon_\nu)$ structure, the SMEFT invariant is already basis dependent: in the up basis one obtains a contraction proportional to $\sim\varepsilon_{pr3}\,y_t^2\delta_{s3}$, whereas in the down basis the same structure transforms as $\sim \varepsilon_{prx}\,y_t^2 V^{*}_{tx}V_{ts}$. Analogously to the $\cO(Y_e^2\Upsilon_\nu)$ invariant, the LEFT coefficients entering $p \to K^+ \nu_r$ in the up basis are given by
\begin{equation}
    [L^{S,LL}_{udd}]_{112r}=-[L^{S,LL}_{udd}]_{121r}=\frac{3y_t^2[\mathcal T_\ell]_r}{\Lambda_\sscript{B}^2}\lzm V_{td}V_{cs}-V_{ts}V_{cd} \dzm\,,
\end{equation}
while in the down basis the corresponding expression is obtained by the substitutions $V_{ts}V_{cd} - V_{td}V_{cs} \to V_{ub}^*$ and $y_t^2 \to y_t^2 |V_{tb}|^2$. Given that $y_t,V_{tb} \simeq 1$, these modifications leave the overall magnitude essentially unchanged. The resulting LEFT coefficients are therefore of the same order as those generated by the $\cO(Y_e^2 \Upsilon_\nu)$ structure, implying comparable sensitivities in the corresponding decay channels (see Table~\ref{tab:dim6bnvtablelim}). In the full numerical analysis, we incorporate the RG evolution between the baryon-violating scale $\Lambda_\sscript{B}$ and the hadronic scale $\mu\simeq 2\,\gev$, as implemented in \texttt{DsixTools}. At tree level, the effective ratio $\cR_{\sscript{BL}}^\sscript{up}\equiv\Lambda^2_\sscript{B}/\Lambda_\sscript{L}$ is approximately $6\times10^6\,\tev$, whereas the RG effects increase this value by a factor of $\sim2$, a typical enhancement for BNV operators whose anomalous dimensions are dominated by self-running~\cite{Beneito:2023xbk}.

%%%%%%%%%%%%%%%%%%%%%%%%%%%%%%
\vspace{0.2cm}
\noindent
%%%%%%%%%%%%%%%%%%%%%%%%%%%%%%
{$\bm{\cO_{duq\ell}.}$} The dominant spurion structure for this operator, arising at $\cO(Y_u Y_d Y_e^2 \Upsilon_\nu)$, already induces BNV at tree level, giving rise to the $p\to K^+\nu_r$ channel, with the corresponding tree-level LEFT WCs evaluated in the up alignment as
\begin{equation}
    \begin{alignedat}{2}
        [L^{S,RL}_{dud}]_{112r}&=-\frac{2m_um_d}{v^2\Lambda^2_\sscript{B}}[\mathcal T_\ell]_r\lzm V_{ts}V_{cd}-V_{cs}V_{td} \dzm\,,
        \\
        [L^{S,RL}_{dud}]_{211r}&=-\frac{2m_um_s}{v^2\Lambda^2_\sscript{B}}[\mathcal T_\ell]_r\lzm V_{td}V_{cs}-V_{ts}V_{cd} \dzm\,.
    \end{alignedat}
\end{equation}
These coefficients remain non-vanishing in either flavor basis, as the CKM factors enter directly through the SMEFT-LEFT matching (see Table~\ref{tab:SMEFT_LEFT_matching} and Eq.~\eqref{eq:up_down_basis}). In addition to the neutrino channel, this operator also generates a charged-lepton mode $p \to K^0 \ell_r^+$, which arises at tree level only in the down alignment:
\begin{equation}
    [L^{S,RL}_{duu}]_{211r}=-\frac{2m_um_s}{v^2\Lambda^2_\sscript{B}}V^*_{ub}[\mathcal T_\ell]_{r}\,,
\end{equation}
while in the up alignment the corresponding coefficient vanishes due to the contraction of the form $\varepsilon_{ab1}[\widehat Y_u]\ud{a}{1}[V\widehat Y_d]\ud{b}{2}$. In contrast, the tree-level amplitude for $p \to \pi^0 \ell_r^+$ vanishes in both flavor bases. However, from a phenomenological standpoint, this mode provides the strongest experimental sensitivity among all two-body decay channels. Although absent at tree level, RG evolution radiatively induces the corresponding effective operators, and the resulting contributions ultimately set the most stringent bounds on this class of interactions. As an explicit example, following the SMEFT-LEFT matching in the up-aligned basis, the non-vanishing coefficients relevant for the $p \to \pi^0 \ell_r^+$ transition in the leading-log approximation can be written as
\begin{equation}
    \begin{alignedat}{2}
    [L^{S,RL}_{duu}]_{111r}&\simeq \frac{m_um_dm_sm_b}{4\pi^2 v^4\Lambda_\sscript{B}^2}[\mathcal T_\ell]_rV_{tb}V_{td}V^*_{us}\log\frac{\mu_f}{\Lambda_\sscript{B}}\,,
    \\
    [L^{S,LL}_{duu}]_{111r}&\simeq-\frac{m_dm_sm_t^2}{2\pi^2v^4\Lambda^2_\sscript{B}}[\mathcal T_\ell]_r V_{cd}V_{cs}V_{td}\log\frac{\mu_f}{\Lambda_\sscript{B}}\,,
    \\
    [L^{S,LR}_{duu}]_{111r}&\simeq \frac{m_um_dm_b m_{\ell_r}}{4\pi^2 v^4\Lambda^2_\sscript{B}}[\mathcal T_\ell]_r V_{cd}V_{tb}V_{td}\log\frac{\mu_f}{\Lambda_\sscript{B}}\,,
    \\
    \end{alignedat}
\end{equation}
where $\mu_f$ denotes the renormalization scale at which the SMEFT is matched onto the LEFT, typically taken to be of $\cO(\mu_{\sscript{EW}})$.

%%%%%%%%%%%%%%%%%%%%%%%%%%%%%%
\vspace{0.2cm}
\noindent
%%%%%%%%%%%%%%%%%%%%%%%%%%%%%%
{$\bm{\cO_{qque}.}$} The leading spurion structure associated with this operator, which is of $\mathcal{O}(Y_u^3 Y_e^3 \Upsilon_\nu)$, involves solely the insertions of up-type Yukawa matrices. Consequently, its phenomenological impact depends explicitly on the choice of flavor basis, exhibiting distinct structures in the up and down alignments. Starting with the down basis, where essentially all flavor components of the operator remain non-vanishing, BNV transition occurs already at tree level. The dominant contribution arises through the  $p \to \pi^0 \ell_r^+$ channel, with the leading LEFT WC given by
\begin{equation}
    [L^{S,LR}_{duu}]_{111r}=-\frac{4\sqrt{2}\,m_t^2m_c}{v^3\Lambda^2_\sscript{B}}V_{td}V_{cs}^*V_{tb}^*V_{ud}^*V_{us}^*[\mathcal T_e]_{r}\,.
\end{equation}
Given that the components of the leptonic spurion satisfy $|[\mathcal T_e]_2/[\mathcal T_e]_1|_{\sscript{NO}(\sscript{IO})} \simeq 70\,(40)$ (see App.~\ref{app:GT_invariants}), the current Super-K limits, whose sensitivities are comparable for final-state $e^+$ and $\mu^+$, imply that the $p \to \pi^0 \mu^+$ channel provides the strongest constraint among the corresponding decay modes~\cite{Super-Kamiokande:2020wjk}.

Conversely, in the up basis, only a limited subset of SMEFT WCs remain non-vanishing at the high scale: 
\begin{equation}
    \begin{alignedat}{2}
        [\mathcal{C}_{qque}]_{123r} &= [\mathcal{C}_{qque}]_{213r}
        =-\frac{\sqrt2 m_t(m_c^2-m_u^2)}{v^3\Lambda^2_\sscript{B}}[\cT_e]_r
        \,,
        \\
        [\mathcal{C}_{qque}]_{132r} &= [\mathcal{C}_{qque}]_{312r}=\frac{\sqrt2 m_c(m_t^2-m_u^2)}{v^3\Lambda^2_\sscript{B}}[\cT_e]_r
        \,,
        \\
        [\mathcal{C}_{qque}]_{321r} &= [\mathcal{C}_{qque}]_{231r}=-\frac{\sqrt2 m_u(m_t^2-m_c^2)}{v^3\Lambda^2_\sscript{B}}[\cT_e]_r\,.
    \end{alignedat}
\end{equation}
Upon SMEFT-LEFT matching, no tree-level LEFT operators mediating two-body proton decay are induced. In the up basis, BNV therefore emerges only via operator mixing, leading to a weaker constraint on $\mathcal{R}_{\sscript{BL}}^{\sscript{up}}$, as shown in Table~\ref{tab:dim6bnvtablelim}. In particular, in the leading-log approximation, the only non-vanishing LEFT coefficient contributing to the $p \to \pi^0 \ell^+_r$ channel is given by
\begin{equation}
    [L^{S,LR}_{duu}]_{111r}\simeq\frac{m_u m_d m_bm_t^2}{4\sqrt2 \pi v^5\Lambda^2_\sscript{B}} [\cT_e]_r V_{cd}^2V_{cb}^*\log\frac{\mu_f}{\Lambda_\sscript{B}}
    \,.
\end{equation}
In conclusion, the $p \to \pi^0 \mu^+$ channel remains the dominant probe in either basis. The additional $Y_e$ factor encoded in $\cT_e$ imposes an extra parametric suppression on the relevant LEFT coefficient, resulting in a comparatively weaker bound.

%%%%%%%%%%%%%%%%%%%%%%%%%%%%%%%%%%%%%%%%%%%%%
\begin{table*}[t]
    \input{tables/LQ_overview_table}
    \caption{Overview of the leptoquark (LQ) mediators considered in our UV analysis, using the notation of Ref.~\cite{deBlas:2017xtg}. The first column lists the mediators and their transformation properties under the SM gauge group. The second column indicates whether the given LQ gauge irrep is a scalar ($S$) or vector ($V$) field. The third column presents the allowed renormalizable interaction terms involving the LQs and SM fermions, written without imposing any flavor symmetry assumptions. The final column lists the BNV operators that result from integrating out the corresponding LQ mediator at tree level. As in Table~\ref{tab:dim6_BNV_SMEFT_Spurions}, Greek indices $\alpha,\beta,\gamma$ label $\SU(3)_\sscript{C}$, while $p,r,s$ represent flavor indices, with $p,r$ for fermions and $s$ for the mediator. In case of $\zeta$ mediator, $A$ denotes the $\SU(2)_\sscript{L}$ adjoint index, whereas the remaining $\SU(2)_\sscript{L}$ contractions are left implicit. For couplings involving identical fermion fields, the associated flavor tensors are subject to additional symmetry relations, which constrain the number of independent flavor structures. In particular, $[y_{\omega_1}^{qq}]^{spr}$ flavor tensor is symmetric, whereas $[y_{\omega_4}^{uu}]^{spr}$ and $[y_\zeta^{qq}]^{spr}$ are antisymmetric under the exchange of quark indices $p\leftrightarrow r$.}
    \label{tab:LQ_overview}
\end{table*}
%%%%%%%%%%%%%%%%%%%%%%%%%%%%%%%%%%%%%%%%%%%%%

%%%%%%%%%%%%%%%%%%%%%%%%%%%%%%
\vspace{0.2cm}
\noindent
%%%%%%%%%%%%%%%%%%%%%%%%%%%%%%
{$\bm{\cO_{duue}.}$}  The lowest-order spurion structure associated with this operator appears at $\mathcal{O}(Y_u Y_d \, Y_e^2\, \Upsilon_\nu)$. An important feature of this combination is that its basis dependence is significantly reduced compared to the spurion structures arising in the other dimension-six BNV operators: the factor $Y_u^\dag Y_d$ transforms covariantly under flavor rotations and, upon switching between the up- and down-aligned bases, it is simply rewritten as $Y_u^\dag Y_d\to \widehat Y_u^\dag V\widehat Y_d$. As a result, the overall spurion structure extracted from this operator is effectively the same in both bases. 

At the level of the low-energy theory, $\cO_{duue}$ matches onto the $\cO^{S,RR}_{duu}$ scalar operator (see Table~\ref{tab:SMEFT_LEFT_matching}), which contributes to the $p\to\pi^0 \ell_r^+$ and $p\to K^+ \ell_r^+$ channels. However, due to the antisymmetric $\varepsilon$-tensor contraction in the SMEFT operator, all tree-level LEFT coefficients leading to two-body proton decay vanish, independently of the chosen flavor basis.\footnote{In configurations involving higher-generation right-handed up-type quarks~\cite{Hou:2005iu} or bottom quarks~\cite{Beneke:2024hox} the heavy quark appears off shell and hadronizes into multi-body final states, giving rise to proton-decay channels that are phase-space suppressed and experimentally subleading. An analogous suppression arises for operators involving tau leptons~\cite{Marciano:1994bg,Heeck:2024jei,Crivellin:2023ter}.} Consequently, any phenomenological impact of this operator arises only radiatively through operator mixing.

In the leading-log approximation, the radiatively generated LEFT coefficients relevant for the two dominant decay channels take different forms depending on the flavor alignment. In particular, in the up-aligned basis, the non-vanishing contributions are 
\begin{equation}
    \begin{alignedat}{2}
        [L^{S,LR}_{duu}]_{111r}&\simeq-\frac{m_dm_cm_bm_t}{4\pi^2v^4\Lambda^2_\sscript{B}}[\cT_e]_rV_{cd}V_{tb}V_{td}\log\frac{\mu_f}{\Lambda_\sscript{B}}\,,
        \\
        [L^{S,LR}_{duu}]_{211r}&\simeq-\frac{m_dm_cm_bm_t}{4\pi^2v^4\Lambda^2_\sscript{B}}[\cT_e]_rV_{cs}V_{tb}V_{td}\log\frac{\mu_f}{\Lambda_\sscript{B}}
        \,,
    \end{alignedat}
\end{equation}
while in the down basis the corresponding expressions read
\begin{equation}
    \begin{alignedat}{2}
        [L^{S,LR}_{duu}]_{111r}&\simeq-\frac{m_um_tm_b^2}{4\pi^2v^4\Lambda^2_\sscript{B}}[\cT_e]_r V_{tb}V_{td}V_{us}^* \log\frac{\mu_f}{\Lambda_\sscript{B}}\,,
        \\
        [L^{S,LR}_{duu}]_{211r}&\simeq-\frac{m_um_tm_b^2}{4\pi^2v^4\Lambda^2_\sscript{B}}[\cT_e]_r V_{tb}V_{ts}V_{us}^* \log\frac{\mu_f}{\Lambda_\sscript{B}}
        \,.
    \end{alignedat}
\end{equation}
Additionally, in the down basis, operator mixing also induces $L^{S,RL}_{duu}$ structures, suppressed by the charged-lepton mass
\begin{equation}
    \begin{alignedat}{2}
        [L^{S,RL}_{duu}]_{111r}&\simeq\frac{m_um_dm_t m_{\ell_r}}{4\pi^2v^4\Lambda^2_\sscript{B}}[\cT_e]_r V_{td}V_{us}^* \log\frac{\mu_f}{\Lambda_\sscript{B}}\,,
        \\
        [L^{S,RL}_{duu}]_{211r}&\simeq\frac{m_u m_s m_t m_{\ell_r}}{4\pi^2v^4\Lambda^2_\sscript{B}}[\cT_e]_r V_{ts}V_{us}^* \log\frac{\mu_f}{\Lambda_\sscript{B}}
        \,.
    \end{alignedat}
\end{equation}  
Evaluating the relevant LEFT coefficients using the full RG evolution implemented in \texttt{DsixTools}, one finds that the resulting limits are driven by the $p \to \pi^{0}\mu^{+}$ and $p \to K^{0}\mu^{+}$ modes. As summarized in Table~\ref{tab:dim6bnvtablelim}, these two channels ultimately provide comparable sensitivity. In addition, as shown in Fig.~\ref{fig:ratios}, for sufficiently large $\Lambda_\sscript{B}$ the RG evolution can alter which proton-decay mode provides the leading constraint. In particular, for $\Lambda_\sscript{B} \gtrsim 3 \times10^4\,\tev$, the $p\to K^+\nu$ decay channel becomes dominant over $p\to \pi^0\mu^+$, yet proton-decay rates near current sensitivities would nevertheless require $\Lambda_\sscript{L}$ to lie close to the Planck scale, rendering the scenario phenomenologically nonviable.

%%%%%%%%%%%%%%%%%%%%%%%%%%%%%%%%%%%%%%%%%%%%%%%%%%%%
\section{BNV beyond the EFT Description}
\label{sec:UV_compl}
%%%%%%%%%%%%%%%%%%%%%%%%%%%%%%%%%%%%%%%%%%%%%%%%%%%%

%%%%%%%%%%%%%%%%%%%%%%%%%%%%%%%%%%%%%%%%%%%%%
%%%%%%%%%%%%%%%%%%%%%%%%%%%%%%%%%%%%%%%%%%%%%
\begin{table*}[t]
    \input{tables/LQ_scalars_flavor}
    \caption{Overview of the leading spurionic flavor invariants for the couplings of the scalar leptoquarks $\omega_1$, $\omega_4$, and $\zeta$. The first column lists representative flavor irreducible representations (irreps) of each LQ. The second and third columns display the flavor structures of the allowed renormalizable couplings to SM fermions, expressed in terms of the relevant flavor spurions (see Table~\ref{tab:LQ_overview} for details). The final two columns report the characteristic spurionic suppression of the SMEFT Wilson coefficients induced at tree level, both for the BNV operators and for a representative baryon-number-conserving operator entering $\mu - e$ conversion. The flavor indices $p$ and $r$ label the SM fermions appearing in the coupling tensors (e.g. two quarks for $y^{qq}_{\omega_1,\zeta}$ or $y^{uu}_{\omega_4}$, and a quark-lepton pair for $y^{q\ell}_{\omega_1,\zeta}$ or $y^{ed}_{\omega_4}$). Each LQ field carries either a single flavor index $s$ when transforming in a fundamental representation of $G_\sscript{F}$, or a pair of indices $(s, s’)$ when transforming as a sextet or bifundamental. We employ the notation $2\,\psi_{(a|}\chi_{|b)}=\psi_a\chi_b+\psi_b\chi_a$ to denote symmetrization over the indices $a$ and $b$.
    }
    \label{tab:LQ_scalar_flavor}
\end{table*}
%%%%%%%%%%%%%%%%%%%%%%%%%%%%%%%%%%%%%%%%%%%%%
%%%%%%%%%%%%%%%%%%%%%%%%%%%%%%%%%%%%%%%%%%%%%

The preceding section presented a systematic study of the flavor structures associated with the dimension-six BNV operators and their impact on proton-decay phenomenology. Following this analysis, we now turn to the question of their UV origin. Our aim is to identify the tree-level UV completions capable of generating the BNV operators analyzed in Sec.~\ref{sec:EFT_analysis_BNV_LFV}, while at the same time going beyond the leading-order EFT treatment.\footnote{For loop-level completions of BNV SMEFT operators see Refs.~\cite{Helo:2019yqp,Gargalionis:2024nij,IBeneito:2025nby}.}

In particular, this section focuses on single-mediator extensions of the SM. Tree-level generation of a dimension-six BNV operator uniquely points to leptoquarks (LQs) as the only viable mediator types~\cite{Dorsner:2016wpm,deBlas:2017xtg}. The five relevant LQ representations, along with their renormalizable couplings and the BNV operators they generate at tree level, are summarized in Table~\ref{tab:LQ_overview}.

Based on the well-established practice of assigning non-trivial flavor charges to new-physics mediators in baryon-number-conserving (BNC) SMEFT~\cite{Greljo:2023adz,Greljo:2023bdy,Palavric:2024gvu,Moreno-Sanchez:2025bzz}, we impose the extended MFV hypothesis at the UV level by allowing the LQ mediators to transform nontrivially under the flavor group. This procedure dictates the spurion transformation properties of the renormalizable UV couplings and enables a systematic classification of the LQ representations consistent with flavor-invariant interaction terms.

Tabs.~\ref{tab:LQ_scalar_flavor} and~\ref{tab:LQ_vectors_flavor} summarize the representative flavor assignments for the scalar and vector LQ mediators considered in this work, each capable of generating BNV SMEFT operators at tree level. Within the extended MFV framework, none of these mediators may transform as a flavor singlet, since the quark sector admits no spurion in the $\rep3_f$ representation of $G_\sscript{F}$ with $f\in\{q,u,d\}$, unlike the lepton sector, where such structures can be constructed (see App.~\ref{app:GT_invariants}). Consequently, UV mediators must carry nontrivial flavor quantum numbers in order to couple consistently to SM fermions, with flavor invariance under $G_{\sscript F}$ achieved through insertions of the SM Yukawa matrices and $\Upsilon_\nu$~\cite{Davidson:2010uu}.

%%%%%%%%%%%%%%%%%%%%%%%%%%%%%%%%%%%%%%%%%%%%%
%%%%%%%%%%%%%%%%%%%%%%%%%%%%%%%%%%%%%%%%%%%%%
\begin{table*}[t]
    \input{tables/LQ_vectors_flavor}
    \caption{Overview of the leading spurionic invariants for the couplings of the vector leptoquarks $\cQ_1$ and $\cQ_5$. The first column lists representative flavor representations of each LQ. The second and third columns display the flavor structures of the allowed renormalizable couplings to SM fermions, written in terms of the relevant flavor spurions (see Table~\ref{tab:LQ_overview} for details). The final two columns report the characteristic spurionic suppression of the SMEFT Wilson coefficients induced at tree level, both for the BNV operators and for a representative baryon-number-conserving operator entering $\mu - e$ conversion. The flavor indices $p$ and $r$ label the SM fermions appearing in the coupling tensors (two quarks for $g^{dq}_{\cQ_1}$ and $g^{uq}_{\cQ_5}$, and a quark-lepton pair for $g^{u\ell}_{\cQ_1}$, $g^{d\ell}_{\cQ_5}$, and $g^{eq}_{\cQ_5}$). Each vector LQ carries either a single flavor index $s$ when transforming in a fundamental representation of $G_\sscript{F}$, or a pair $(s,s’)$ when transforming as a sextet or a bifundamental. We employ the notation $2\,\psi_{(a|}\chi_{|b)}=\psi_a\chi_b+\psi_b\chi_a$ to denote symmetrization over the indices $a$ and $b$.}
    \label{tab:LQ_vectors_flavor}
\end{table*}
%%%%%%%%%%%%%%%%%%%%%%%%%%%%%%%%%%%%%%%%%%%%%
%%%%%%%%%%%%%%%%%%%%%%%%%%%%%%%%%%%%%%%%%%%%%

For the mediator representations included here, the induced Wilson coefficients exhibit distinct spurionic suppression patterns, reflecting the transformation properties of the UV couplings under $G_\sscript{F}$. Among these possibilities, mediators transforming as $\rep3_q$ generically yield the least-suppressed contributions to all BNV operators generated at tree level. This makes the $\rep3_q$ assignment particularly relevant, as it provides UV completions whose spurionic behavior aligns with the dominant structures identified in the EFT analysis of the previous section (see Table~\ref{tab:dim6_BNV_SMEFT_Spurions}).

Beyond the fundamental $\rep3_f$ assignment, the extended MFV framework also allows the UV mediators to transform in $\repbar6_f$ and bifundamental representations, in particular of the form $(\rep 3_{f_q},\rep 3_{f_\ell})$ or $(\rep3_{f_q},\repbar3_{f_\ell})$, where $f_q$ ($f_\ell$) factor is associated with the quark (lepton) sector. By contrast, $\rep8_f$ representations are forbidden by group-theoretical considerations. The spurionic suppression arising from the bifundamental representations shows a more diverse structure compared to the fundamental assignment. For example, in the case of $\omega_1$ mediator, the $(\rep3_q,\rep3_\ell)$ assignment generates contributions to the BNV operator $\cO_{qqq\ell}$ with essentially the same spurionic scaling as the fundamental $\rep3_q$ representation. At the same time, the corresponding baryon-number-conserving operators, such as $\mathcal{O}_{\ell q}^{(1,3)}$, may exhibit additional spurionic suppression, reflecting the need to contract both flavor indices of the bifundamental mediator through spurions belonging to different sectors. A similar pattern appears for the other LQ representations: while certain BNV operators induced by bifundamental representations retain a scaling comparable to the $\rep3_q$ case, the accompanying baryon-number-conserving operators typically exhibit a stronger degree of spurionic suppression.

It is noteworthy that some bifundamental representations introduce an additional layer of spurionic suppression for the BNV operators relative to the leading-order EFT treatment. This can be demonstrated by considering a few specific mediator assignments. For instance, adopting a UV completion in which the scalar mediators $\omega_1$ or $\zeta$ transform as $(\rep3_q,\repbar3_\ell)$ leads to the bound of the form
\begin{equation}
    \Lambda_\sscript{B}\gtrsim2\,\mathrm{TeV}\lzm \dfrac{\Lambda_\sscript{L}}{10^3\,\mathrm{TeV}}\dzm\,,
\end{equation}
which, when compared with the bound for $\cO_{qqq\ell}$ in Table~\ref{tab:dim6bnvtablelim}, demonstrates the additional suppression introduced by extra insertion of $\Upsilon_\nu$, thereby further linking the proton-decay rate to the smallness of the neutrino masses.

Similarly, an even more pronounced example arises in the case of vector mediator $\cQ_1$ transforming as $(\rep3_u,\repbar3_\ell)$, for which the bound reads
\begin{equation}
    \Lambda_\sscript{B}\gtrsim1\,\mathrm{TeV}\lzm \dfrac{\Lambda_\sscript{L}}{10^7\,\mathrm{TeV}}\dzm\,,
\end{equation}
which is substantially weaker than the corresponding limit for $\cO_{duq\ell}$ in Table~\ref{tab:dim6bnvtablelim}, again due to the enhanced suppression in $\Upsilon_\nu$.

These examples demonstrate how bifundamental assignments can introduce additional suppression, leading to weaker proton-decay bounds than those obtained in the minimally suppressed scenario. While such representations often entail more intricate UV dynamics and are less typical in minimal model-building approaches, they nonetheless demonstrate the importance of exploring explicit UV completions, as they can depart substantially from the expectations derived from the EFT spurion analysis alone.

In addition to the proton-decay limits, constraints on the baryon-number-conserving operators generated by these UV mediators provide a complementary probe of their parameter space. Current bounds obtained from the updated MFV SMEFT global fit of Ref.~\cite{Bartocci:2023nvp} typically place the scalar and vector LQs in the $1-3\,\tev$ mass range. Future sensitivities based on FCC projections without flavor assumptions~\cite{Allwicher:2024sso,terHoeve:2025gey} are expected to extend this reach up to the $10-20\,\tev$ regime. Although these bounds provide useful complementary information, they are significantly weaker than those obtained from proton-decay searches, even when considering scenarios where additional spurion insertions reduce the latter.

Lastly, a further class of baryon-number-conserving observables is provided by cLFV processes, which are unavoidably induced once the lepton-sector spurions are switched on in the extended MFV framework. All five leptoquark mediators considered here generate the operators relevant for $\mu- e$ conversion in nuclei, whereas none of them induces the structures required for the $\mu\to 3e$ and $\mu\to e\gamma$ transitions at tree level. The corresponding amplitudes involve at least one insertion of the neutrino-mass spurion or the associated leptonic flavor invariants, and are therefore suppressed by the smallness of the neutrino masses and the charged-lepton Yukawas. As shown in App.~\ref{app:cLFV_ex_MFV}, the resulting limits offer complementary sensitivity to the leptonic spurion structure and to the flavor assignments of the mediators. However, throughout the parameter space relevant for our analysis, these cLFV bounds remain well below those implied by proton-decay searches. They therefore serve primarily as a consistency check of the extended MFV framework, ensuring that the leptonic spurion structure implied by the UV completion remains compatible with existing limits on cLFV transitions.

%%%%%%%%%%%%%%%%%%%%%%%%%%%%%%%%%%%%%%%%%%%%%%%%%%%%
\section{Reduced Flavor Symmetries}
\label{sec:MFV_alter}
%%%%%%%%%%%%%%%%%%%%%%%%%%%%%%%%%%%%%%%%%%%%%%%%%%%%

Up to this point our analysis has relied on the extended MFV assumption, under which the full quark-lepton flavor symmetry is broken exclusively by the Yukawa matrices and the neutrino-mass spurion. Although MFV provides a coherent and systematically organized framework, its restrictiveness becomes apparent in the third-generation sector, where the large top Yukawa coupling induces a substantial breaking of the full $G_\sscript{F}$ symmetry and may challenge the convergence of the spurion expansion. This observation has motivated increasing interest in flavor frameworks that relax MFV, particularly those in which new physics couples preferentially to third-generation fermions~\cite{Barbieri:2011ci,Isidori:2012ts,Barbieri:2012uh,Allwicher:2023shc}. Such constructions arise naturally across several classes of UV models, ranging from supersymmetric models~\cite{Barbieri:2011ci,Papucci:2011wy,Larsen:2012rq}, composite Higgs setups~\cite{Barbieri:2012tu,Matsedonskyi:2014iha,Panico:2016ull}, and more recently in flavor non-universal gauge theories~\cite{Allwicher:2020esa,Fuentes-Martin:2020bnh,Fuentes-Martin:2022xnb,Fuentes-Martin:2024fpx,Greljo:2018tuh,Barbieri:2023qpf,Barbieri:2024zkh,Bordone:2017bld,Davighi:2023evx,Davighi:2023iks,Covone:2024elw,Lizana:2024jby,FernandezNavarro:2023rhv,FernandezNavarro:2024hnv,FernandezNavarro:2025zmb,Isidori:2025rci,Greljo:2025mwj}. These considerations motivate the exploration of flavor symmetries in which the full $G_\sscript{F}$ structure is partially reduced. This section focuses on three representative realizations of reduced flavor symmetries: $\U(2)^5$, $\U(2)_{q,u}^2\times \U(3)^3_{d,\ell,e}$ and $\U(2)_{q,u,d}^3\times \U(3)^2_{\ell,e}$.

In the following, we briefly characterize each of the reduced flavor-symmetry groups, specifying the transformation properties of the fermion fields and identifying the corresponding spurion structures. The present section does not attempt a full spurion analysis of the BNV operators or a derivation of the complete proton-decay bounds for each reduced-symmetry framework. Instead, we focus on the leading-order EFT structures permitted by these symmetries and examine how the reduced flavor assumptions modify the spurionic suppressions relative to extended MFV. This approach highlights the principal ways in which these frameworks depart from the MFV conclusions, both in the spurion hierarchies and in the resulting BNV amplitudes, providing a basis for assessing the phenomenological implications.

%%%%%%%%%%%%%%%%%%%%%%%%%%%%%%%%%%%%%%%%%%%%%%%%%%%%
\subsection{\texorpdfstring{$\bm{\U(2)^5}$}{x}}
%%%%%%%%%%%%%%%%%%%%%%%%%%%%%%%%%%%%%%%%%%%%%%%%%%%%

%%%%%%%%%%%%%%%%%%%%%%%%%%%%%%%%%%%%%%%%%%%%%
\begin{table*}[t]
    \input{tables/Dim6_BNV_SMEFT_Spurion_Exp_U2}
    \caption{Dimension-six BNV SMEFT operators and the corresponding flavor-invariant structures in the $\U(2)^5$ framework. The first column lists the relevant operator structures, while the second one denotes the corresponding $\cO(1)$ flavor invariants. The third column specifies the operator component following the classification of Ref.~\cite{Alonso:2014zka}, with S, A and M denoting the symmetric, antisymmetric, and mixed-symmetry components, respectively. The last two columns report the resulting lower bounds on $\Lambda_{\sscript B}^{\sscript{up}}$ and $\Lambda_{\sscript B}^{\sscript{down}}$. All structures are predominantly constrained by the single proton-decay channel $p \to K^+\nu$, in agreement with Ref.~\cite{Gisbert:2024sjw}. Decays into third-generation charged leptons are kinematically forbidden and can proceed only through off-shell $\tau$ exchange, which is far less constraining due to double Fermi constant insertions, additional phase-space suppression, and the comparatively weaker experimental sensitivities~\cite{Crivellin:2023ter,Heeck:2024jei,Marciano:1994bg}.
    }
    \label{tab:dim6_BNV_SMEFT_Spurions_U2}
\end{table*}
%%%%%%%%%%%%%%%%%%%%%%%%%%%%%%%%%%%%%%%%%%%%%

A convenient starting point for reduced flavor symmetries is the $\U(2)^5$ framework, in which the first two generations of each fermion species transform collectively as doublets while the third generation is treated as a singlet. Concretely, the flavor symmetry is taken to be
\begin{equation}
    G_\sscript{F}\equiv \U(2)_q\times \U(2)_u\times \U(2)_d\times \U(2)_\ell\times \U(2)_e\,,
\end{equation}
under which the quark and lepton fields decompose as $\psi = [ \psi^a~\psi_3 ]^\intercal$, with $a=1,2$ and $\psi=\{q,u,d,\ell,e\}$. Here $\psi^a$ transforms as a doublet of the corresponding $\U(2)_\psi$ factor, while $\psi_3$ is a singlet.

The breaking of $\U(2)^5$ is parametrized by a minimal set of spurions that connect the doublet and singlet sectors and reproduce the observed fermion masses and mixing. In the quark sector, the breaking is encoded in a minimal set of spurions consisting of a doublet $V_q\sim\rep2_q$ linking the light and third generations, and two bifundamental $2\times2$ matrices $\Delta_{u,d}\sim(\rep2_q,\repbar2_{u,d})$, which encode the breaking within the first two generations of the up- and down-type sectors, respectively. Similarly, in the lepton sector the breaking of $\U(2)_\ell\times\U(2)_e$ is encoded in a single bifundamental spurion $\Delta_e\sim(\rep2_\ell,\repbar2_e)$.\footnote{Note that a doublet spurion $V_\ell \sim \rep2_\ell$ may be introduced in general, but it is not needed in the minimal parametrization of the charged-lepton Yukawa matrix.} With this spurion content, the quark and charged-lepton Yukawa matrices take the block form~\cite{Greljo:2022cah}
\begin{equation}\label{eq:U2_spurions}
    Y_d=\begin{bmatrix}\Delta_d&0\\0&y_b\end{bmatrix}\,,
    \quad
    Y_u=\begin{bmatrix}\Delta_u & V_q\\0&y_t\end{bmatrix}\,,
    \quad
    Y_e=\begin{bmatrix}\Delta_e&0\\0&y_\tau\end{bmatrix}\,,
\end{equation}
where $y_b$, $y_t$, and $y_\tau$ denote the third-generation Yukawa couplings of the bottom quark, top quark, and tau lepton, respectively. A similar approach applies to the neutrino sector, in which the coefficient of the Weinberg operator in Eq.~\eqref{eq:Weinberg_op} gives rise to the LNV spurion $\Upsilon_\nu$, transforming as a symmetric tensor in lepton-flavor space. In the $\U(2)^5$ framework, this spurion admits the block decomposition
\begin{equation}
    \Upsilon_\nu=\begin{bmatrix}\Delta_\nu & V_\nu \\ V_\nu^\intercal & y_\nu\end{bmatrix}\,,
    \quad
    \Delta_\nu\sim\rep3_\ell\,,
    \quad
    V_\nu\sim \rep2_\ell\,,
    \quad
    y_\nu\in\mathbb R\,,
\end{equation}
where $\Delta_\nu$ corresponds to the symmetric triplet component arising from the decomposition $\rep2\otimes\rep2=\rep3\oplus\rep1$.

Having outlined the essential features of the $\U(2)^5$ flavor framework, we now turn to its implications for the dimension-six BNV operators. Our approach is analogous to that adopted in the extended MFV framework: we construct the flavor invariants associated with each BNV operator, taking into account the intrinsic permutation symmetries of their Wilson coefficients. A key difference emerges already at this stage: contrary to MFV, where the absence of fermion-sector singlets forces every BNV structure to involve spurion insertions, the $\U(2)^5$ symmetry admits fully $\cO(1)$ flavor invariants constructed solely from the singlet components of the fermion fields. Consequently, leading BNV structures can be written without any spurion suppressions, provided they are consistent with the operator’s internal symmetry relations. As a result, this directly implies significantly stronger bounds on $\Lambda_\sscript{B}$.

In the present analysis we focus on these $\cO(1)$ contributions. Extending the construction to next-to-leading order would require insertions of the doublet spurions, and already at $\cO(V_q)$ one finds sixteen additional independent structures. While such an extension is certainly feasible, it lies beyond the scope of this work. Phenomenologically,
an immediate implication of the absence of spurion insertions is that the leading contributions do not depend on the LNV scale $\Lambda_\sscript{L}$, as no neutrino-sector spurions enter the construction. The resulting limits are therefore expressed directly in terms of $\Lambda_\sscript{B}$, in contrast to the extended MFV case where the ratio $\cR_{\sscript{BL}}=\Lambda_\sscript{B}^2/\Lambda_\sscript{L}$ is constrained. The complete set of leading $\cO(1)$ flavor invariants for the dimension-six BNV operators, together with the resulting limits on $\Lambda_\sscript{B}$, is presented in Table~\ref{tab:dim6_BNV_SMEFT_Spurions_U2}. As anticipated, the absence of spurionic suppression leads to significantly stronger bounds on $\Lambda_\sscript{B}$ than in the extended MFV case, where even the most favorable scenarios require $\Lambda_\sscript{B}\gtrsim 10^6\,\tev$. Notably, in these scenarios proton decay is induced only at the two-loop level; nevertheless, this loop suppression alone is insufficient to lower the BNV scale to the multi-TeV regime~\cite{Gisbert:2024sjw}.

The strength of these bounds, driven by the lack of spurionic suppression, provides a clear motivation to investigate other partially reduced flavor symmetries. Such frameworks may permit controlled suppressions of the BNV operators and thereby offer a more nuanced phenomenological picture, situated between the restrictive extended MFV case and the unsuppressed $\U(2)^5$ limit. Accordingly, the remainder of the section will examine the intermediate frameworks $\U(2)_{q,u}^2\times \U(3)^3_{d,\ell,e}$ and $\U(2)_{q,u,d}^3\times \U(3)^2_{\ell,e}$, which represent concrete realizations of partially reduced flavor symmetries with distinct predicted suppression patterns.

Before concluding the discussion of the $\U(2)^5$ framework, it is useful to comment briefly on the possible UV realizations of the leading BNV structures. Since $\U(2)^5$ admits fermion singlets, all unsuppressed $\cO(1)$ flavor invariants identified in Table~\ref{tab:dim6_BNV_SMEFT_Spurions_U2} can be generated by appropriate UV mediators transforming as flavor singlets. As a concrete illustration, a scalar mediator such as $\omega_1$, when taken to transform as a singlet under $\U(2)^5$, can induce the symmetric component of $\cO_{duue}$. Likewise, a singlet $\omega_4$, whose up-type coupling $[y_{\omega_4}^{uu}]^{pr}$ is intrinsically antisymmetric, naturally generates the antisymmetric component of the same operator.

Additionally, beyond singlet assignments, one may also consider UV mediators transforming in nontrivial representations of $\U(2)^5$. In such cases, flavor invariance requires the renormalizable interactions to involve the appropriate spurion insertions. For instance, if $\omega_4$ transforms as $\rep2_u$, the coupling to two up-type quarks can be written as $[y_{\omega_4}^{uu}]^{spr}= (\Delta_u^\dagger V_q)^s\varepsilon^{pr}$, while the right-handed leptons and down quarks is given by $[y_{\omega_4}^{ed}]\ud{s}{pr}= (\Delta_u^\dagger V_q)^s \,\delta_{p3}\delta_{r3}$. Integrating out the mediator in this flavor representation at tree level generates the BNV operator $\cO_{duue}$ with coefficient $[\cC_{duue}]_{prst} \sim (V_q^\dagger \Delta_u)( \Delta_u^\dagger V_q)\,\delta_{p3} \varepsilon_{rs} \delta_{t3}$, which is suppressed by $\cO(V_q^2\Delta_u^2)$~\cite{deBlas:2017xtg}. Numerically, one finds $(V_q^\dagger \Delta_u)(\Delta_u^\dagger V_q)\simeq10^{\eminus8}$~\cite{Greljo:2022cah}, implying a corresponding lower bound on the mediator mass of $M_{\omega_4}\gtrsim1.5\times10^3\,\tev$.

%%%%%%%%%%%%%%%%%%%%%%%%%%%%%%%%%%%%%%%%%%%%%%%%%%%%
\subsection{\texorpdfstring{$\bm{\U(2)_{q,u,d}^3\times \U(3)^2_{\ell,e}}$}{x} and \texorpdfstring{$\bm{\U(2)_{q,u}^2\times \U(3)^3_{d,\ell,e}}$}{x}}
%%%%%%%%%%%%%%%%%%%%%%%%%%%%%%%%%%%%%%%%%%%%%%%%%%%%

%%%%%%%%%%%%%%%%%%%%%%%%%%%%%%%%%%%%%%%%%%%%%%%%%%%%
\begin{table*}[t]
    \input{tables/Dim6_BNV_SMEFT_Spurion_Exp_relaxed}
    \caption{Dimension-six BNV SMEFT operators and the corresponding flavor-invariant structures for the $\U(2)_{q,u,d}^3\times \U(3)_{\ell,e}^2$ (upper panel) and $\U(2)_{q,u}^2\times \U(3)_{d,\ell,e}^3$ (lower panel) framework. The first column lists the relevant operator structures, while the second displays the least suppressed spurion insertions required to render each operator invariant under the full flavor symmetry group. The third column specifies the operator component following the classification of Ref.~\cite{Alonso:2014zka}, with S, A and M denoting the symmetric, antisymmetric, and mixed-symmetry components, respectively. The fourth (sixth) column reports the bounds on the ratio $\cR_{\sscript{BL}}\equiv\Lambda_\sscript B^2 / \Lambda_\sscript L$ in the up (down) basis, denoted as $\cR_{\sscript{BL}}^{\sscript{up}}$ ($\cR_{\sscript{BL}}^{\sscript{down}}$), while the adjacent columns indicate the decay channel that provides the most stringent limit in each case.}
    \label{tab:Dim6_BNV_relaxed_flavor}
\end{table*}
%%%%%%%%%%%%%%%%%%%%%%%%%%%%%%%%%%%%%%%%%%%%%%%%%%%%

We next consider two intermediate flavor symmetries that interpolate between the $\U(2)^5$ and extended MFV construction, namely $\U(2)_{q,u,d}^3\times \U(3)^2_{\ell,e}$ and $\U(2)_{q,u}^2\times \U(3)^3_{d,\ell,e}$. The first symmetry choice is fully specified by the $\U(2)^3$ spurions in the quark sector (see Eq.~\eqref{eq:U2_spurions}) together with the $\U(3)^2$ lepton-sector assignments specified in Sec.~\ref{sec:MFV_overview}. The construction of the BNV flavor invariants in the quark sector proceeds exactly as in the $\U(2)^5$ case, while, in contrast, the invariance under $\U(3)^2_{\ell,e}$ is achieved through the inclusion of the appropriate lepton-sector spurions. Consequently, the lepton-sector invariants necessarily acquire spurion suppressions leading to a characteristic scaling with the neutrino masses.

The second framework of interest involves $\U(2)_{q,u}^2\times \U(3)^3_{d,\ell,e}$ symmetry. Restoring $\U(3)_d$ symmetry has an immediate consequence for the spurion content: since $q$ and $u$ retain their $\U(2)$ transformation properties, whereas $d$ now transforms as a $\U(3)_d$ triplet, the symmetry breaking is controlled by three spurions, in particular, $\Delta_u\sim(\rep2_q,\repbar2_u)$, $\Sigma_d\sim(\rep2_q,\repbar3_d)$ and $\Lambda_d\sim\rep3_d$.\footnote{A doublet spurion $V_q\sim\rep2_q$ may be included to account for additional mixing between the third generation and the light quarks, but it is not required for the minimal parametrization of the Yukawa matrices.} With this spurion content, the minimal parametrization of the quark Yukawa matrices takes the form~\cite{Greljo:2022cah}
\begin{equation}
    Y_u=\begin{bmatrix} \Delta_u&0\\0&y_t \end{bmatrix}\,,
    \qquad
    Y_d=\begin{bmatrix}\Sigma_d\\\Lambda_d^\dag\end{bmatrix}\,,
    \qquad
    y_t\in\mathbb{R}\,.
\end{equation}
Restoring $\U(3)_d$ further implies that any BNV operator involving down quarks requires insertion of the $\Lambda_d\simeq 10^{\eminus2}$ spurion, introducing additional suppression compared to the previous case.

The resulting flavor-invariant structures for both symmetry assumptions, together with the corresponding decay channels and bounds on the ratio $\Lambda_\sscript{B}^2/\Lambda_\sscript{L}$, are collected in Table~\ref{tab:Dim6_BNV_relaxed_flavor}. In addition to the entries shown, it is worth noting that several operators give rise to multiple proton-decay channels with comparable sensitivity.

For the $\U(2)_{q,u,d}^3 \times \U(3)_{\ell,e}^2$ symmetry, the operator $\cO_{qque}$ in the down basis is constrained not only by the quoted $p \to K^0 \mu^+$ channel but also by $p \to \pi^0 \mu^+$ and $p \to K^+ \nu$, which provide limits of similar strength. Likewise, the bounds on the antisymmetric component of $\cO_{duue}$ (in both the up and down bases) are accompanied by the mode $p \to K^0 \mu^+$, again yielding comparable sensitivity. Lastly, for the antisymmetric component of $\cO_{qqq\ell}$ in the down basis, the quoted $p \to K^+ \nu$ constraint is representative of a broader set of channels exhibiting similar reach.

A similar situation occurs in the $\U(2)_{q,u}^2 \times \U(3)_{d,\ell,e}^3$ framework. For the antisymmetric component of $\cO_{duue}$, the channel $p \to K^0 \mu^+$ provides constraints comparable to those from $p \to \pi^0 \mu^+$. The operator $\cO_{duq\ell}$ in the down basis receives contributions from several additional modes, including $p \to K^0 e^+$, $p \to K^+\nu$, $p \to \eta^0 e^+$, and $p \to K^0 \mu^+$, all of which probe similar scales. Similarly, the antisymmetric component of $\cO_{qqq\ell}$ in the down basis yields a variety of competitive channels, such as $p \to K^0 e^+$, $p \to \pi^0 e^+$, $n \to K^0 \nu$, $p \to \eta^0 e^+$, and $p \to K^0 \mu^+$.

Finally, for both flavor assumptions, the constraint arising from the fully symmetric component of $\cO_{qqq\ell}$ is weaker by about three orders of magnitude relative to those obtained from the antisymmetric or mixed components. This reduction stems from the fact that the symmetric structure involves only third-generation quarks and contributes to proton decay only at two loops, leading to a parametrically suppressed rate~\cite{Gisbert:2024sjw}.

Taken together, the two intermediate flavor frameworks yield proton-decay limits that allow a multi-TeV BNV scale for all dimension-six operators aside from $\cO_{qqq\ell}$, consistent with the pattern observed in the MFV analysis. The bounds associated with $\cO_{qqq\ell}$ and $\cO_{qque}$ closely track their MFV counterparts, whereas those for $\cO_{duq\ell}$ and $\cO_{duue}$ are enhanced (reduced) by about an order of magnitude in the $\U(2)_{q,u,d}^{3}\times\U(3)_{\ell,e}^{2}$ ($\U(2)_{q,u}^{2}\times\U(3)_{d,\ell,e}^{3}$) framework, where the weakening in the latter case follows from the additional insertion of the down-sector spurion $\Lambda_d$.

%\AP{Paragraph on two new plots... Similarly as in Sec.~MFV, we also construct the $r$ ratios for the two alternatives...}

As a final point, we note that the leading BNV structures listed in Table~\ref{tab:Dim6_BNV_relaxed_flavor} can, as in the $\U(2)^5$ case, be generated by flavor-singlet UV mediators, in contrast to extended MFV where mediators need to transform in non-trivial flavor representations. From a model-building perspective, this makes the present flavor assumptions considerably simpler than in the MFV case. Allowing non-singlet mediators again yields subleading, spurion-suppressed contributions.

%%%%%%%%%%%%%%%%%%%%%%%%%%%%%%%%%%%%%%%%%%%%%%%%%%%%
\section{Conclusions}
\label{sec:conc}
%%%%%%%%%%%%%%%%%%%%%%%%%%%%%%%%%%%%%%%%%%%%%%%%%%%%

The hierarchical pattern of fermion masses and mixings in the SM suggests that flavor symmetries may play an essential role in structuring new physics, as emphasized in MFV and related frameworks. Proton decay, in turn, offers a uniquely sensitive probe of BNV and constitutes one of the clearest possible signals of physics beyond the SM. Previous work examined BNV within MFV settings~\cite{Smith:2011rp,Helset:2019eyc}, showing that the scale of new physics mediating BNV can be lowered by many orders of magnitude relative to the naive expectation $\Lambda_\sscript{B}\gtrsim10^{16}\,\gev$, though these works relied on schematic estimates rather than a full phenomenological analysis.

In this work we carried out a detailed EFT analysis of BNV under several flavor-symmetry assumptions within the SMEFT framework. In the MFV case, we confirmed the unavoidable link between BNV and LNV, showing that proton-decay amplitudes acquire a characteristic suppression proportional to $m_\nu^2$. This relation both explains why the BNV scale $\Lambda_\sscript{B}$ may lie far below the naive expectation and implies that an eventual observation of proton decay would indirectly point to a Majorana origin of neutrino masses. 

For each of the four independent dimension-six BNV operators, we derived bounds on $\Lambda_\sscript{B}$ (and their implicit dependence on the LNV scale $\Lambda_\sscript{L}$) and identified the dominant decay channels. Our analysis combines a full numerical SMEFT-LEFT evolution supplemented with an analytic leading-log discussion, further clarifying the origin of relevant effects. We show that the operator $\cO_{qqq\ell}$ generates proton decay at tree level and thus yields the strongest bounds, whereas, $\cO_{duq\ell}$, $\cO_{qque}$ and $\cO_{duue}$ allow for much weaker limits. Aside from $\cO_{qqq\ell}$, all operators remain compatible with $\Lambda_\sscript{B}$ in the multi-TeV range for suitable choices of $\Lambda_\sscript{L}$. We note that the required values for $\Lambda_\sscript{L}$ may lie well below the naive expectation for type-I seesaw scale, while remaining naturally accommodated in type-II or inverse seesaw realizations. Finally, we identified correlations among the dominant proton-decay channels that depend only on $\Lambda_\sscript{B}$, allowing potential future measurements to discriminate which dimension-six BNV operator is preferred.

Beyond the EFT analysis, we also examined the set of possible single-particle UV completions. In the MFV framework, such mediators necessarily transform nontrivially under the flavor group, and certain representations were found to introduce an additional spurionic suppression, enhancing the neutrino-mass dependence to $\Gamma_p\propto m_\nu^4$. While these constructions are not meant to represent the fully developed UV scenarios, they serve to illustrate how going beyond a leading-order EFT description can open qualitatively different possibilities.

In addition, we explored several reduced flavor assumptions. Starting with the $\U(2)^5$ framework, at leading order, proton decay no longer inherits the neutrino-mass suppression present in MFV, leading to substantially stronger bounds on the BNV scale. In turn, the intermediate symmetries $\U(2)_{q,u,d}^3 \times \U(3)_{\ell,e}^2$ and $\U(2)_{q,u}^2 \times \U(3)_{d,\ell,e}^3$ restore the MFV-like suppression pattern and permit multi-TeV values of $\Lambda_\sscript{B}$. These frameworks additionally permit flavor-singlet UV mediators for the leading BNV structures, offering simpler UV completions.

Looking ahead, the present analysis suggests several directions for future work. From a phenomenological perspective, a natural next step involves broadening the EFT treatment beyond the leading structures considered here, as well as extending the study to higher-dimensional BNV operators, which can mediate qualitatively distinct effects. In parallel, another complementary direction involves exploring more complete UV realizations beyond the minimal setups discussed in this work.

In conclusion, we have shown that the interplay between flavor symmetries and BNV yields a rich and predictive structure, with direct implications for proton-decay searches, offering clear criteria for interpreting possible future signals. Upcoming experimental advances are expected to provide further valuable insights into these underlying mechanisms.

%%%%%%%%%%%%%%%%%%%%%%%%%%%%%%
\vspace{0.2cm}
\noindent
%%%%%%%%%%%%%%%%%%%%%%%%%%%%%%
{\textbf{Acknowledgments.}}
We thank Juan Herrero-García and Andreas Crivellin for helpful discussions, and John Gargalionis and Arcadi Santamaria for reading the manuscript. ABB also thanks the University of Basel and Prof. Admir Greljo’s group, and to the Universit\`a di Padova and INFN Sezione di Padova, for their warm hospitality at the various stages of this project. The work of ABB is funded by the grant CIACIF/2021/061 of the ``Generalitat Valenciana'' and also by the Spanish ``Agencia Estatal de Investigaci\'on'' through MICIN/AEI/10.13039/501100011033. The work of AP is supported by MCIU/AEI/10.13039/501100011033 (grants CEX2023-001292-S and PID2023-146220NB-I00). The work of AS received funding from the INFN Iniziativa Specifica APINE.

%%%%%%%%%%%%%%%%%%%%%%%%%%%%%%%%%%%%%%%%%%%%%%%%%%%%%%%%%%%%%%%%%%%%%%%%%%%
\appendix
%%%%%%%%%%%%%%%%%%%%%%%%%%%%%%%%%%%%%%%%%%%%%%%%%%%%%%%%%%%%%%%%%%%%%%%%%%%

%%%%%%%%%%%%%%%%%%%%%%%%%%%%%%%%%%%%%%%%%%%%%%%%%%%%%%%%%%%%%%%%%%%%%%
\section{Group Theory of Flavor Invariants}
\label{app:GT_invariants}
%%%%%%%%%%%%%%%%%%%%%%%%%%%%%%%%%%%%%%%%%%%%%%%%%%%%%%%%%%%%%%%%%%%%%%
\noindent
This appendix presents the group-theoretical construction of lepton and quark flavor invariants in the extended MFV framework, with emphasis on identifying the least-suppressed invariant structures.

%%%%%%%%%%%%%%%%%%%%%%%%%%%%%%%%%%%%%%%%%%%%%%%%%%%%%%%%%%%%%%%%%%%%%%
\subsection{Lepton Sector}
%%%%%%%%%%%%%%%%%%%%%%%%%%%%%%%%%%%%%%%%%%%%%%%%%%%%%%%%%%%%%%%%%%%%%%
\noindent
We consider the subgroup $G_\sscript{F}\supset \U(3)_\ell\times \U(3)_e$, under which the SM lepton fields transform canonically as triplets under the corresponding $\U(3)$ factors. The relevant spurions are the charged-lepton Yukawa matrix $Y_e\sim(\rep{3}_\ell,\repbar{3}_e)$, and the symmetric neutrino mass spurion $\Upsilon_\nu\sim\rep{6}_\ell$. To identify the leading MFV structures, we examine the allowed flavor-singlet contractions formed from the spurions and the lepton fields. Since each dimension-six BNV operator includes only one lepton-field insertion, the $\ell$ and $e$ contractions are treated separately.
\\[5pt]
For $\ell\sim \rep{3}_\ell$, starting with the decomposition of the form
\begin{equation}
    Y_e Y_e^\dag\sim(\rep1,\rep1)\oplus(\rep8_\ell,\rep1)\oplus(\rep1,\rep8_e)\oplus(\rep8_\ell,\rep8_e)\,,
\end{equation}
it follows that the $(Y_eY_e^\dag)\ell$ product contains
\begin{equation}
    (Y_eY_e^\dag)\ell\supset \rep8_\ell\otimes \rep3_\ell=\rep3_\ell\oplus\repbar6_\ell\oplus\rep{15}_\ell\,.
\end{equation}
Lastly, taking the tensor product of this structure with the $\Upsilon_\nu\sim\rep6_\ell$ yields
\begin{equation}
    \Upsilon_\nu (Y_eY_e^\dag)\ell\supset \rep6_\ell\otimes(\rep3_\ell\oplus\repbar6_\ell\oplus\rep{15}_\ell)\supset \rep6_\ell\otimes \repbar6_\ell\supset\rep1\,.
\end{equation}
Following these decompositions, the least suppressed flavor invariant reads
\begin{equation} \label{eq:leptonflavorstructure1}
    \cI_F^\ell\sim \varepsilon_{p r s}[\Upsilon_\nu]^{p t}[Y_e^\dag]\ud{m}{t}[Y_e]\ud{r}{m}\ell^{s} \,,
\end{equation}
with the contractions in flavor space made explicit.
\\[5pt]
In case of $e\sim\rep3_e$, following the similar approach as for the $\ell$ case, our starting point is the decomposition
\begin{equation}
    Y_e^\dag Y_e^\dag\sim(\rep3_\ell,\repbar3_e)\oplus(\rep3_\ell,\rep6_e)\oplus(\repbar6_\ell,\repbar3_e)\oplus(\repbar6_\ell,\rep6_e)\,,
\end{equation}
from which we can write
\begin{equation}
    Y_e^\dag Y_e^\dag e\supset (\repbar6_\ell,\repbar3_e)\otimes\rep3_e\supset (\repbar6_\ell,\rep1)\,.
\end{equation}
Lastly, contracting with $\Upsilon_\nu$, we obtain
\begin{equation}
    \Upsilon_\nu Y_e^\dag Y_e^\dag e\supset \rep6_\ell\otimes(\repbar6_\ell,\rep1)\supset\rep1\,.
\end{equation}
With these decompositions in mind, the flavor invariant takes the form
\begin{equation}\label{eq:e_van_inv}
    \cI_F^e\sim [\Upsilon_\nu]^{p r} \varepsilon_{s t m} [Y_e^\dag]\ud{s}{(p|}[Y_e^\dag]\ud{t}{|r)} e^{m}\,,
\end{equation}
where $2\,\psi_{(a|}\chi_{|b)}=\psi_a\chi_b+\psi_b\chi_a$ denotes the symmetrized sum. However, as can be seen from Eq.~\eqref{eq:e_van_inv}, the contraction involves an antisymmetric tensor $\varepsilon_{s t m}$ together with a spurion combination that is symmetric under the exchange $s\leftrightarrow t$. As a result, the entire expression vanishes identically. This observation is consistent with the findings of Ref.~\cite{Smith:2011rp}, which notes that BNV operators involving $e$ fields are more suppressed by flavor considerations than those involving $\ell$. Our derivation, however, is based on an explicit group-theoretical analysis and operator construction. Furthermore, this implies that constructing a non-vanishing invariant necessitates the insertion of an additional spurion, leading to a minimally suppressed structure that can be written as
\begin{equation} \label{eq:leptonflavorstructure2}
    \cI_F^e\sim \varepsilon_{p r s}[\Upsilon_\nu]^{p t} [Y_e^\dag]\ud{m}{t}[Y_e]\ud{r}{m}[Y_e]\ud{s}{n}e^{n}\,.
\end{equation}
Given that the purely leptonic spurion structures defined in Eqs.~\eqref{eq:leptonflavorstructure1} and~\eqref{eq:leptonflavorstructure2} systematically enter the construction of BNV operators within the extended MFV framework (see Sec.~\ref{sec:flavor_invariant_BNV_ops}), we introduce the following shorthand notation in order to streamline the presentation of these recurrent combinations
\begin{equation} \label{eq:flavorYLdefinition}
    [\mathcal T_\ell]_{s} \equiv  \varepsilon_{p r s}[\Upsilon_\nu]^{p t} [Y_e Y_e^\dag]\ud{r}{t} \sim \repbar3_\ell\, ,
\end{equation}
and
\begin{equation} \label{eq:flavorYEdefinition}
    [\mathcal T_e]_{n} \equiv  \varepsilon_{p r s}[\Upsilon_\nu]^{p t}[Y_e Y_e^\dag]\ud{r}{t}[Y_e]\ud{s}{n}\sim \repbar3_e \, .
\end{equation}
The flavor invariants derived in this section have been independently verified using the \texttt{Sym2Int} package~\cite{Fonseca:2017lem,Fonseca:2019yya}.
\\[5pt]
Lastly, we summarize the numerical inputs relevant for the lepton-sector invariants. Using the parametrization of $\Upsilon_\nu$ in terms of current neutrino-oscillation parameters detailed in Sec.~\ref{sec:MFV_overview}, along with the measured values for the charged-lepton Yukawa matrix, we find the following numerical values for $\mathcal T_\ell$ in the Normal Ordering (NO) and Inverted Ordering (IO) 
\begin{equation}\label{eq:numvalTl}
    \begin{alignedat}{2}
    \mathcal T_\ell^{\sscript{NO}} &= \frac{\Lambda_{\sscript{L}}}{\rm TeV} \begin{bmatrix}
        7.2\\
        2.5\\
        (-3+i)\times 10^{-3}
    \end{bmatrix} \times 10^{\eminus 17} \, ,
    \\
    \mathcal T_\ell^{\sscript{IO}} &= \frac{\Lambda_{\sscript{L}}}{\rm TeV} \begin{bmatrix}
        -8.3+0.1\,i\\
        -1.5-0.5\,i\\
         (7+2\,i)\times 10^{-3}
    \end{bmatrix} \times 10^{\eminus 17} \, ,
    \end{alignedat}
\end{equation}
while for the additionally spurion-suppressed $\mathcal T_e$ we find
\begin{equation} \label{eq:numvalTe}
    \begin{alignedat}{2}
    \mathcal T_e^{\sscript{NO}} &= \frac{\Lambda_{\sscript{L}}}{\rm TeV} \begin{bmatrix}
        0.02\\
        1.5-0.2\,i\\
        -0.03
    \end{bmatrix} \times 10^{\eminus 20} \, , 
    \\
    \mathcal T_e^{\sscript{IO}} &= \frac{\Lambda_{\sscript{L}}}{\rm TeV} \begin{bmatrix}
        -0.2\\
        -9.0-2.9\,i\\
        0.7+0.2\,i
    \end{bmatrix} \times 10^{\eminus 21} \, .
    \end{alignedat}
\end{equation}
As input parameters, we use the central values from the latest global fit to neutrino-oscillation data~\cite{Esteban:2024eli}, together with the most recent PDG determinations of the charged-lepton masses~\cite{ParticleDataGroup:2024cfk}.

%%%%%%%%%%%%%%%%%%%%%%%%%%%%%%%%%%%%%%%%%%%%%%%%%%%%%%%%%%%%%%%%%%%%%%
\subsection{Quark Sector}
%%%%%%%%%%%%%%%%%%%%%%%%%%%%%%%%%%%%%%%%%%%%%%%%%%%%%%%%%%%%%%%%%%%%%%
\noindent
The construction of flavor invariants in the quark sector proceeds in close analogy with the lepton-sector analysis. The similar group-theoretical logic applies, with the spurion content now involving the two quark Yukawa matrices $Y_{u,d}\sim(\rep3_q,\repbar3_{u,d})$, and all contractions need to be invariant under the quark subgroup $G_\sscript{F}\supset \U(3)_q\times \U(3)_u\times \U(3)_d$.
\\[5pt]
We start by classifying the structures generated by two spurion insertions
\begin{align}
    Y_uY_u^\dag&\sim(\rep1,\rep1)\oplus(\rep8_q,\rep1)\oplus(\rep1,\rep8_u)\oplus(\rep8_q,\rep8_u)\,,
    \nonumber\\[2pt]
    Y_uY_u&\sim(\repbar3_q,\rep3_u)\oplus(\rep6_q,\rep3_u)\oplus(\repbar3_q,\repbar6_u)\oplus(\rep6_q,\repbar6_u)\,,
    \nonumber\\[2pt]
    Y_uY_d^\dag&\sim(\rep1,\repbar3_u\otimes\rep3_d)\oplus(\rep8_q,\repbar3_u\otimes\rep3_d)\,,
    \nonumber\\[2pt]
    Y_uY_d&\sim(\repbar3_q,\repbar3_u\otimes\repbar3_d)\oplus(\rep6_q,\repbar3_u\otimes\repbar3_d)\,.
\end{align}
The remaining structures follow straightforwardly by taking conjugates or by substituting $u\to d$. We further note the additional decompositions that prove useful in constructing the invariants
\begin{equation}\label{eq:4_decs}
    \begin{alignedat}{4}
        \rep8\otimes \rep3 &= \rep3\oplus \repbar6\oplus \rep{15}\,,
        &\qquad
        \rep6\otimes \rep3 &= \rep8\oplus \rep{10}\,,
        \\
        \rep8\otimes \repbar3 &= \repbar3\oplus \rep6\oplus \repbar{15}\,,
        &\qquad
        \rep6\otimes \repbar3 &= \rep3\oplus \rep{15}\,.
    \end{alignedat}
\end{equation}
These decompositions provide the building blocks for assembling all relevant flavor invariants, both at the operator level within SMEFT and in the corresponding UV interactions.
\\[5pt]
At the operator level, this procedure can be illustrated using the example of $\cO_{qqq\ell}$, which contains three insertions of $q\sim\rep3_q$. At leading order, the flavor contraction follows from the decomposition
\begin{equation}
    \rep3_q \otimes \rep3_q \otimes \rep3_q
    =\rep3_q \otimes (\rep6_q \oplus \repbar3_q)
    \supset \rep3_q \otimes \repbar3_q
    = \rep1 \oplus \rep8_q\,,
\end{equation}
so a flavor singlet can be formed directly using the invariant tensor $\varepsilon_{prs}$. At next-to-leading order in the spurion expansion, another invariant can be formed by inserting $Y_uY_u^\dag\supset (\rep8_q,\rep1)$. Selecting the adjoint $\rep8_q$ component, the decomposition 
\begin{equation}
    \rep3_q\otimes\rep3_q\otimes\rep3_q\supset \rep3_q\otimes\rep6_q\supset \rep8_q
\end{equation}
provides an octet that can be contracted with the octet from $Y_u Y_u^\dag$ to form a singlet. With the explicit contractions in the flavor space, this invariant reads $\cI_{F}^{qqq}\sim\varepsilon_{prx}[Y_uY_u^\dagger]\ud{x}{s}$.
\\[5pt]
A similar group-theoretical strategy applies to operators involving quarks in different flavor representations. For instance, $\cO_{duq\ell}$, which transforms as $\rep3_d\otimes\rep3_u\otimes\rep3_q$, does not admit a flavor singlet at leading order. However, at next-to-leading order the decomposition of product $Y_uY_d$ contains a $(\repbar3_q,\repbar3_u\otimes\repbar3_d)$ component, which combines with $\cO_{duq\ell}\sim \rep3_d\otimes\rep3_u\otimes\rep3_q$ to produce a flavor singlet. The flavor invariant then reads $\cI_F^{duq}\sim \varepsilon_{abs}[Y_u]\ud{a}{r}[Y_d]\ud{b}{p}$.
\\[5pt]
Furthermore, flavor invariants involving the UV mediators are obtained in complete analogy with those constructed at the operator level. In this context, if the UV mediator transforms as a flavor singlet, no invariant involving a single quark field can be formed. Indeed, a quark transforming as $\rep3_x$ with $x\in\{q,u,d\}$ would require a compensating spurion in the conjugate representation $\repbar3_x$, which cannot be generated from quark Yukawa insertions alone. Since Yukawa matrices transform as bifundamentals, any combination of them yields representations containing an even number of fundamental indices. This conclusion is also not altered by determinant insertions, as the symmetrization of two Yukawa matrices of the same type removes the would-be $\repbar3_x$ spurion.

To this end, a useful illustration is obtained by introducing a new spurion $\Omega$ (representing a UV mediator) transforming in a nontrivial quark-flavor representation and asking under which conditions a flavor-singlet object involving a single quark field can be constructed. If $\Omega\sim\rep3_x$ with $x\in\{q,u,d\}$, the construction is straightforward: a single insertion of the appropriate Yukawa matrix suffices to form an invariant. On the other hand, for $\Omega\sim \rep6_q$ the relevant decomposition becomes
\begin{equation}
\Omega\, (Y_uY_u^\dag)\, q
\;\supset\;
\rep6_q \otimes (\rep8_q \otimes \rep3_q)
\;\supset\;
\rep6_q \otimes \repbar6_q
\;\supset\; \rep1\,,
\end{equation}
where we select the adjoint component of $Y_uY_u^\dag$ product. Similarly, for $\Omega\sim \rep6_x$ with $x\in\{u,d\}$, singlet constructions remain possible, although they may require additional Yukawa insertions.
\\[5pt]
Next, considering the case $\Omega \sim \rep8_q$, one finds that no flavor singlet can be built from a single insertion of $q\sim\rep3_q$, even after allowing for two Yukawa insertions. As follows from the decompositions given by Eq.~\eqref{eq:4_decs}, the products $\rep8_q \otimes \rep3_q$ and $\rep8_q \otimes \repbar3_q$ contain no singlet. Moreover, extracting an octet from $Y_{u,d}Y_{u,d}^\dagger$ does not alter this conclusion, since the subsequent tensor products likewise fail to yield a singlet. A similar argument applies when considering three Yukawa insertions. From the decomposition of $Y_{u,d}^3$, one may isolate the $Y_{u,d}^3\supset (\rep3_q\otimes\rep3_q\otimes\rep3_q,\rep1)$ components, where the singlet arises through fully antisymmetric $\varepsilon$-contractions in the $u$- or $d$-sector. However, on the $q$-side the relevant products $\rep3_q\otimes \rep3_q\otimes \rep3_q\otimes \rep3_q$ and $\rep3_q\otimes \rep3_q\otimes \rep3_q\otimes \repbar3_q$ contain no $\rep8_q$ representation that could be paired with $\Omega\sim\rep8_q$. As a consequence, even allowing for three Yukawa insertions, no flavor singlet involving a single $q$ field can be formed when $\Omega$ transforms as a flavor octet. Similar reasoning applies to $\Omega\sim \rep8_x$ with $x\in\{u,d\}$, where no singlet constructions are possible.

%%%%%%%%%%%%%%%%%%%%%%%%%%%%%%%%%%%%%%%%%%%%%%%%%%%%%%%%%%%%%%%%%%%%%%%%%%%%%%
\section{Extended MFV and cLFV} 
\label{app:cLFV_ex_MFV}
%%%%%%%%%%%%%%%%%%%%%%%%%%%%%%%%%%%%%%%%%%%%%%%%%%%%%%%%%%%%%%%%%%%%%%%%%%%%%%
\noindent
Charged-lepton flavor violation (cLFV) arises in the extended MFV framework through the same leptonic spurions that enter the BNV analysis. In the SM these effects are suppressed by neutrino masses, so any observable signal would be a clear indication of new physics. This appendix summarizes the leading MFV contributions to a representative set of low-energy cLFV processes: $\mu \to e\gamma$, $\mu \to 3e$, and $\mu- e$ conversion in nuclei.

%%%%%%%%%%%%%%%%%%%%%%%%%%%%%%%%%%%%%%%%%%%%%%%%%%%%
\vspace{0.2cm}
\noindent
{$\bm{\mu\to e\gamma}$\textbf{.}} The radiative decay $\mu\to e\gamma$ provides one of the strongest constraints on cLFV, with $\mathrm{BR}(\mu\to e\gamma)<4.2\times10^{\eminus13}$~\cite{MEG:2016leq} and an expected order-of-magnitude improvement from MEG II~\cite{MEG:2016leq,MEGII:2018kmf}. In the EFT framework, the dominant contribution arises from dipole operators of the form $\bar\ell\sigma^{\mu\nu}e$. Under extended MFV, the bilinear $\bar\ell\,e\sim \repbar3_\ell\otimes\rep3_e$ must be dressed with appropriate spurion insertions to form a flavor singlet. A single insertion of $Y_e$ produces only flavor-aligned structures and does not induce cLFV. Off-diagonal transitions require additional spurions, with the leading contribution generated by $\Upsilon_\nu \Upsilon_\nu^\dag Y_e$. Using the decomposition $\Upsilon_\nu \Upsilon_\nu^\dag=\rep1\oplus \rep 8_\ell\oplus\rep{27}_\ell$, the least-suppressed LFV invariant takes the form
\begin{equation}
    \cI_F^{\mu\to e\gamma}\sim \bar\ell_p [\Upsilon_\nu]^{ps}[\Upsilon_\nu^\dag]_{st}[Y_e]\ud{t}{r}e^r\,.
\end{equation}
The induced dipole Wilson coefficient is therefore
\begin{equation}
    [\cC_{e\gamma}]\ud{p}{r}=\frac{v}{\sqrt2}\frac{1}{\Lambda_{\sscript{LFV}}^2}[\Upsilon_\nu]^{p s}[\Upsilon_\nu^\dag]_{st}[Y_e]\ud{t}{r}\,,
\end{equation}
with $\Lambda_{\sscript{LFV}}$ the characteristic LFV scale. Inserting the spurion parametrization outlined in App.~\ref{app:GT_invariants}, the resulting branching ratio scales as
\begin{equation}
    \mathrm{BR}(\mu\to 3e)\approx\mathcal N^{(1)}_\nu\times \lzm\frac{\Lambda_{\sscript{L}}}{\Lambda_\sscript{LFV}}\dzm^4\times 10^{\eminus50}\,,
\end{equation}
where $\mathcal N^{(1)}_\nu\approx 90\,(30)$ for the case of NO (IO). Saturating the experimental bound yields
\begin{equation}\label{eq:mu_egamma_scales}
    \frac{\Lambda_\sscript{L}}{\Lambda_\sscript{LFV}}\Bigg\lvert_{\sscript{NO}}\lesssim 8\times10^8\,,
    \qquad
    \frac{\Lambda_\sscript{L}}{\Lambda_\sscript{LFV}}\Bigg\lvert_{\sscript{IO}}\lesssim10\times 10^8\,.
\end{equation}

%%%%%%%%%%%%%%%%%%%%%%%%%%%%%%%%%%%%%%%%%%%%%%%%%%%%
\vspace{0.1cm}
\noindent 
{$\bm{\mu\to 3e}$\textbf{.}} The three-body decay $\mu \to 3e$ probes a broader set of effective interactions in comparison to $\mu\to e\gamma$. Besides dipole operators, this process receives direct contributions from four-fermion contact operators, as well as the operators contributing to the corrections to the gauge boson vertices. The current bound is $\mathrm{BR}(\mu \to 3e) < 1.0 \times 10^{-12}$~\cite{SINDRUM:1987nra}, with future experiments such as Mu3e aiming to improve this sensitivity by several orders of magnitude~\cite{Blondel:2013ia}.
\\[5pt]
In MFV, the relevant contributions can be classified according to the flavor structure of the lepton bilinear that induces the transition. For instance, the $\bar\ell\gamma_\mu \ell$ bilinear contains an adjoint of $\U(3)_\ell$ in the decomposition, so the least-suppressed MFV invariant is obtained from $\Upsilon_\nu\,\Upsilon_\nu^\dag \supset \rep8_\ell$, implying a suppression of order $\cO(\Upsilon_\nu^2)$. An analogous construction applies to $\bar e\gamma_\mu e$ bilinear. Since $\Upsilon_\nu$ transforms as a $\U(3)_e$ singlet, the leading invariant requires both $Y_e^\dag$ and $Y_e$, producing an overall suppression $\cO(Y_e^2 \Upsilon_\nu^2)$. For dipole operators, the leading MFV invariant is $Y_e \Upsilon_\nu \Upsilon_\nu^\dag$, however, as in $\mu\to e\gamma$, the effective scaling of the branching ratio is ultimately controlled by $\Upsilon_\nu^2$ due to the cancellation of the explicit chiral factor in $\mathrm{BR}(\mu \to 3e)$.
\\[5pt]
After inserting the spurion parametrization, the branching ratio takes the form
\begin{equation}\label{eq:mu_to_3e_via_lambdas}
    \mathrm{BR}(\mu\to 3e)\approx\mathcal N_\nu^{(2)}\times \lzm\frac{\Lambda_{\sscript{L}}}{\Lambda_\sscript{LFV}}\dzm^4\times 10^{\eminus51}\,,
\end{equation}
where $\mathcal N_\nu^{(2)}\approx 2\,(50)$ for NO (IO). Imposing the experimental limit yields
\begin{equation}\label{eq:mu_3e_scales}
    \frac{\Lambda_\sscript{L}}{\Lambda_\sscript{LFV}}\Bigg\lvert_{\sscript{NO}}\lesssim 5\times10^9\,,
    \qquad
    \frac{\Lambda_\sscript{L}}{\Lambda_\sscript{LFV}}\Bigg\lvert_{\sscript{IO}}\lesssim 2\times 10^9\,.
\end{equation}

%%%%%%%%%%%%%%%%%%%%%%%%%%%%%%%%%%%%%%%%%%%%%%%%%%%%
\vspace{0.2cm}
\noindent
{$\bm{\mathrm{CR}(\mu- e,\,\mathrm{N})}$\textbf{.}} Unlike the purely leptonic processes $\mu \to e \gamma$ and $\mu \to 3e$, $\mu - e$ conversion is mediated by effective interactions involving both leptonic and quark currents. These include the vector-type, scalar-type as well as the dipole operators~\cite{Raidal:1997hq,Calibbi:2021pyh}. The current bound from SINDRUM II~\cite{SINDRUMII:2006dvw} is $\mathrm{CR}(\mu- e,\,\mathrm{Au})<7.0\times10^{\eminus13}$, with Mu2e~\cite{Mu2e:2014fns} and COMET~\cite{Fujii:2023vgo} expected to improve sensitivity down to $\cO(10^{\eminus17})$. As in the $\mu\to 3e$ case, the contributing SMEFT structures can be grouped according to the lepton bilinear, leading to the same MFV spurion power counting. 
\\[5pt]
The resulting conversion rate in gold can be written as
\begin{equation}
    \mathrm{CR}(\mu-e,\,\mathrm{Au})\approx\mathcal N_\nu^{(3)}\times\lzm\frac{\Lambda_{\sscript{L}}}{\Lambda_\sscript{LFV}}\dzm^4\times 10^{\eminus50}\,,
    \vspace{+0.2cm}
\end{equation}
where $\mathcal N^{(3)}_\nu\approx 10\,(4)$ for the case of NO (IO). Lastly, for the ratio of $\Lambda_\sscript{L}$ and $\Lambda_\sscript{LFV}$ scales we obtain
\begin{equation}\label{eq:mu_e_conv_scales}
    \frac{\Lambda_\sscript{L}}{\Lambda_\sscript{LFV}}\Bigg\lvert_{\sscript{NO}}\lesssim 10\times10^8\,,
    \qquad
    \frac{\Lambda_\sscript{L}}{\Lambda_\sscript{LFV}}\Bigg\lvert_{\sscript{IO}}\lesssim10\times 10^8\,.
\end{equation}

\bibliographystyle{JHEP}
\bibliography{main}

\end{document}

%% file: tables/Dim6_BNV_SMEFT_Spurion_Exp.tex
\centering
\scalebox{0.87}{
\begin{tabular}{c@{\hspace{1.0cm}}c@{\hspace{1.0cm}}c@{\hspace{0.8cm}}c}
\toprule
\textbf{Operator}
&\textbf{Definition} 
&\textbf{Spurion expansion}
&\textbf{Order}
\\
\midrule
\addlinespace[0.2cm] 
\multirow{2}{*}{\vspace{-0.5cm}$[\cO_{qqq\ell}]^{prst}$}
&\multirow{2}{*}{\vspace{-0.5cm}$\varepsilon_{\alpha\beta\gamma}[i\sig_2]_{\dot a\dot d}[i\sig_2]_{\dot b\dot e}(\bar q^{c\,\alpha,\dot a,p}q^{\beta,\dot b,r})(\bar q^{c\,\gamma,\dot e,s}\ell^{\dot d,t})$}
&$\frac{1}{\Lambda_\sscript{B}^2}\varepsilon_{prs} [\mathcal T_\ell]_t$
&$\cO(Y_e^2\Upsilon_\nu)$
\\[0.2cm]
&
&$\frac{1}{\Lambda_\sscript{B}^2}\lzv \varepsilon_{prx} [Y_u Y_u^\dagger]\ud{x}{s} 
    + \varepsilon_{pxs} [Y_u Y_u^\dagger]\ud{x}{r} + \varepsilon_{xrs} [Y_u Y_u^\dagger]\ud{x}{p} \dzv [\mathcal T_\ell]_t$
&$\cO(Y_u^2Y_e^2\Upsilon_\nu)$
\\[0.0cm]
\noalign{\vskip 4pt}
% \cmidrule{2-4}
\cdashline{1-4}[.4pt/2pt]
\noalign{\vskip 7pt}
$[\cO_{duq\ell}]^{prst}$
&$\varepsilon_{\alpha\beta\gamma}(\bar d^{c\,\alpha,p}u^{\beta,r})(\bar q^{c\,\gamma,s}i\sigma_2 \ell^t)$
&$\frac{1}{\Lambda_\sscript{B}^2}\varepsilon_{abs}[Y_u]\ud{a}{r}[Y_d]\ud{b}{p} [\mathcal T_\ell]_t$
&$\cO(Y_u Y_d Y_e^2 \Upsilon_\nu)$
\\[0.0cm]
\noalign{\vskip 4pt}
% \cmidrule{2-4}
\cdashline{1-4}[.4pt/2pt]
\noalign{\vskip 7pt}
$[\cO_{qque}]^{prst}$
&$\varepsilon_{\alpha\beta\gamma}(\bar q^{c\,\alpha,p}i\sigma_2 q^{\beta,r})(\bar u^{c\,\gamma,s}e^t)$
&$\frac{1}{2\Lambda_\sscript{B}^2}\lzv \varepsilon_{abp}[Y_uY_u^\dagger]\ud{b}{r} + \varepsilon_{abr}[Y_uY_u^\dagger]\ud{b}{p} \dzv[Y_u]\ud{a}{s} [\mathcal T_e]_t$
&$\cO(Y_u^3 Y_e^3 \Upsilon_\nu)$
\\[0.0cm]
\noalign{\vskip 4pt}
% \cmidrule{2-4}
\cdashline{1-4}[.4pt/2pt]
\noalign{\vskip 7pt}
\multirow{1}{*}{$[\cO_{duue}]^{prst}$}
&\multirow{1}{*}{$\varepsilon_{\alpha\beta\gamma}(\bar d^{c\,\alpha,p}u^{\beta,r})(\bar u^{c\,\gamma,s}e^t)$}
&$\frac{1}{\Lambda_\sscript{B}^2}\varepsilon_{ars}[Y_u^\dagger Y_d]\ud{a}{p}[\mathcal T_e]_t$
&$\cO(Y_uY_dY_e^3\Upsilon_\nu)$
\\[0.2cm]
\bottomrule
\end{tabular}
}

%% file: tables/Dim6_LEFT_matching.tex
\centering
\scalebox{0.77}{
\begin{tabular}{c@{\hspace{0.47cm}}c@{\hspace{0.47cm}}c}
\toprule
\textbf{Operator}
&\textbf{Definition} 
&\textbf{Tree-level matching}
\\

\midrule
\addlinespace[0.2cm]
$[\cO^{S,LL}_{udd}]_{prst}$
&$\varepsilon_{\alpha\beta\gamma}(\bar u_{L,p}^{c,\alpha}d_{L,r}^{\beta})(\bar d_{L,s}^{c,\gamma}\nu_{L,t})$
&$[\cC_{qqq\ell}]_{srpt}-[\cC_{qqq\ell}]_{rspt}+[\cC_{qqq\ell}]_{rpst}$
\\[0.3cm]
$[\cO^{S,LL}_{duu}]_{prst}$
&$\varepsilon_{\alpha\beta\gamma}(\bar d_{L,p}^{c,\alpha}u_{L,r}^\beta)(\bar u_{L,r}^{c,\gamma} e_{L,t})$
&$[\cC_{qqq\ell}]_{srpt}-[\cC_{qqq\ell}]_{rspt}+[\cC_{qqq\ell}]_{rpst}$
\\[0.3cm]
$[\cO^{S,LR}_{duu}]_{prst}$
&$\varepsilon_{\alpha\beta\gamma}(\bar d_{L,p}^{c,\alpha}u_{L,r}^\beta)(\bar u_{R,s}^{c,\gamma}e_{R,t})$
&$-[\cC_{qque}]_{prst}-[\cC_{qque}]_{rpst}$
\\[0.3cm]
$[\cO^{S,RL}_{duu}]_{prst}$
&$\varepsilon_{\alpha\beta\gamma}(\bar d_{R,p}^{c,\alpha}u_{R,r}^\beta)(\bar u_{L,s}^{c,\gamma}e_{L,t})$
&$[\cC_{duq\ell}]_{prst}$
\\[0.3cm]
$[\cO^{S,RL}_{dud}]_{prst}$
&$\varepsilon_{\alpha\beta\gamma}(\bar d_{R,p}^{c,\alpha}u_{R,r}^\beta)(\bar d_{L,s}^{c,\gamma}\nu_{L,t})$
&$-[\cC_{duq\ell}]_{prst}$
\\[0.3cm]
$[\cO^{S,RR}_{duu}]_{prst}$
&$\varepsilon_{\alpha\beta\gamma}(\bar d_{R,p}^{c,\alpha}u_{R,r}^\beta)(\bar u_{R,r}^{c,\gamma} e_{R,t})$
&$[\cC_{duue}]_{prst}$
\\[0.2cm]
\bottomrule
\end{tabular}
}

% \centering
% \scalebox{0.857}{
% \begin{tabular}{c@{\hspace{0.4cm}}c@{\hspace{0.4cm}}c@{\hspace{0.4cm}}c@{\hspace{0.4cm}}c@{\hspace{0.4cm}}c}
% \toprule
% \multirow{1}{*}{\textbf{Operator}}
% &\multirow{1}{*}{\textbf{Definition}} 
% &\textbf{Matching}
% &\multirow{1}{*}{\textbf{Operator}}
% &\multirow{1}{*}{\textbf{Definition} }
% &\textbf{Matching}
% \\
% \midrule
% $[\cO^{S,LL}_{udd}]_{prst}$
% &$\varepsilon_{\alpha\beta\gamma}(\bar u_{L,p}^{c,\alpha}d_{L,r}^{\beta})(\bar d_{L,s}^{c,\gamma}\nu_{L,t})$
% &$[\cC_{qqq\ell\ell}]_{srpt}-[\cC_{qqq\ell\ell}]_{rspt}+[\cC_{qqq\ell\ell}]_{rpst}$
% &$[\cO^{S,LR}_{duu}]_{prst}$
% &$\varepsilon_{\alpha\beta\gamma}(\bar d_{L,p}^{c,\alpha}u_{L,r}^\beta)(\bar u_{R,s}^{c,\gamma}e_{R,t})$
% &$-[\cC_{qquee}]_{prst}-[\cC_{qquee}]_{rpst}$
% \\[0.3cm]
% $[\cO^{S,LL}_{duu}]_{prst}$
% &$\varepsilon_{\alpha\beta\gamma}(\bar d_{L,p}^{c,\alpha}u_{L,r}^\beta)(\bar u_{L,r}^{c,\gamma} e_{L,t})$
% &$[\cC_{qqq\ell\ell}]_{srpt}-[\cC_{qqq\ell\ell}]_{rspt}+[\cC_{qqq\ell\ell}]_{rpst}$
% &$[\cO^{S,RL}_{duu}]_{prst}$
% &$\varepsilon_{\alpha\beta\gamma}(\bar d_{R,p}^{c,\alpha}u_{R,r}^\beta)(\bar u_{L,s}^{c,\gamma}e_{L,t})$
% &$[\cC_{duq\ell\ell}]_{prst}$
% \\[0.3cm]
% $[\cO^{S,RL}_{dud}]_{prst}$
% &$\varepsilon_{\alpha\beta\gamma}(\bar d_{R,p}^{c,\alpha}u_{R,r}^\beta)(\bar d_{L,s}^{c,\gamma}\nu_{L,t})$
% &$-[\cC_{duq\ell\ell}]_{prst}$
% &$[\cO^{S,RR}_{duu}]_{prst}$
% &$\varepsilon_{\alpha\beta\gamma}(\bar d_{R,p}^{c,\alpha}u_{R,r}^\beta)(\bar u_{R,r}^{c,\gamma} e_{R,t})$
% &$[\cC_{duue}]_{prst}$
% \\[0.2cm]
% \bottomrule
% \end{tabular}
% }

%% file: tables/Dim6_BNV_table_bounds.tex
\centering
\scalebox{0.94}{
\begin{tabular}{c@{\hspace{0.5cm}}c@{\hspace{0.5cm}}c@{\hspace{0.5cm}}c@{\hspace{0.5cm}}c@{\hspace{0.5cm}}cc}
\toprule
\textbf{Operator}
&\textbf{Spurion expansion}
& \textbf{$\bm{\cR_{\sscript{BL}}^{\sscript{up}}\,[\tev]}$}
&\textbf{Channel}
& \textbf{$\bm{\cR_{\sscript{BL}}^{\sscript{down}}\,[\tev]}$}
&\textbf{Channel}
&
\\
\midrule
\addlinespace[0.2cm] 
\multirow{2}{*}{\vspace{-0.5cm}$[\cO_{qqq\ell}]^{prst}$}
& $\frac{1}{\Lambda_\sscript{B}^2}\varepsilon_{prs} [\mathcal T_\ell]_t$
& $1\times10^7$
& $p \to K^+\nu$
& $1\times10^7$
& $p \to K^+\nu$
\\ [0.2cm]

& $\frac{1}{\Lambda_\sscript{B}^2}\lzv \varepsilon_{prx} [Y_u Y_u^\dagger]\ud{x}{s} 
    + \varepsilon_{pxs} [Y_u Y_u^\dagger]\ud{x}{r} + \varepsilon_{xrs} [Y_u Y_u^\dagger]\ud{x}{p} \dzv [\mathcal T_\ell]_t$
&$1\times10^7$
& $p \to K^+\nu$
&$1\times10^7$
& $p \to K^+\nu$
\\[0.2cm]
\noalign{\vskip 2pt}
\cdashline{1-6}[.4pt/2pt]
\noalign{\vskip 5pt}

$[\cO_{duq\ell}]^{prst}$
& $\frac{1}{\Lambda_\sscript{B}^2}\varepsilon_{abs}[Y_u]\ud{a}{r}[Y_d]\ud{b}{p} [\mathcal T_\ell]_t$
&$2\times10^{\eminus1}$
&$p\to \pi^0 e^+$
&$2\times10^{\eminus1}$
&$p\to \pi^0 e^+$
\\[0.0cm]
\noalign{\vskip 4pt}
% \cmidrule{2-4}
\cdashline{1-6}[.4pt/2pt]
\noalign{\vskip 7pt}
$[\cO_{qque}]^{prst}$
& $\frac{1}{2\Lambda_\sscript{B}^2}\lzv \varepsilon_{abp}[Y_uY_u^\dagger]\ud{b}{r} + \varepsilon_{abr}[Y_uY_u^\dagger]\ud{b}{p} \dzv[Y_u]\ud{a}{s} [\mathcal T_e]_t$
&$2\times10^{\eminus4}$
& $p \to \pi^0 \mu^+$
&$3\times10^{\eminus2}$
& ${\color{black}{p \to \pi^0 \mu^+}}$
\\[0.0cm]
\noalign{\vskip 4pt}
% \cmidrule{2-4}
\cdashline{1-6}[.4pt/2pt]
\noalign{\vskip 7pt}
$[\cO_{duue}]^{prst}$
& $\frac{1}{\Lambda_\sscript{B}^2}\varepsilon_{ars}[Y_u^\dagger Y_d]\ud{a}{p}[\mathcal T_e]_t$
&$2\times10^{\eminus7}$
& $p \to \pi^0 \mu^+$
& $7\times10^{\eminus6}$
& $p \to K^0 \mu^+$
\\[0.1cm]

\bottomrule
\end{tabular}
}

%% file: tables/LQ_overview_table.tex
\centering
\scalebox{0.81}{
\begin{tabular}{cccc@{\hspace{0.5cm}}c}
\toprule
\multicolumn{2}{c}{\textbf{Field}}
% &\textbf{Gauge irrep}
&\textbf{Type}
&\textbf{UV Lagrangian} 
&\textbf{BNV Operators generated}
\\
\midrule
$\omega_1$
&$(\rep{3},\rep{1})_{-1/3}$
&$S$
&$[y_{\omega_1}^{q\ell}]\ud{s}{pr}\omega_{1s}^{\dag}\bar q^{c\,p}i\sig_2\ell^{r}
			+[y_{\omega_1}^{qq}]^{spr}\omega_{1s}^{\alpha\dag}\varepsilon_{\alpha\beta\gamma}\bar q^\beta_{p}i\sig_2q^{c\gamma}_{r}
			+[y_{\omega_1}^{eu}]\ud{s}{pr}\omega_{1s}^{\dag}\bar e^{c\,p}u^{r}+[y_{\omega_1}^{du}]^{spr}\omega_{1s}^{\alpha\dag}\varepsilon_{\alpha\beta\gamma}\bar d^\beta_{p}u^{c\gamma}_{r}
			+\hermc$
&$\cO_{duq\ell}$, $\cO_{qque}$, $\cO_{qqq\ell}$, $\cO_{duue}$
\\[0.2cm]
$\omega_4$
&$(\rep{3},\rep{1})_{-4/3}$
&$S$
&$[y_{\omega_4}^{ed}]\ud{s}{pr}\omega_{4s}^{\dag}\bar e^{c\,p}d^{r}
		+
		[y_{\omega_4}^{uu}]^{spr}\omega_{4s}^{\alpha\dag}\varepsilon_{\alpha\beta\gamma}\bar u_{p}^\beta u_{r}^{c\gamma}
		+\hermc$
&$\cO_{duue}$
\\[0.2cm]
$\zeta$
&$(\rep{3},\rep{3})_{-1/3}$
&$S$
&$[y_\zeta^{q\ell}]\ud{s}{pr}\zeta^{A\dag}_s\bar q^{c\,p}i\sig_2\sig^A\ell^{r}
		+
		[y_\zeta^{qq}]^{spr}\zeta^{A\alpha\dag}_s\varepsilon_{\alpha\beta\gamma}\bar q^\beta_{p}\sig^Ai\sig_2q^{c\gamma}_{r}
		+\hermc$
&$\cO_{qqq\ell}$
\\[0.1cm]
\noalign{\vskip 2pt}
% \cmidrule{1-4}
\cdashline{1-5}[.4pt/2pt]
\noalign{\vskip 2pt}
$\cQ_1$
&$(\rep{3},\rep{2})_{1/6}$
&$V$
&$[g_{\cQ_1}^{u\ell}]\ud{s}{pr} \cQ_{1s}^{\mu\dag}\bar u^{c\,p}\gamma_\mu \ell^{r}
		+[g_{\cQ_1}^{dq}]^{spr}\cQ_{1s}^{\mu \alpha\dag}\varepsilon_{\alpha\beta\gamma}\bar d^\beta_{p}\gamma_\mu i\sig_2q^{c\gamma}_{r}
		+\hermc$
&$\cO_{duq\ell}$
\\[0.2cm]
$\cQ_5$
&$(\rep{3},\rep{2})_{-5/6}$
&$V$
&$[g_{\cQ_5}^{d\ell}]\ud{s}{pr}\cQ_{5s}^{\mu\dag}\bar d^{c\,p}\gamma_\mu \ell^{r}
			+
			[g_{\cQ_5}^{eq}]\ud{s}{pr}\cQ_{5s}^{\mu\dag}\bar e^{c\,p}\gamma_\mu q^{r}
			+
			[g_{\cQ_5}^{uq}]^{spr}\cQ_{5s}^{\mu\alpha\dag}\varepsilon_{\alpha\beta\gamma}\bar u^\beta_{p}\gamma_\mu q^{c\gamma}_{r}+\hermc$
&$\cO_{duq\ell}$, $\cO_{qque}$
\\
\bottomrule
\end{tabular}
}

%% file: tables/LQ_scalars_flavor.tex
\centering
\scalebox{0.93}{
\begin{tabular}{c@{\hspace{1.5cm}}c@{\hspace{1.cm}}c@{\hspace{1.5cm}}c@{\hspace{1.cm}}c}
\toprule
$\bm{\omega_1}$\textbf{ irreps}
&$\bm{y_{\omega_1}^{q\ell}}$
&$\bm{y_{\omega_1}^{qq}}$
&$\bm{\cO_{qqq\ell}}$
&$\bm{\cO_{duq\ell}}$
\\
\midrule
$\rep3_q$
&$\delta\ud{s}{p}[\mathcal{T}_\ell]_r$
&$\varepsilon^{sx(p|}[Y_u Y_u^\dag]\ud{|r)}{x}$
& $\cO(Y_u^2Y_e^2\Upsilon_\nu)$ & $\cO(Y_uY_dY_e^2\Upsilon_\nu)$
\\[0.2cm]
$\rep3_u$
& $[Y_u^\dag]\ud{s}{p}[\mathcal{T}_\ell]_r$ 
& $[Y_u^\dag]\ud{s}{y}\varepsilon^{yx(p|}[Y_u Y_u^\dag]\ud{|r)}{x}$
% & ${\color{red}{\varepsilon^{sxy}[Y_u]\ud{(p|}{x}[Y_u]\ud{|r)}{y}}}$
& $\cO(Y_u^{{\color{black}{4}}}Y_e^2\Upsilon_\nu)$ & $\cO(Y_u^2Y_dY_e^2\Upsilon_\nu)$
\\[0.2cm]
$\repbar6_q$
&$\varepsilon_{yp(s|}[Y_uY_u^\dag]\ud{y}{|s')}[\mathcal{T}_\ell]_r$
&${\color{black}{\delta\ud{(p|}{s}\delta\ud{|r)}{s'}}}$
& $\cO(Y_u^2Y_e^2\Upsilon_\nu)$ & $\cO(Y_u^3Y_dY_e^2\Upsilon_\nu)$
\\[0.2cm]
$(\rep3_q,\rep3_\ell)$
&$\delta\ud{s}{p}\,(\delta\ud{s'}{r}+[\Upsilon_\nu \Upsilon_\nu^\dag]\ud{s'}{r})$ 
&$[\mathcal{T}_\ell^\dag]^{s'}\varepsilon^{sx(p|}[Y_u Y_u^\dag]\ud{|r)}{x}$
& $\cO(Y_u^2Y_e^2\Upsilon_\nu)$ & $\cO(Y_uY_dY_e^2\Upsilon_\nu)$
\\[0.2cm]
$(\rep3_u,\rep3_\ell)$
&$[Y_u^\dag]\ud{s}{p}\,(\delta\ud{s'}{r}+[\Upsilon_\nu \Upsilon_\nu^\dag]\ud{s'}{r})$ 
&$[\mathcal{T}_\ell^\dag]^{s'}[Y_u^\dag]\ud{s}{y}\varepsilon^{yx(p|}[Y_u Y_u^\dag]\ud{|r)}{x}$
% &${\color{red}{[\mathcal{T}_\ell^\dag]^{s'}\varepsilon^{sxy}[Y_u]\ud{(p|}{x}[Y_u]\ud{|r)}{y}}}$
& $\cO(Y_u^{{\color{black}{4}}}Y_e^2\Upsilon_\nu)$ & $\cO(Y_u^2Y_dY_e^2\Upsilon_\nu)$
\\[0.2cm]
$(\rep3_q,\repbar3_\ell)$
&$\delta\ud{s}{p}[\Upsilon_\nu^\dag]_{s'r}$ 
&$[\mathcal{T}_\ell]_{s'}\varepsilon^{sx(p|}[Y_u Y_u^\dag]\ud{|r)}{x}$
& $\cO(Y_u^2Y_e^2\Upsilon_\nu^2)$ & $\cO(Y_uY_dY_e^2\Upsilon_\nu^2)$
\\[0.1cm]
\midrule
$\bm{\omega_1}$\textbf{ irreps}
&$\bm{y_{\omega_1}^{eu}}$
&$\bm{y_{\omega_1}^{du}}$
&$\bm{\cO_{\ell q}^{(1,3)}}$
&$\bm{\cO_{e u}}$
\\
\midrule
$\rep3_q$
&$[Y_u]\ud{s}{r}[\mathcal{T}_e]_p$ 
& $\varepsilon^{sxy}[Y_d^\dag]\ud{p}{x}[Y_u^\dag]\ud{r}{y}$
& $\cO(Y_e^4\Upsilon_\nu^2)$ & $\cO(Y_u^2Y_e^6\Upsilon_\nu^2)$
\\[0.2cm]
$\rep3_u$
&$\delta\ud{s}{r}[\mathcal{T}_e]_{p}$ 
&$\varepsilon^{srx}[Y_d^\dag Y_u]\ud{p}{x}$
& $\cO(Y_u^2Y_e^4\Upsilon_\nu^2)$ & $\cO(Y_e^6\Upsilon_\nu^2)$
\\[0.2cm]
$\repbar6_q$
&$\varepsilon_{yx(s|}[Y_uY_u^\dag]\ud{y}{|s')}[Y_u]\ud{x}{r}[\mathcal{T}_e]_p$ 
% &${\color{red}{\varepsilon_{rxy}[Y_u^\dag]\ud{x}{(s|}[Y_u^\dag]\ud{y}{|s')}[\cT_e]_p  }}$ 
& $[Y_d^\dag]\ud{p}{(s|}[Y_u^\dag]\ud{r}{|s')}$
& $\cO(Y_u^4Y_e^4\Upsilon_\nu^2)$ & $\cO(Y_{u}^{{\color{black}{6}}}Y_e^6\Upsilon_\nu^2)$
\\[0.2cm]
$(\rep3_q,\rep3_\ell)$
& $[Y_u]\ud{s}{r}(\delta\ud{s'}{x}+[\Upsilon_\nu \Upsilon_\nu^\dag]\ud{s'}{x})[Y_e]\ud{x}{p}$ & $\varepsilon^{sxy}[Y_d^\dag]\ud{p}{x}[Y_u^\dag]\ud{r}{y}[\mathcal{T}_\ell^\dag]^{s'}$
& $\cO(\Upsilon_\nu^2)$ & $\cO(Y_u^2Y_e^2\Upsilon_\nu^2)$
\\[0.2cm]
$(\rep3_u,\rep3_\ell)$
&$\delta\ud{s}{r}(\delta\ud{s'}{x}+[\Upsilon_\nu \Upsilon_\nu^\dag]\ud{s'}{x})[Y_e]\ud{x}{p}$ 
&$\varepsilon^{srx}[Y_d^\dag Y_u]\ud{p}{x}[\mathcal{T}_\ell^\dag]^{s'}$
& $\cO(Y_u^2\Upsilon_\nu^2)$ & $\cO(Y_e^2\Upsilon_\nu^2)$
\\[0.2cm]
$(\rep3_q,\repbar3_\ell)$
&$[Y_u]\ud{s}{r}[\Upsilon_\nu^\dag Y_e]_{s'p}$ 
&$\varepsilon^{sxy}[Y_d^\dag]\ud{p}{x}[Y_u^\dag]\ud{r}{y}[\mathcal{T}_\ell]_{s'}$
& $\cO(\Upsilon_\nu^2)$ & $\cO(Y_u^2Y_e^2\Upsilon_\nu^2)$
\\[0.1cm]
\midrule
\midrule
$\bm{\omega_4}$\textbf{ irreps}
&$\bm{y_{\omega_4}^{ed}}$
&$\bm{y_{\omega_4}^{uu}}$
&$\bm{\cO_{duue}}$
&$\bm{\cO_{ed}}$
\\
\midrule
$\rep3_q$ & $[Y_d]\ud{s}{r}[\mathcal{T}_e]_p$ & $[Y_u]\ud{s}{x}\varepsilon^{xpr}$ & $\cO(Y_u Y_dY_e^3\Upsilon_\nu)$ & $\cO (Y_d^2Y_e^6\Upsilon_\nu^2)$
\\[0.2cm]
$\rep3_d$ & $\delta\ud{s}{r}[\mathcal{T}_e]_p$ & $[Y_d^\dag Y_u]\ud{s}{x}\varepsilon^{xpr}$ & $\cO(Y_u Y_dY_e^3\Upsilon_\nu)$ & $\cO (Y_e^6\Upsilon_\nu^2)$
\\[0.2cm]
$(\rep3_q,\rep3_\ell)$ & $[Y_d]\ud{s}{r}(\delta\ud{s'}{y}+[\Upsilon_\nu \Upsilon_\nu^\dag ]\ud{s'}{y})[Y_e]\ud{y}{p}$ & $[\mathcal{T}_\ell^\dag]^{s'}[Y_u]\ud{s}{x}\varepsilon^{xpr}$ & $\cO(Y_uY_dY_e^3\Upsilon_\nu)$ & $\cO(Y_d^2Y_e^2\Upsilon_\nu^2)$
\\[0.2cm]
$(\rep3_d,\rep3_\ell)$ & $\delta\ud{s}{r}(\delta\ud{s'}{y}+[\Upsilon_\nu \Upsilon_\nu^\dag]\ud{s'}{y})[Y_e]\ud{y}{p}$ & $[\mathcal{T}_\ell^\dag]^{s'}[Y_d^\dag Y_u]\ud{s}{x}\varepsilon^{xpr}$ & $\cO(Y_uY_dY_e^3\Upsilon_\nu)$ & $\cO(Y_e^2\Upsilon_\nu^2)$
\\[0.2cm]
$(\rep3_d,\repbar3_\ell)$ & $\delta\ud{s}{r}[\Upsilon_\nu^\dag Y_e]_{s'p}$ & $[\mathcal{T}_\ell]_{s'}[Y_d^\dag Y_u]\ud{s}{x}\varepsilon^{xpr}$ & $\cO(Y_uY_dY_e^3\Upsilon_\nu^2)$ & $\cO(Y_e^2\Upsilon_\nu^2)$
\\[0.1cm]
\midrule
\midrule
$\bm{\zeta}$\textbf{ irreps}
&$\bm{y_{\zeta}^{q\ell}}$
&$\bm{y_{\zeta}^{qq}}$
&$\bm{\cO_{qqq\ell}}$
&$\bm{\cO_{\ell q}^{(1,3)}}$
\\
\midrule
$\rep3_q$ & $\delta\ud{s}{p}[\mathcal{T}_\ell]_r$ & $\varepsilon^{spr}$ & $\cO(Y_e^2\Upsilon_\nu)$ & $\cO (Y_e^4\Upsilon_\nu^2)$
\\[0.2cm]
$\rep3_u$ & $[Y_u^\dag]\ud{s}{p}[\mathcal{T}_\ell]_r$ & $[Y_u^\dag]\ud{s}{y}\varepsilon^{ypr}$ & $\cO(Y_u^2Y_e^2\Upsilon_\nu)$ & $\cO(Y_u^2Y_e^4\Upsilon_\nu^2)$
\\[0.2cm]
$(\rep3_q,\rep3_\ell)$ & $\delta\ud{s}{p}( \delta\ud{s'}{r}+[\Upsilon_\nu \Upsilon_\nu^\dag]\ud{s'}{r} )$ & $[\mathcal{T}_\ell^\dag]^{s'}\varepsilon^{spr}$ & $\cO(Y_e^2\Upsilon_\nu)$ & $\cO(\Upsilon_\nu^2)$ 
\\[0.2cm]
$(\rep3_u,\rep3_\ell)$ & $[Y_u^\dag]\ud{s}{p}( \delta\ud{s'}{r}+[\Upsilon_\nu \Upsilon_\nu^\dag]\ud{s'}{r} )$ & $[\mathcal{T}_\ell^\dag]^{s'}[Y_u^\dag]\ud{s}{y}\varepsilon^{ypr}$ & $\cO(Y_u^2Y_e^2\Upsilon_\nu)$ & $\cO(Y_u^2\Upsilon_\nu^2)$
\\[0.2cm]
$(\rep3_q,\repbar3_\ell)$ & $\delta\ud{s}{p}[\Upsilon_\nu^\dag]_{s'r}$ & $[\mathcal{T}_\ell]_{s'}\varepsilon^{spr}$ & $\cO(Y_e^2\Upsilon_\nu^2)$ & $\cO(\Upsilon_\nu^2)$
\\[0.1cm]
\bottomrule
\end{tabular}
}

%% file: tables/LQ_vectors_flavor.tex
\centering
\scalebox{0.93}{
\begin{tabular}{c@{\hspace{1.5cm}}c@{\hspace{1cm}}c@{\hspace{1.5cm}}c@{\hspace{1cm}}c}
\toprule
$\bm{\cQ_1}$\textbf{ irreps}
&$\bm{g_{\cQ_1}^{u\ell}}$
&$\bm{g_{\cQ_1}^{dq}}$
&$\bm{\cO_{duq\ell}}$
&$\bm{\cO_{\ell u}}$
\\
\midrule
$\rep3_q$ & $[Y_u]\ud{s}{p}[\mathcal{T}_\ell]_r$ & $[Y_d^\dag]\ud{p}{x}\varepsilon^{sxr}$ & $\cO(Y_u Y_dY_e^2\Upsilon_\nu)$ & $\cO(Y_u^2Y_e^4\Upsilon_\nu^2)$
\\[0.2cm]
$\rep3_u$ & $\delta\ud{s}{p}[\mathcal{T}_\ell]_r$ & $[Y_u^\dag]\ud{s}{x}[Y_d^\dag]\ud{p}{y}\varepsilon^{xyr}$ & $\cO(Y_u Y_dY_e^2\Upsilon_\nu)$ & $\cO (Y_e^4\Upsilon_\nu^2)$
\\[0.2cm]
$\repbar6_q$ 
&$\varepsilon_{yx(s|}[Y_uY_u^\dag]\ud{x}{|s')}[Y_u]\ud{y}{p}[\mathcal{T}_\ell]_r$ 
% &${\color{red}{\varepsilon^{pxy}[Y_u^\dag]\ud{x}{(s|}[Y_u^\dag]\ud{y}{|s')}[\mathcal{T}_\ell]_r}}$ 
& $[Y_d^\dag]\ud{p}{(s|}\delta\ud{r}{|s')}$ 
& $\cO(Y_u^{{\color{black}{3}}} Y_dY_e^2\Upsilon_\nu)$ 
& $\cO(Y_u^{{\color{black}{6}}}Y_e^4\Upsilon_\nu^2)$
\\[0.2cm]
$(\rep3_q,\rep3_\ell)$ & $[Y_u^\dag]\ud{s}{p}(\delta\ud{s'}{r}+[\Upsilon_\nu \Upsilon_\nu^\dag]\ud{s'}{r})$ & $[\mathcal{T}_\ell^\dag]^{s'}[Y_d^\dag]\ud{p}{x}\varepsilon^{sxr}$ & $\cO(Y_uY_dY_e^2\Upsilon_\nu)$ & $\cO(Y_u^2\Upsilon_\nu^2)$
\\[0.2cm]
$(\rep3_u,\rep3_\ell)$ & $\delta\ud{s}{p}(\delta\ud{s'}{r}+[\Upsilon_\nu \Upsilon_\nu^\dag]\ud{s'}{r})$ & $[\mathcal{T}_\ell^\dag]^{s'}[Y_u^\dag]\ud{s}{x}[Y_d^\dag]\ud{p}{y}\varepsilon^{xyr}$ & $\cO(Y_uY_dY_e^2\Upsilon_\nu)$ & $\cO(\Upsilon_\nu^2)$ 
\\[0.2cm]
$(\rep3_u,\repbar3_\ell)$ & $\delta\ud{s}{p}[\Upsilon_\nu^\dag]_{s'r}$ & $[\mathcal{T}_\ell]_{s'}[Y_u^\dag]\ud{s}{x}[Y_d^\dag]\ud{p}{y}\varepsilon^{xyr}$ & $\cO(Y_uY_dY_e^2\Upsilon_\nu^2)$ & $\cO(\Upsilon_\nu^2)$
\\[0.1pt]
\midrule
\midrule
$\bm{\cQ_5}$\textbf{ irreps}
&$\bm{g_{\cQ_5}^{d\ell}}$
&$\bm{g_{\cQ_5}^{eq}}$
&$\bm{\cO_{duq\ell}}$
&$\bm{\cO_{qque}}$
\\
\midrule
$\rep3_q$ & $[Y_d]\ud{s}{p}[\mathcal{T}_\ell]_r$ & $\delta\ud{s}{r}[\mathcal{T}_e]_p$ &
$\cO(Y_uY_dY_e^2\Upsilon_\nu)$ & $\cO(Y_uY_e^3\Upsilon_\nu)$
\\[0.2cm]
$\rep3_d$ &$\delta\ud{s}{p}[\mathcal{T}_\ell]_r$ & $[Y_d^\dag]\ud{s}{r}[\mathcal{T}_e]_p$
& $\cO(Y_uY_dY_e^2\Upsilon_\nu)$ & $\cO(Y_uY_d^2Y_e^3\Upsilon_\nu)$
\\[0.2cm]
$\repbar6_q$ 
& $\varepsilon_{yx(s|}[Y_uY_u^\dag]\ud{x}{|s')}[Y_d]\ud{y}{p}[\mathcal{T}_\ell]_r$ 
% &${\color{red}{\varepsilon^{pxy}[Y_d^\dag]\ud{x}{(s|}[Y_d^\dag]\ud{y}{|s')}[\mathcal{T}_\ell]_r}}$
& $\varepsilon_{rx(s|}[Y_uY_u^\dag]\ud{x}{|s')}[\mathcal{T}_e]_p$ 
&$\cO(Y_u^3Y_dY_e^2\Upsilon_\nu)$ 
& $\cO(Y_u^3Y_e^3\Upsilon_\nu)$
\\[0.2cm]
$(\rep3_q,\rep3_\ell)$&$[Y_d]\ud{s}{p}\delta\ud{s'}{r}$ & $\delta\ud{s}{r}(\delta\ud{s'}{y}+[\Upsilon_\nu \Upsilon_\nu^\dag ]\ud{s'}{y})[Y_e]\ud{y}{p}$
& $\cO(Y_uY_dY_e^2\Upsilon_\nu)$ & $\cO(Y_uY_e^3\Upsilon_\nu)$
\\[0.2cm]
$(\rep3_d,\rep3_\ell)$
& $\delta\ud{s}{p}\delta\ud{s'}{r}$ & $[Y_d^\dag]\ud{s}{r}(\delta\ud{s'}{y}+[\Upsilon_\nu \Upsilon_\nu^\dag ]\ud{s'}{y})[Y_e]\ud{y}{p}$
& $\cO(Y_uY_dY_e^2\Upsilon_\nu)$ & $\cO(Y_uY_d^2Y_e^3\Upsilon_\nu)$
\\[0.2cm]
$(\rep3_q,\repbar3_\ell)$
& $[Y_d]\ud{s}{p}[\Upsilon_\nu^\dag]_{s'r}$ & $\delta\ud{s}{r}[\Upsilon_\nu^\dag Y_e]_{s'p}$
& $\cO(Y_uY_dY_e^2\Upsilon_\nu^2)$ & $\cO(Y_uY_e^3\Upsilon_\nu^2)$
\\[0.1cm]
\midrule
$\bm{\cQ_5}$\textbf{ irreps}
&$\bm{g_{\cQ_5}^{uq}}$
&
&$\bm{\cO_{qu}^{(1,8)}}$
&$\bm{\cO}_{qe}$
\\
\midrule
$\rep3_q$ & $[Y_u^\dag]\ud{p}{x}\varepsilon^{sxr}$
& 
&$\cO(Y_u^2)$ & $\cO(Y_e^6\Upsilon_\nu^2)$
\\[0.2cm]
$\rep3_d$& $[Y_u^\dag]\ud{p}{x}[Y_d^\dag]\ud{s}{y}\varepsilon^{xyr}$
&& $\cO(Y_u^2Y_d^2)$ & $\cO(Y_d^2Y_e^6\Upsilon_\nu^2)$
\\[0.2cm]
$\repbar6_q$ & $[Y_u^\dag]\ud{p}{(s|}\delta\ud{r}{|s')}$
& 
&$\cO(Y_u^2)$ &$\cO(Y_u^4Y_e^6\Upsilon_\nu^2)$
\\[0.2cm]
$(\rep3_q,\rep3_\ell)$& $[\mathcal{T}_\ell^\dag]^{s'}[Y_u^\dag]\ud{p}{x}\varepsilon^{sxr}$
&& $\cO(Y_u^2Y_e^4\Upsilon_\nu^2)$ & $\cO(Y_e^2\Upsilon_\nu^2)$
\\[0.2cm]
$(\rep3_d,\rep3_\ell)$
& $[\mathcal{T}_\ell^\dag]^{s'}[Y_u^\dag]\ud{p}{x}[Y_d^\dag]\ud{s}{y}\varepsilon^{xyr}$
&& $\cO(Y_u^2Y_d^2Y_e^4\Upsilon_\nu^2)$ & $\cO(Y_d^2Y_e^2\Upsilon_\nu^2)$
\\[0.2cm]
$(\rep3_q,\repbar3_\ell)$
& $[\mathcal{T}_\ell]_{s'}[Y_u^\dag]\ud{p}{x}\varepsilon^{sxr}$
&& $\cO(Y_u^2Y_e^4\Upsilon_\nu^2)$ &$\cO(Y_e^2\Upsilon_\nu^2)$
\\[0.1cm]
\bottomrule
\end{tabular}
}

%% file: tables/Dim6_BNV_SMEFT_Spurion_Exp_U2.tex
\centering
\scalebox{1.0}{
\begin{tabular}{c@{\hspace{1.5cm}}c@{\hspace{1.5cm}}c@{\hspace{1.3cm}}c@{\hspace{1.3cm}}c}
\toprule
\textbf{Operator}
&\textbf{Invariant} 
&\textbf{Component} 
&\textbf{$\bm{\Lambda_{\sscript B}^{\sscript{up}}}$ [TeV]}
& \textbf{$\bm{\Lambda_{\sscript B}^{\sscript{down}}}$ [TeV]}
\\
\midrule
\addlinespace[0.2cm] 
\multirow{3}{*}{\vspace{-0.5cm}$[\cO_{qqq\ell}]^{prst}$}
&$\frac{1}{\Lambda_\sscript{B}^2}\delta_{p3} \delta_{r3}\delta_{s3}\delta_{t3}$ 
&S
& $2 \times 10^{9}$
& $2 \times 10^{8}$
\\[0.2cm]
&$\frac{1}{\Lambda_\sscript{B}^2} \left[\varepsilon_{pr} \delta_{s3}+\varepsilon_{sp} \delta_{r3}+\varepsilon_{rs} \delta_{p3}\right]\delta_{t3} $ 
&A
&$1 \times 10^{11}$
&$1 \times 10^{11}$
\\[0.2cm]
&$\frac{1}{\Lambda_\sscript{B}^2}\left[\varepsilon_{pr} \delta_{s3} + \varepsilon_{sr} \delta_{p3}\right]\delta_{t3}$ 
&M
&$3 \times 10^{11}$&$2 \times 10^{11}$
\\[0.0cm]
\noalign{\vskip 4pt}
\cdashline{1-5}[.4pt/2pt]
\noalign{\vskip 7pt}
$[\cO_{duq\ell}]^{prst}$
&$\frac{1}{\Lambda_\sscript{B}^2}\delta_{p3} \delta_{r3}\delta_{s3}\delta_{t3}$
&-
& $4 \times 10^8$ & $3 \times 10^7$
\\[0.0cm]
\noalign{\vskip 4pt}
\cdashline{1-5}[.4pt/2pt]
\noalign{\vskip 7pt}
$[\cO_{qque}]^{prst}$
&$\frac{1}{\Lambda_\sscript{B}^2}\delta_{p3} \delta_{r3}\delta_{s3}\delta_{t3}$
&-
& $1 \times 10^6$ & $5 \times 10^7$
\\[0.0cm]
\noalign{\vskip 4pt}
\cdashline{1-5}[.4pt/2pt]
\noalign{\vskip 7pt}
\multirow{3}{*}{$[\cO_{duue}]^{prst}$}
&$\frac{1}{\Lambda_\sscript{B}^2}\delta_{p3} \delta_{r3}\delta_{s3}\delta_{t3}$
&S
& $ 1\times10^7$ & $1\times10^7$
\\[0.2cm]
&$\frac{1}{\Lambda_\sscript{B}^2}\delta_{p3} \varepsilon_{rs}\delta_{t3}$
&A
&$2\times 10^7$&$2\times 10^7$
\\[0.0cm]
\bottomrule
\end{tabular}
}

%% file: tables/Dim6_BNV_SMEFT_Spurion_Exp_relaxed.tex
\centering
\scalebox{0.90}{
\begin{tabular}
{c@{\hspace{1.0cm}}c@{\hspace{0.7cm}}c@{\hspace{0.8cm}}c@{\hspace{0.8cm}}c@{\hspace{0.8cm}}c@{\hspace{0.8cm}}c}
\toprule
\textbf{Operator}
&\textbf{Spurion expansion}
&\textbf{Component}
&\textbf{$\bm{\cR_{\sscript{BL}}^{\sscript{up}}\,[\tev]}$} 
& \textbf{Channel} 
& \textbf{$\bm{\cR_{\sscript{BL}}^{\sscript{down}}\,[\tev]}$} 
& \textbf{Channel}
\\
\midrule
\addlinespace[0.2cm] 
\multirow{3}{*}{\vspace{-0.5cm}$[\cO_{qqq\ell}]^{prst}$}
&$\frac{1}{\Lambda_\sscript{B}^2}\delta_{p3} \delta_{r3}\delta_{s3}[\mathcal{T}_\ell]_{t}$ 
&S
& $80$ & $p \to K^+\nu$ & $7 $ & $p \to \pi^0 e^+$
\\[0.2cm]
&$\frac{1}{\Lambda_\sscript{B}^2} \left[\varepsilon_{pr} \delta_{s3}+\varepsilon_{sp} \delta_{r3}+\varepsilon_{rs} \delta_{p3}\right][\mathcal{T}_\ell]_{t}$
&A
& $10^7$ & $p \to K^+\nu$ & $10^7$ & $p \to K^+\nu$
\\[0.2cm]
&$\frac{1}{\Lambda_\sscript{B}^2} \left[\varepsilon_{pr} \delta_{s3} + \varepsilon_{sr} \delta_{p3}\right][\mathcal{T}_\ell]_{t}$ 
&M
& $10^7$ & $p \to K^+\nu$ & $9 \times 10^6$ & $p \to K^+\nu$
\\[0.0cm]
\noalign{\vskip 4pt}
\cdashline{1-7}[.4pt/2pt]
\noalign{\vskip 7pt}
$[\cO_{duq\ell}]^{prst}$
&$\frac{1}{\Lambda_\sscript{B}^2}\delta_{p3} \delta_{r3}\delta_{s3}[\mathcal{T}_\ell]_{t}$
&-
& 1 & $p \to K^+\nu$ & $10^{\eminus1}$ & $p \to \pi^0 e^+$
\\[0.0cm]
\noalign{\vskip 4pt}
\cdashline{1-7}[.4pt/2pt]
\noalign{\vskip 7pt}
$[\cO_{qque}]^{prst}$
&$\frac{1}{\Lambda_\sscript{B}^2}\delta_{p3} \delta_{r3}\delta_{s3}[\mathcal{T}_e]_{t}$
&-
& $10^{\eminus4}$ & $p \to K^+\nu$ & $10^{\eminus1}$ & $p \to K^0 \mu^+$
\\[0.0cm]
\noalign{\vskip 4pt}
\cdashline{1-7}[.4pt/2pt]
\noalign{\vskip 7pt}
\multirow{3}{*}{$[\cO_{duue}]^{prst}$}
&$\frac{1}{\Lambda_\sscript{B}^2}\delta_{p3} \delta_{r3}\delta_{s3}[\mathcal{T}_e]_{t}$
&S
& $10^{\eminus8}$ & $p \to K^+\nu$ & $10^{\eminus8}$ & $p \to K^+\nu$
\\[0.2cm]
&$\frac{1}{\Lambda_\sscript{B}^2}\delta_{p3} \varepsilon_{rs}[\mathcal{T}_e]_{t}$
&A
&$6 \times 10^{\eminus4}$ & $p \to \pi^0 \mu^+$ & $6 \times 10^{\eminus4}$ &  $p \to \pi^0 \mu^+$
\\[0.2cm]
\midrule
\midrule
\addlinespace[0.2cm]
\multirow{3}{*}{\vspace{-0.5cm}$[\cO_{qqq\ell}]^{prst}$}
&$\frac{1}{\Lambda_\sscript{B}^2}\delta_{p3} \delta_{r3}\delta_{s3}[\mathcal{T}_\ell]_{t}$ 
&S
& $80$ & $p \to K^+\nu$ & $7 $ & $p \to \pi^0 e^+$
\\[0.2cm]
&$\frac{1}{\Lambda_\sscript{B}^2} \left[\varepsilon_{pr} \delta_{s3}+\varepsilon_{sp} \delta_{r3}+\varepsilon_{rs} \delta_{p3}\right][\mathcal{T}_\ell]_{t}$ 
&A
& $10^7$ & $p \to K^+\nu$ & $10^7$ & $p \to K^+\nu$
\\[0.2cm]
&$\frac{1}{\Lambda_\sscript{B}^2} \left[\varepsilon_{pr} \delta_{s3} + \varepsilon_{sr} \delta_{p3}\right][\mathcal{T}_\ell]_{t}$ 
&M
& $10^7$ & $p \to K^+\nu$ & $9 \times 10^6$ & $p \to K^+\nu$
\\[0.0cm]
\noalign{\vskip 4pt}
\cdashline{1-7}[.4pt/2pt]
\noalign{\vskip 7pt}
$[\cO_{duq\ell}]^{prst}$
&$\frac{1}{\Lambda_\sscript{B}^2}[\Lambda_d^\dag]_p \delta_{r3}\delta_{s3}[\mathcal{T}_\ell]_{t}$
&-
& $2\times 10^{\eminus2}$ & $p\to K^+\nu$ & $4 \times 10^{\eminus3}$ & $p\to \pi^0 e^+$
\\[0.0cm]
\noalign{\vskip 4pt}
\cdashline{1-7}[.4pt/2pt]
\noalign{\vskip 7pt}
$[\cO_{qque}]^{prst}$
&$\frac{1}{\Lambda_\sscript{B}^2}\delta_{p3} \delta_{r3}\delta_{s3}[\mathcal{T}_e]_{t}$
&-
& $10^{\eminus4}$ & $p \to K^+\nu$ & $10^{\eminus1}$ & $p \to K^0 \mu^+$
\\[0.0cm]
\noalign{\vskip 4pt}
\cdashline{1-7}[.4pt/2pt]
\noalign{\vskip 7pt}
\multirow{3}{*}{$[\cO_{duue}]^{prst}$}
&$\frac{1}{\Lambda_\sscript{B}^2}[\Lambda_d^\dag]_p \delta_{r3}\delta_{s3}[\mathcal{T}_e]_{t}$
&S
& $10^{\eminus10}$ & $p\to K^+\nu$ & $10^{\eminus10}$ & $p\to K^+\nu$
\\[0.2cm]
&$\frac{1}{\Lambda_\sscript{B}^2}[\Lambda_d^\dag]_p \varepsilon_{rs}[\mathcal{T}_e]_{t}$
&A
& $7 \times 10^{\eminus6}$ & $p\to \pi^0 \mu^+$ & $7 \times 10^{\eminus6}$ & $p\to \pi^0 \mu^+$
\\[0.2cm]
\bottomrule
\end{tabular}
}

%% file: main.bib
@article{DAmbrosio:2002vsn,
    author = "D'Ambrosio, G. and Giudice, G. F. and Isidori, G. and Strumia, A.",
    title = "{Minimal flavor violation: An Effective field theory approach}",
    eprint = "hep-ph/0207036",
    archivePrefix = "arXiv",
    reportNumber = "CERN-TH-2002-147, IFUP-TH-2002-17",
    doi = "10.1016/S0550-3213(02)00836-2",
    journal = "Nucl. Phys. B",
    volume = "645",
    pages = "155--187",
    year = "2002"
}

@article{Greljo:2022cah,
    author = "Greljo, Admir and Palavri\'c, Ajdin and Thomsen, Anders Eller",
    title = "{Adding Flavor to the SMEFT}",
    eprint = "2203.09561",
    archivePrefix = "arXiv",
    primaryClass = "hep-ph",
    doi = "10.1007/JHEP10(2022)005",
    journal = "JHEP",
    volume = "10",
    pages = "010",
    year = "2022"
}

@article{Kumar:2025aek,
    author = "Kumar, Ranjeet and Srivastava, Rahul",
    title = "{Dark Matter Induced Proton Decays}",
    eprint = "2506.04370",
    archivePrefix = "arXiv",
    primaryClass = "hep-ph",
    month = "6",
    year = "2025"
}

@article{Banik:2025wpi,
    author = "Banik, Sumit and Crivellin, Andreas and Naterop, Luca and Stoffer, Peter",
    title = "{Two-loop anomalous dimensions for baryon-number-violating operators in SMEFT}",
    eprint = "2510.08682",
    archivePrefix = "arXiv",
    primaryClass = "hep-ph",
    reportNumber = "ZU-TH 61/25",
    month = "10",
    year = "2025"
}

@article{Raidal:1997hq,
    author = "Raidal, Martti and Santamaria, Arcadi",
    title = "{Muon electron conversion in nuclei versus mu ---{\ensuremath{>}} e gamma: An Effective field theory point of view}",
    eprint = "hep-ph/9710389",
    archivePrefix = "arXiv",
    reportNumber = "FTUV-97-56, IFIC-97-58",
    doi = "10.1016/S0370-2693(98)00020-3",
    journal = "Phys. Lett. B",
    volume = "421",
    pages = "250--258",
    year = "1998"
}

@article{Greljo:2023adz,
    author = "Greljo, Admir and Palavri\'c, Ajdin",
    title = "{Leading directions in the SMEFT}",
    eprint = "2305.08898",
    archivePrefix = "arXiv",
    primaryClass = "hep-ph",
    doi = "10.1007/JHEP09(2023)009",
    journal = "JHEP",
    volume = "09",
    pages = "009",
    year = "2023"
}

@article{MEG:2016leq,
    author = "Baldini, A. M. and others",
    collaboration = "MEG",
    title = "{Search for the lepton flavour violating decay $\mu ^+ \rightarrow \mathrm {e}^+ \gamma $ with the full dataset of the MEG experiment}",
    eprint = "1605.05081",
    archivePrefix = "arXiv",
    primaryClass = "hep-ex",
    doi = "10.1140/epjc/s10052-016-4271-x",
    journal = "Eur. Phys. J. C",
    volume = "76",
    number = "8",
    pages = "434",
    year = "2016"
}

@article{MEGII:2018kmf,
    author = "Baldini, A. M. and others",
    collaboration = "MEG II",
    title = "{The design of the MEG II experiment}",
    eprint = "1801.04688",
    archivePrefix = "arXiv",
    primaryClass = "physics.ins-det",
    doi = "10.1140/epjc/s10052-018-5845-6",
    journal = "Eur. Phys. J. C",
    volume = "78",
    number = "5",
    pages = "380",
    year = "2018"
}

@article{SINDRUM:1987nra,
    author = "Bellgardt, U. and others",
    collaboration = "SINDRUM",
    title = "{Search for the Decay $\mu^+ \to e^+ e^+ e^-$}",
    reportNumber = "SIN-PR-87-09",
    doi = "10.1016/0550-3213(88)90462-2",
    journal = "Nucl. Phys. B",
    volume = "299",
    pages = "1--6",
    year = "1988"
}

@article{Blondel:2013ia,
    author = "Blondel, A. and others",
    title = "{Research Proposal for an Experiment to Search for the Decay $\mu \to eee$}",
    eprint = "1301.6113",
    archivePrefix = "arXiv",
    primaryClass = "physics.ins-det",
    month = "1",
    year = "2013"
}

@article{SINDRUMII:2006dvw,
    author = "Bertl, Wilhelm H. and others",
    collaboration = "SINDRUM II",
    title = "{A Search for muon to electron conversion in muonic gold}",
    doi = "10.1140/epjc/s2006-02582-x",
    journal = "Eur. Phys. J. C",
    volume = "47",
    pages = "337--346",
    year = "2006"
}

@article{Fujii:2023vgo,
    author = "Fujii, Yuki",
    collaboration = "COMET",
    title = "{A search for a muon to electron conversion in COMET}",
    eprint = "2308.14275",
    archivePrefix = "arXiv",
    primaryClass = "hep-ex",
    doi = "10.1088/1748-0221/18/10/C10010",
    journal = "JINST",
    volume = "18",
    number = "10",
    pages = "C10010",
    year = "2023"
}

@article{Mu2e:2014fns,
    author = "Bartoszek, L. and others",
    collaboration = "Mu2e",
    title = "{Mu2e Technical Design Report}",
    eprint = "1501.05241",
    archivePrefix = "arXiv",
    primaryClass = "physics.ins-det",
    reportNumber = "FERMILAB-TM-2594, FERMILAB-DESIGN-2014-01",
    doi = "10.2172/1172555",
    month = "10",
    year = "2014"
}

@article{Palavric:2024gvu,
    author = "Palavri{\'c}, Ajdin",
    title = "{Discrete leptonic flavor symmetries: UV mediators and phenomenology}",
    eprint = "2408.16044",
    archivePrefix = "arXiv",
    primaryClass = "hep-ph",
    doi = "10.1103/PhysRevD.110.115025",
    journal = "Phys. Rev. D",
    volume = "110",
    number = "11",
    pages = "115025",
    year = "2024"
}

@article{Fuentes-Martin:2020zaz,
    author = "Fuentes-Martin, Javier and Ruiz-Femenia, Pedro and Vicente, Avelino and Virto, Javier",
    title = "{DsixTools 2.0: The Effective Field Theory Toolkit}",
    eprint = "2010.16341",
    archivePrefix = "arXiv",
    primaryClass = "hep-ph",
    reportNumber = "MITP/20-061, IFIC/20-50",
    doi = "10.1140/epjc/s10052-020-08778-y",
    journal = "Eur. Phys. J. C",
    volume = "81",
    number = "2",
    pages = "167",
    year = "2021"
}

@article{Celis:2017hod,
    author = "Celis, Alejandro and Fuentes-Martin, Javier and Vicente, Avelino and Virto, Javier",
    title = "{DsixTools: The Standard Model Effective Field Theory Toolkit}",
    eprint = "1704.04504",
    archivePrefix = "arXiv",
    primaryClass = "hep-ph",
    reportNumber = "LMU-ASC-24-17, IFIC-17-18",
    doi = "10.1140/epjc/s10052-017-4967-6",
    journal = "Eur. Phys. J. C",
    volume = "77",
    number = "6",
    pages = "405",
    year = "2017"
}

@article{Moreno-Sanchez:2025bzz,
    author = "Moreno-S{\'a}nchez, Adri{\'a}n and Palavri{\'c}, Ajdin",
    title = "{Leptonic Flavor from Modular $A_4$: UV Mediators and SMEFT Realizations}",
    eprint = "2505.01535",
    archivePrefix = "arXiv",
    primaryClass = "hep-ph",
    month = "5",
    year = "2025"
}

@article{Nath:2006ut,
    author = "Nath, Pran and Fileviez Perez, Pavel",
    title = "{Proton stability in grand unified theories, in strings and in branes}",
    eprint = "hep-ph/0601023",
    archivePrefix = "arXiv",
    doi = "10.1016/j.physrep.2007.02.010",
    journal = "Phys. Rept.",
    volume = "441",
    pages = "191--317",
    year = "2007"
}

@article{Greljo:2023bdy,
    author = "Greljo, Admir and Palavri\'c, Ajdin and Smolkovi\v{c}, Aleks",
    title = "{Leading directions in the SMEFT: Renormalization effects}",
    eprint = "2312.09179",
    archivePrefix = "arXiv",
    primaryClass = "hep-ph",
    doi = "10.1103/PhysRevD.109.075033",
    journal = "Phys. Rev. D",
    volume = "109",
    number = "7",
    pages = "075033",
    year = "2024"
}

@article{Aoki:2008ku,
    author = "Aoki, Y. and Boyle, P. and Cooney, P. and Del Debbio, L. and Kenway, R. and Maynard, C. M. and Soni, A. and Tweedie, R.",
    collaboration = "RBC-UKQCD",
    title = "{Proton lifetime bounds from chirally symmetric lattice QCD}",
    eprint = "0806.1031",
    archivePrefix = "arXiv",
    primaryClass = "hep-lat",
    reportNumber = "EDINBURGH-2008-19, RBRC-728, BNL-HET-08-13",
    doi = "10.1103/PhysRevD.78.054505",
    journal = "Phys. Rev. D",
    volume = "78",
    pages = "054505",
    year = "2008"
}

@article{ParticleDataGroup:2024cfk,
    author = "Navas, S. and others",
    collaboration = "Particle Data Group",
    title = "{Review of particle physics}",
    doi = "10.1103/PhysRevD.110.030001",
    journal = "Phys. Rev. D",
    volume = "110",
    number = "3",
    pages = "030001",
    year = "2024"
}

@article{Bali:2022qja,
    author = {Bali, Gunnar S. and Collins, Sara and S{\"o}ldner, Wolfgang and Weish{\"a}upl, Simon},
    collaboration = "RQCD",
    title = "{Leading order mesonic and baryonic SU(3) low energy constants from Nf=3 lattice QCD}",
    eprint = "2201.05591",
    archivePrefix = "arXiv",
    primaryClass = "hep-lat",
    doi = "10.1103/PhysRevD.105.054516",
    journal = "Phys. Rev. D",
    volume = "105",
    number = "5",
    pages = "054516",
    year = "2022"
}

@article{Cirigliano:2005ck,
    author = "Cirigliano, Vincenzo and Grinstein, Benjamin and Isidori, Gino and Wise, Mark B.",
    title = "{Minimal flavor violation in the lepton sector}",
    eprint = "hep-ph/0507001",
    archivePrefix = "arXiv",
    reportNumber = "UCSD-PTH-05-11, CALT-68-2566",
    doi = "10.1016/j.nuclphysb.2005.08.037",
    journal = "Nucl. Phys. B",
    volume = "728",
    pages = "121--134",
    year = "2005"
}

@article{Gavela:2009cd,
    author = "Gavela, M. B. and Hambye, T. and Hernandez, D. and Hernandez, P.",
    title = "{Minimal Flavour Seesaw Models}",
    eprint = "0906.1461",
    archivePrefix = "arXiv",
    primaryClass = "hep-ph",
    reportNumber = "FTUAM-09-09, IFT-UAM-CSIC-09-27, ULB-TH-09-15, IFIC-09-22, FTUV-09-0607",
    doi = "10.1088/1126-6708/2009/09/038",
    journal = "JHEP",
    volume = "09",
    pages = "038",
    year = "2009"
}

@article{Nikolidakis:2007fc,
    author = "Nikolidakis, Emanuel and Smith, Christopher",
    title = "{Minimal Flavor Violation, Seesaw, and R-parity}",
    eprint = "0710.3129",
    archivePrefix = "arXiv",
    primaryClass = "hep-ph",
    doi = "10.1103/PhysRevD.77.015021",
    journal = "Phys. Rev. D",
    volume = "77",
    pages = "015021",
    year = "2008"
}

@article{Davidson:2010uu,
    author = "Davidson, Sacha and Descotes-Genon, Sebastien",
    title = "{Minimal Flavour Violation for Leptoquarks}",
    eprint = "1009.1998",
    archivePrefix = "arXiv",
    primaryClass = "hep-ph",
    reportNumber = "LPT-ORSAY-10-72",
    doi = "10.1007/JHEP11(2010)073",
    journal = "JHEP",
    volume = "11",
    pages = "073",
    year = "2010"
}

@article{Buchmuller:1985jz,
    author = "Buchmuller, W. and Wyler, D.",
    title = "{Effective Lagrangian Analysis of New Interactions and Flavor Conservation}",
    reportNumber = "CERN-TH-4254/85",
    doi = "10.1016/0550-3213(86)90262-2",
    journal = "Nucl. Phys. B",
    volume = "268",
    pages = "621--653",
    year = "1986"
}

@article{Claudson:1981gh,
    author = "Claudson, Mark and Wise, Mark B. and Hall, Lawrence J.",
    title = "{Chiral Lagrangian for Deep Mine Physics}",
    reportNumber = "HUTP-81/A036",
    doi = "10.1016/0550-3213(82)90401-1",
    journal = "Nucl. Phys. B",
    volume = "195",
    pages = "297--307",
    year = "1982"
}

@article{Weinberg:1979sa,
    author = "Weinberg, Steven",
    title = "{Baryon and Lepton Nonconserving Processes}",
    reportNumber = "HUTP-79-A050",
    doi = "10.1103/PhysRevLett.43.1566",
    journal = "Phys. Rev. Lett.",
    volume = "43",
    pages = "1566--1570",
    year = "1979"
}

@article{Yoo:2021gql,
    author = "Yoo, Jun-Sik and Aoki, Yasumichi and Boyle, Peter and Izubuchi, Taku and Soni, Amarjit and Syritsyn, Sergey",
    title = "{Proton decay matrix elements on the lattice at physical pion mass}",
    eprint = "2111.01608",
    archivePrefix = "arXiv",
    primaryClass = "hep-lat",
    reportNumber = "RBRC-1333, KEK-CP-0385",
    doi = "10.1103/PhysRevD.105.074501",
    journal = "Phys. Rev. D",
    volume = "105",
    number = "7",
    pages = "074501",
    year = "2022"
}

@article{Weinberg:1980bf,
    author = "Weinberg, Steven",
    title = "{Varieties of Baryon and Lepton Nonconservation}",
    reportNumber = "HUTP-80/A023",
    doi = "10.1103/PhysRevD.22.1694",
    journal = "Phys. Rev. D",
    volume = "22",
    pages = "1694",
    year = "1980"
}

@article{Grzadkowski:2010es,
    author = "Grzadkowski, B. and Iskrzynski, M. and Misiak, M. and Rosiek, J.",
    title = "{Dimension-Six Terms in the Standard Model Lagrangian}",
    eprint = "1008.4884",
    archivePrefix = "arXiv",
    primaryClass = "hep-ph",
    reportNumber = "IFT-9-2010, TTP10-35",
    doi = "10.1007/JHEP10(2010)085",
    journal = "JHEP",
    volume = "10",
    pages = "085",
    year = "2010"
}

@article{Smith:2011rp,
    author = "Smith, Christopher",
    title = "{Proton stability from a fourth family}",
    eprint = "1105.1723",
    archivePrefix = "arXiv",
    primaryClass = "hep-ph",
    doi = "10.1103/PhysRevD.85.036005",
    journal = "Phys. Rev. D",
    volume = "85",
    pages = "036005",
    year = "2012"
}

@article{Minkowski:1977sc,
    author = "Minkowski, Peter",
    title = "{$\mu \to e\gamma$ at a Rate of One Out of $10^{9}$ Muon Decays?}",
    reportNumber = "Print-77-0182 (BERN)",
    doi = "10.1016/0370-2693(77)90435-X",
    journal = "Phys. Lett. B",
    volume = "67",
    pages = "421--428",
    year = "1977"
}

@article{Mohapatra:1980yp,
    author = "Mohapatra, Rabindra N. and Senjanovic, Goran",
    title = "{Neutrino Masses and Mixings in Gauge Models with Spontaneous Parity Violation}",
    reportNumber = "FERMILAB-PUB-80-061-THY, FERMILAB-PUB-80-061-T",
    doi = "10.1103/PhysRevD.23.165",
    journal = "Phys. Rev. D",
    volume = "23",
    pages = "165",
    year = "1981"
}

@article{Foot:1988aq,
    author = "Foot, Robert and Lew, H. and He, X. G. and Joshi, Girish C.",
    title = "{Seesaw Neutrino Masses Induced by a Triplet of Leptons}",
    reportNumber = "UM-P-88/89, OZ-P-88/7",
    doi = "10.1007/BF01415558",
    journal = "Z. Phys. C",
    volume = "44",
    pages = "441",
    year = "1989"
}

@article{Alonso:2014zka,
    author = "Alonso, Rodrigo and Chang, Hsi-Ming and Jenkins, Elizabeth E. and Manohar, Aneesh V. and Shotwell, Brian",
    title = "{Renormalization group evolution of dimension-six baryon number violating operators}",
    eprint = "1405.0486",
    archivePrefix = "arXiv",
    primaryClass = "hep-ph",
    doi = "10.1016/j.physletb.2014.05.065",
    journal = "Phys. Lett. B",
    volume = "734",
    pages = "302--307",
    year = "2014"
}

@article{Greljo:2025mwj,
    author = "Greljo, Admir and Palavri{\'c}, Ajdin and Stefanek, Ben A.",
    title = "{Minimal Flavor Protection for TeV-scale New Physics}",
    eprint = "2512.04159",
    archivePrefix = "arXiv",
    primaryClass = "hep-ph",
    month = "12",
    year = "2025"
}

@article{Allwicher:2023shc,
    author = "Allwicher, Lukas and Cornella, Claudia and Isidori, Gino and Stefanek, Ben A.",
    title = "{New physics in the third generation. A comprehensive SMEFT analysis and future prospects}",
    eprint = "2311.00020",
    archivePrefix = "arXiv",
    primaryClass = "hep-ph",
    reportNumber = "ZU-TH 71/23, MITP-23-060, KCL-PH-TH/2023-59",
    doi = "10.1007/JHEP03(2024)049",
    journal = "JHEP",
    volume = "03",
    pages = "049",
    year = "2024"
}

@article{Super-Kamiokande:2020wjk,
    author = "Takenaka, A. and others",
    collaboration = "Super-Kamiokande",
    title = "{Search for proton decay via $p\to e^+\pi^0$ and $p\to \mu^+\pi^0$ with an enlarged fiducial volume in Super-Kamiokande I-IV}",
    eprint = "2010.16098",
    archivePrefix = "arXiv",
    primaryClass = "hep-ex",
    doi = "10.1103/PhysRevD.102.112011",
    journal = "Phys. Rev. D",
    volume = "102",
    number = "11",
    pages = "112011",
    year = "2020"
}

@article{Super-Kamiokande:2013rwg,
    author = "Abe, K. and others",
    collaboration = "Super-Kamiokande",
    title = "{Search for Nucleon Decay via $n \to \bar{\nu} \pi^{0}$ and $p \to \bar{\nu} \pi^{+}$ in Super-Kamiokande}",
    eprint = "1305.4391",
    archivePrefix = "arXiv",
    primaryClass = "hep-ex",
    doi = "10.1103/PhysRevLett.113.121802",
    journal = "Phys. Rev. Lett.",
    volume = "113",
    number = "12",
    pages = "121802",
    year = "2014"
}

@article{Calibbi:2021pyh,
    author = "Calibbi, Lorenzo and Marcano, Xabier and Roy, Joydeep",
    title = "{Z lepton flavour violation as a probe for new physics at future $e^+e^-$ colliders}",
    eprint = "2107.10273",
    archivePrefix = "arXiv",
    primaryClass = "hep-ph",
    reportNumber = "TUM-HEP 1352/21",
    doi = "10.1140/epjc/s10052-021-09777-3",
    journal = "Eur. Phys. J. C",
    volume = "81",
    number = "12",
    pages = "1054",
    year = "2021"
}

@article{Gisbert:2024sjw,
    author = "Gisbert, Hector and Rodr{\'\i}guez-S{\'a}nchez, Antonio and Vale Silva, Luiz",
    title = "{Constraints on baryon-number-violating top-quark operators in standard model effective field theory}",
    eprint = "2409.00218",
    archivePrefix = "arXiv",
    primaryClass = "hep-ph",
    doi = "10.1103/f81b-gry8",
    journal = "Phys. Rev. D",
    volume = "112",
    number = "1",
    pages = "015026",
    year = "2025"
}

@article{Chivukula:1987py,
    author = "Chivukula, R. Sekhar and Georgi, Howard",
    title = "{Composite Technicolor Standard Model}",
    reportNumber = "BUHEP-87-2, HUTP-87/A003",
    doi = "10.1016/0370-2693(87)90713-1",
    journal = "Phys. Lett. B",
    volume = "188",
    pages = "99--104",
    year = "1987"
}

@article{Marciano:1994bg,
    author = "Marciano, William J.",
    editor = "Rolandi, Gigi",
    title = "{Tau physics: A Theoretical perspective}",
    reportNumber = "BNL-61141",
    doi = "10.1016/0920-5632(95)00126-T",
    journal = "Nucl. Phys. B Proc. Suppl.",
    volume = "40",
    pages = "3--15",
    year = "1995"
}

@article{Dong:2011rh,
    author = "Dong, Zhe and Durieux, Gauthier and Gerard, Jean-Marc and Han, Tao and Maltoni, Fabio",
    title = "{Baryon number violation at the LHC: the top option}",
    eprint = "1107.3805",
    archivePrefix = "arXiv",
    primaryClass = "hep-ph",
    reportNumber = "CP3-11-24, MADPH-11-1573",
    doi = "10.1103/PhysRevD.85.016006",
    journal = "Phys. Rev. D",
    volume = "85",
    pages = "016006",
    year = "2012"
}

@article{Helo:2019yqp,
    author = "Helo, Juan Carlos and Hirsch, Martin and Ota, Toshihiko",
    title = "{Proton decay at one loop}",
    eprint = "1904.00036",
    archivePrefix = "arXiv",
    primaryClass = "hep-ph",
    doi = "10.1103/PhysRevD.99.095021",
    journal = "Phys. Rev. D",
    volume = "99",
    number = "9",
    pages = "095021",
    year = "2019"
}

@article{Heeck:2024jei,
    author = "Heeck, Julian and Watkins, Dima",
    title = "{Baryon number violation involving tau leptons}",
    eprint = "2405.18478",
    archivePrefix = "arXiv",
    primaryClass = "hep-ph",
    doi = "10.1007/JHEP07(2024)170",
    journal = "JHEP",
    volume = "07",
    pages = "170",
    year = "2024"
}

@article{Super-Kamiokande:2022egr,
    author = "Matsumoto, R. and others",
    collaboration = "Super-Kamiokande",
    title = "{Search for proton decay via $p\rightarrow \mu^+K^0$ in 0.37 megaton-years exposure of Super-Kamiokande}",
    eprint = "2208.13188",
    archivePrefix = "arXiv",
    primaryClass = "hep-ex",
    doi = "10.1103/PhysRevD.106.072003",
    journal = "Phys. Rev. D",
    volume = "106",
    number = "7",
    pages = "072003",
    year = "2022"
}

@article{BaBar:2011yks,
    author = "del Amo Sanchez, P. and others",
    collaboration = "BaBar",
    title = "{Searches for the baryon- and lepton-number violating decays $B^0\rightarrow\Lambda_c^+\ell^-$, $B^-\rightarrow\Lambda\ell^-$, and $B^-\rightarrow\bar{\Lambda}\ell^-$}",
    eprint = "1101.3830",
    archivePrefix = "arXiv",
    primaryClass = "hep-ex",
    reportNumber = "SLAC-PUB-14360, BABAR-PUB-10-030",
    doi = "10.1103/PhysRevD.83.091101",
    journal = "Phys. Rev. D",
    volume = "83",
    pages = "091101",
    year = "2011"
}

@article{Wilczek:1979hc,
    author = "Wilczek, Frank and Zee, A.",
    title = "{Operator Analysis of Nucleon Decay}",
    reportNumber = "Print-79-0709 (PRINCETON)",
    doi = "10.1103/PhysRevLett.43.1571",
    journal = "Phys. Rev. Lett.",
    volume = "43",
    pages = "1571--1573",
    year = "1979"
}

@article{Abbott:1980zj,
    author = "Abbott, L. F. and Wise, Mark B.",
    title = "{The Effective Hamiltonian for Nucleon Decay}",
    reportNumber = "SLAC-PUB-2487",
    doi = "10.1103/PhysRevD.22.2208",
    journal = "Phys. Rev. D",
    volume = "22",
    pages = "2208",
    year = "1980"
}

@article{LHCb:2022wro,
    author = "Aaij, Roel and others",
    collaboration = "LHCb",
    title = "{Search for the baryon- and lepton-number violating decays B0{\textrightarrow}p{\ensuremath{\mu}}- and Bs0{\textrightarrow}p{\ensuremath{\mu}}-}",
    eprint = "2210.10412",
    archivePrefix = "arXiv",
    primaryClass = "hep-ex",
    reportNumber = "CERN-EP-2022-195, LHCb-PAPER-2022-022",
    doi = "10.1103/PhysRevD.108.012021",
    journal = "Phys. Rev. D",
    volume = "108",
    number = "1",
    pages = "012021",
    year = "2023"
}

@article{Hou:2005iu,
    author = "Hou, Wei-Shu and Nagashima, Makiko and Soddu, Andrea",
    title = "{Baryon number violation involving higher generations}",
    eprint = "hep-ph/0509006",
    archivePrefix = "arXiv",
    reportNumber = "SLAC-PUB-11461",
    doi = "10.1103/PhysRevD.72.095001",
    journal = "Phys. Rev. D",
    volume = "72",
    pages = "095001",
    year = "2005"
}

@article{Beneke:2024hox,
    author = "Beneke, Martin and Finauri, Gael and Petrov, Alexey A.",
    title = "{Indirect constraints on third generation baryon number violation}",
    eprint = "2404.09642",
    archivePrefix = "arXiv",
    primaryClass = "hep-ph",
    reportNumber = "TUM-HEP-1504/24, USC-TH-2024-01",
    doi = "10.1007/JHEP09(2024)090",
    journal = "JHEP",
    volume = "09",
    pages = "090",
    year = "2024"
}

@article{Crivellin:2023ter,
    author = "Crivellin, Andreas and Hoferichter, Martin",
    title = "{Rescattering effects in nucleon-to-meson form factors and application to tau-lepton-induced proton decay}",
    eprint = "2302.01939",
    archivePrefix = "arXiv",
    primaryClass = "hep-ph",
    reportNumber = "PSI-PR-23-03, PSI-PR-23-01, ZU-TH 08/23",
    doi = "10.1016/j.physletb.2023.138169",
    journal = "Phys. Lett. B",
    volume = "845",
    pages = "138169",
    year = "2023"
}

@article{Alonso:2011jd,
    author = "Alonso, Rodrigo and Isidori, Gino and Merlo, Luca and Munoz, Luis Alfredo and Nardi, Enrico",
    title = "{Minimal flavour violation extensions of the seesaw}",
    eprint = "1103.5461",
    archivePrefix = "arXiv",
    primaryClass = "hep-ph",
    reportNumber = "FTUAM-11-42, IFT-UAM-CSIC-11-16, TUM-HEP-797-11",
    doi = "10.1007/JHEP06(2011)037",
    journal = "JHEP",
    volume = "06",
    pages = "037",
    year = "2011"
}

@article{Gargalionis:2024nij,
    author = "Gargalionis, John and Herrero-Garc\'\i{}a, Juan and Schmidt, Michael A.",
    title = "{Model-independent estimates for loop-induced baryon-number-violating nucleon decays}",
    eprint = "2401.04768",
    archivePrefix = "arXiv",
    primaryClass = "hep-ph",
    reportNumber = "CPPC-2024-02",
    doi = "10.1007/JHEP06(2024)182",
    journal = "JHEP",
    volume = "06",
    pages = "182",
    year = "2024"
}

@article{Jenkins:2017jig,
    author = "Jenkins, Elizabeth E. and Manohar, Aneesh V. and Stoffer, Peter",
    title = "{Low-Energy Effective Field Theory below the Electroweak Scale: Operators and Matching}",
    eprint = "1709.04486",
    archivePrefix = "arXiv",
    primaryClass = "hep-ph",
    doi = "10.1007/JHEP03(2018)016",
    journal = "JHEP",
    volume = "03",
    pages = "016",
    year = "2018",
    note = "[Erratum: JHEP 12, 043 (2023)]"
}

@article{Beneito:2023xbk,
    author = "Beneito, I, Arnau Bas and Gargalionis, John and Herrero-Garcia, Juan and Santamaria, Arcadi and Schmidt, Michael A.",
    title = "{An EFT approach to baryon number violation: lower limits on the new physics scale and correlations between nucleon decay modes}",
    eprint = "2312.13361",
    archivePrefix = "arXiv",
    primaryClass = "hep-ph",
    reportNumber = "IFIC/23-52, CPPC-2023-12",
    doi = "10.1007/JHEP07(2024)004",
    journal = "JHEP",
    volume = "07",
    pages = "004",
    year = "2024"
}

@article{Davidson:2006bd,
    author = "Davidson, Sacha and Palorini, Federica",
    title = "{Various definitions of Minimal Flavour Violation for Leptons}",
    eprint = "hep-ph/0607329",
    archivePrefix = "arXiv",
    doi = "10.1016/j.physletb.2006.09.016",
    journal = "Phys. Lett. B",
    volume = "642",
    pages = "72--80",
    year = "2006"
}

@article{Hyper-Kamiokande:2018ofw,
    author = "Abe, K. and others",
    collaboration = "Hyper-Kamiokande",
    title = "{Hyper-Kamiokande Design Report}",
    journal="",
    eprint = "1805.04163",
    archivePrefix = "arXiv",
    primaryClass = "physics.ins-det",
    month = "5",
    year = "2018"
}

@article{DUNE:2020ypp,
    author = "Abi, Babak and others",
    collaboration = "DUNE",
    title = "{Deep Underground Neutrino Experiment (DUNE), Far Detector Technical Design Report, Volume II: DUNE Physics}",
    eprint = "2002.03005",
    journal="",
    archivePrefix = "arXiv",
    primaryClass = "hep-ex",
    reportNumber = "FERMILAB-PUB-20-025-ND, FERMILAB-DESIGN-2020-02",
    month = "2",
    year = "2020"
}

@article{Isidori:2012ts,
    author = "Isidori, Gino and Straub, David M.",
    title = "{Minimal Flavour Violation and Beyond}",
    eprint = "1202.0464",
    archivePrefix = "arXiv",
    primaryClass = "hep-ph",
    doi = "10.1140/epjc/s10052-012-2103-1",
    journal = "Eur. Phys. J. C",
    volume = "72",
    pages = "2103",
    year = "2012"
}

@article{CMS:2024dzv,
    author = "Hayrapetyan, Aram and others",
    collaboration = "CMS",
    title = "{Search for baryon number violation in top quark production and decay using proton-proton collisions at $\sqrt{s}$ = 13 TeV}",
    eprint = "2402.18461",
    archivePrefix = "arXiv",
    primaryClass = "hep-ex",
    reportNumber = "CMS-TOP-22-003, CERN-EP-2024-027",
    doi = "10.1103/PhysRevLett.132.241802",
    journal = "Phys. Rev. Lett.",
    volume = "132",
    pages = "241802",
    year = "2024"
}

@article{Barbieri:2011ci,
    author = "Barbieri, Riccardo and Isidori, Gino and Jones-Perez, Joel and Lodone, Paolo and Straub, David M.",
    title = "{$U(2)$ and Minimal Flavour Violation in Supersymmetry}",
    eprint = "1105.2296",
    archivePrefix = "arXiv",
    primaryClass = "hep-ph",
    doi = "10.1140/epjc/s10052-011-1725-z",
    journal = "Eur. Phys. J. C",
    volume = "71",
    pages = "1725",
    year = "2011"
}

@article{IBeneito:2025nby,
    author = "I Beneito, Arnau Bas and Gargalionis, John and Herrero-Garcia, Juan and Schmidt, Michael A.",
    title = "{Squeezing Proton Decay and Neutrino Masses: Upper Bounds on Standard Model Extensions}",
    eprint = "2503.20928",
    archivePrefix = "arXiv",
    primaryClass = "hep-ph",
    month = "3",
    year = "2025"
}

@article{Dorsner:2012nq,
    author = "Dorsner, Ilja and Fajfer, Svjetlana and Kosnik, Nejc",
    title = "{Heavy and light scalar leptoquarks in proton decay}",
    eprint = "1204.0674",
    archivePrefix = "arXiv",
    primaryClass = "hep-ph",
    reportNumber = "LAL-12-111",
    doi = "10.1103/PhysRevD.86.015013",
    journal = "Phys. Rev. D",
    volume = "86",
    pages = "015013",
    year = "2012"
}

@article{deBlas:2017xtg,
    author = "de Blas, J. and Criado, J. C. and Perez-Victoria, M. and Santiago, J.",
    title = "{Effective description of general extensions of the Standard Model: the complete tree-level dictionary}",
    eprint = "1711.10391",
    archivePrefix = "arXiv",
    primaryClass = "hep-ph",
    reportNumber = "CERN-TH-2017-251",
    doi = "10.1007/JHEP03(2018)109",
    journal = "JHEP",
    volume = "03",
    pages = "109",
    year = "2018"
}

@article{Dorsner:2016wpm,
    author = "Dor\v{s}ner, I. and Fajfer, S. and Greljo, A. and Kamenik, J. F. and Ko\v{s}nik, N.",
    title = "{Physics of leptoquarks in precision experiments and at particle colliders}",
    eprint = "1603.04993",
    archivePrefix = "arXiv",
    primaryClass = "hep-ph",
    doi = "10.1016/j.physrep.2016.06.001",
    journal = "Phys. Rept.",
    volume = "641",
    pages = "1--68",
    year = "2016"
}

@article{Barbieri:2012uh,
    author = "Barbieri, Riccardo and Buttazzo, Dario and Sala, Filippo and Straub, David M.",
    title = "{Flavour physics from an approximate $U(2)^3$ symmetry}",
    eprint = "1203.4218",
    archivePrefix = "arXiv",
    primaryClass = "hep-ph",
    doi = "10.1007/JHEP07(2012)181",
    journal = "JHEP",
    volume = "07",
    pages = "181",
    year = "2012"
}

@article{ThomasArun:2025dav,
    author = "Thomas Arun, Mathew and M, Shyam and Pal, Ritik",
    title = "{RG evolution and effect of intermediate new-physics on $\Delta B=1$ four-fermion operators}",
    journal="",
    eprint = "2511.06106",
    archivePrefix = "arXiv",
    primaryClass = "hep-ph",
    month = "11",
    year = "2025"
}

@article{Helset:2019eyc,
    author = "Helset, Andreas and Kobach, Andrew",
    title = "{Baryon Number, Lepton Number, and Operator Dimension in the SMEFT with Flavor Symmetries}",
    eprint = "1909.05853",
    archivePrefix = "arXiv",
    primaryClass = "hep-ph",
    doi = "10.1016/j.physletb.2019.135132",
    journal = "Phys. Lett. B",
    volume = "800",
    pages = "135132",
    year = "2020"
}

@article{Aoki:2017puj,
    author = "Aoki, Yasumichi and Izubuchi, Taku and Shintani, Eigo and Soni, Amarjit",
    title = "{Improved lattice computation of proton decay matrix elements}",
    eprint = "1705.01338",
    archivePrefix = "arXiv",
    primaryClass = "hep-lat",
    reportNumber = "RBRC-1235, KEK-CP-358",
    doi = "10.1103/PhysRevD.96.014506",
    journal = "Phys. Rev. D",
    volume = "96",
    number = "1",
    pages = "014506",
    year = "2017"
}

@article{Fonseca:2017lem,
    author = "Fonseca, Renato M.",
    editor = "Grzadkowski, Bohdan and Kalinowski, Jan and Krawczyk, Maria",
    title = "{The Sym2Int program: going from symmetries to interactions}",
    eprint = "1703.05221",
    archivePrefix = "arXiv",
    primaryClass = "hep-ph",
    doi = "10.1088/1742-6596/873/1/012045",
    journal = "J. Phys. Conf. Ser.",
    volume = "873",
    number = "1",
    pages = "012045",
    year = "2017"
}

@article{Fonseca:2019yya,
    author = "Fonseca, Renato M.",
    title = "{Enumerating the operators of an effective field theory}",
    eprint = "1907.12584",
    archivePrefix = "arXiv",
    primaryClass = "hep-ph",
    doi = "10.1103/PhysRevD.101.035040",
    journal = "Phys. Rev. D",
    volume = "101",
    number = "3",
    pages = "035040",
    year = "2020"
}

@article{Bartocci:2023nvp,
    author = {Bartocci, Riccardo and Biek{\"o}tter, Anke and Hurth, Tobias},
    title = "{A global analysis of the SMEFT under the minimal MFV assumption}",
    eprint = "2311.04963",
    archivePrefix = "arXiv",
    primaryClass = "hep-ph",
    reportNumber = "MITP/23-063",
    doi = "10.1007/JHEP05(2024)074",
    journal = "JHEP",
    volume = "05",
    pages = "074",
    year = "2024"
}

@article{terHoeve:2025gey,
    author = "ter Hoeve, Jaco and Mantani, Luca and Rojo, Juan and Rossia, Alejo N. and Vryonidou, Eleni",
    title = "{Connecting scales: RGE effects in the SMEFT at the LHC and future colliders}",
    eprint = "2502.20453",
    archivePrefix = "arXiv",
    primaryClass = "hep-ph",
    doi = "10.1007/JHEP06(2025)125",
    journal = "JHEP",
    volume = "06",
    pages = "125",
    year = "2025"
}

@article{Allwicher:2024sso,
    author = "Allwicher, Lukas and McCullough, Matthew and Renner, Sophie",
    title = "{New physics at Tera-Z: precision renormalised}",
    eprint = "2408.03992",
    archivePrefix = "arXiv",
    primaryClass = "hep-ph",
    reportNumber = "CERN-TH-2024-133, ZU-TH 39/24",
    doi = "10.1007/JHEP02(2025)164",
    journal = "JHEP",
    volume = "02",
    pages = "164",
    year = "2025"
}

@article{Greljo:2018tuh,
    author = "Greljo, Admir and Stefanek, Ben A.",
    title = "{Third family quark\textendash{}lepton unification at the TeV scale}",
    eprint = "1802.04274",
    archivePrefix = "arXiv",
    primaryClass = "hep-ph",
    reportNumber = "MITP-18-012",
    doi = "10.1016/j.physletb.2018.05.033",
    journal = "Phys. Lett. B",
    volume = "782",
    pages = "131--138",
    year = "2018"
}

@article{Barbieri:2023qpf,
    author = "Barbieri, Riccardo and Isidori, Gino",
    title = "{Minimal flavour deconstruction}",
    eprint = "2312.14004",
    archivePrefix = "arXiv",
    primaryClass = "hep-ph",
    doi = "10.1007/JHEP05(2024)033",
    journal = "JHEP",
    volume = "05",
    pages = "033",
    year = "2024"
}

@article{Barbieri:2024zkh,
    author = "Barbieri, Riccardo",
    title = "{Phenomenology of Minimal Flavour Deconstruction at the lowest new scale}",
    eprint = "2409.08657",
    archivePrefix = "arXiv",
    primaryClass = "hep-ph",
    month = "9",
    year = "2024"
}

@article{Davighi:2023iks,
    author = "Davighi, Joe and Isidori, Gino",
    title = "{Non-universal gauge interactions addressing the inescapable link between Higgs and flavour}",
    eprint = "2303.01520",
    archivePrefix = "arXiv",
    primaryClass = "hep-ph",
    doi = "10.1007/JHEP07(2023)147",
    journal = "JHEP",
    volume = "07",
    pages = "147",
    year = "2023"
}

@article{Bordone:2017bld,
    author = "Bordone, Marzia and Cornella, Claudia and Fuentes-Martin, Javier and Isidori, Gino",
    title = "{A three-site gauge model for flavor hierarchies and flavor anomalies}",
    eprint = "1712.01368",
    archivePrefix = "arXiv",
    primaryClass = "hep-ph",
    reportNumber = "ZU-TH-36-17",
    doi = "10.1016/j.physletb.2018.02.011",
    journal = "Phys. Lett. B",
    volume = "779",
    pages = "317--323",
    year = "2018"
}

@article{Fuentes-Martin:2022xnb,
    author = "Fuentes-Martin, Javier and Isidori, Gino and Lizana, Javier M. and Selimovic, Nudzeim and Stefanek, Ben A.",
    title = "{Flavor hierarchies, flavor anomalies, and Higgs mass from a warped extra dimension}",
    eprint = "2203.01952",
    archivePrefix = "arXiv",
    primaryClass = "hep-ph",
    reportNumber = "ZU-TH-08/22",
    doi = "10.1016/j.physletb.2022.137382",
    journal = "Phys. Lett. B",
    volume = "834",
    pages = "137382",
    year = "2022"
}

@article{Allwicher:2020esa,
    author = "Allwicher, Lukas and Isidori, Gino and Thomsen, Anders Eller",
    title = "{Stability of the Higgs Sector in a Flavor-Inspired Multi-Scale Model}",
    eprint = "2011.01946",
    archivePrefix = "arXiv",
    primaryClass = "hep-ph",
    reportNumber = "ZU-TH-41/20",
    doi = "10.1007/JHEP01(2021)191",
    journal = "JHEP",
    volume = "01",
    pages = "191",
    year = "2021"
}

@article{Covone:2024elw,
    author = "Covone, Sebastiano and Davighi, Joe and Isidori, Gino and Pesut, Marko",
    title = "{Flavour deconstructing the composite Higgs}",
    eprint = "2407.10950",
    archivePrefix = "arXiv",
    primaryClass = "hep-ph",
    reportNumber = "CERN-TH-2024-112",
    doi = "10.1007/JHEP01(2025)041",
    journal = "JHEP",
    volume = "01",
    pages = "041",
    year = "2025"
}

@article{FernandezNavarro:2023rhv,
    author = "Fern{\'a}ndez Navarro, Mario and King, Stephen F.",
    title = "{Tri-hypercharge: a separate gauged weak hypercharge for each fermion family as the origin of flavour}",
    eprint = "2305.07690",
    archivePrefix = "arXiv",
    primaryClass = "hep-ph",
    doi = "10.1007/JHEP08(2023)020",
    journal = "JHEP",
    volume = "08",
    pages = "020",
    year = "2023"
}

@article{FernandezNavarro:2024hnv,
    author = "Fern{\'a}ndez Navarro, Mario and King, Stephen F. and Vicente, Avelino",
    title = "{Minimal complete tri-hypercharge theories of flavour}",
    eprint = "2404.12442",
    archivePrefix = "arXiv",
    primaryClass = "hep-ph",
    doi = "10.1007/JHEP07(2024)147",
    journal = "JHEP",
    volume = "07",
    pages = "147",
    year = "2024"
}

@article{FernandezNavarro:2025zmb,
    author = "Fern{\'a}ndez Navarro, Mario and King, Stephen F. and Vicente, Avelino",
    title = "{Natural neutrino mass hierarchy in a theory of gauge flavour deconstruction}",
    eprint = "2506.21687",
    archivePrefix = "arXiv",
    primaryClass = "hep-ph",
    month = "6",
    year = "2025"
}

@article{Fuentes-Martin:2024fpx,
    author = "Fuentes-Mart{\'\i}n, Javier and Lizana, Javier M.",
    title = "{Deconstructing flavor anomalously}",
    eprint = "2402.09507",
    archivePrefix = "arXiv",
    primaryClass = "hep-ph",
    reportNumber = "IFT-UAM/CSIC-24-21",
    doi = "10.1007/JHEP07(2024)117",
    journal = "JHEP",
    volume = "07",
    pages = "117",
    year = "2024"
}

@article{Davighi:2023evx,
    author = "Davighi, Joe and Stefanek, Ben A.",
    title = "{Deconstructed hypercharge: a natural model of flavour}",
    eprint = "2305.16280",
    archivePrefix = "arXiv",
    primaryClass = "hep-ph",
    reportNumber = "ZU-TH 24/23",
    doi = "10.1007/JHEP11(2023)100",
    journal = "JHEP",
    volume = "11",
    pages = "100",
    year = "2023"
}

@article{Fuentes-Martin:2020bnh,
    author = "Fuentes-Mart{\'\i}n, Javier and Stangl, Peter",
    title = "{Third-family quark-lepton unification with a fundamental composite Higgs}",
    eprint = "2004.11376",
    archivePrefix = "arXiv",
    primaryClass = "hep-ph",
    reportNumber = "ZU-TH-08/20, LAPTH-014/20",
    doi = "10.1016/j.physletb.2020.135953",
    journal = "Phys. Lett. B",
    volume = "811",
    pages = "135953",
    year = "2020"
}

@article{Lizana:2024jby,
    author = "Lizana, Javier M.",
    title = "{A common origin of the Higgs boson and the flavor hierarchies}",
    eprint = "2412.14243",
    archivePrefix = "arXiv",
    primaryClass = "hep-ph",
    reportNumber = "IFT-UAM/CSIC-24-185",
    doi = "10.1007/JHEP05(2025)176",
    journal = "JHEP",
    volume = "05",
    pages = "176",
    year = "2025"
}

@article{Isidori:2025rci,
    author = "Isidori, Gino and Paradisi, Paride and Sainaghi, Andrea and Selimovic, Nudzeim",
    title = "{Anarchic neutrinos from flavor deconstruction: phenomenology of the lepton sector}",
    eprint = "2510.23703",
    archivePrefix = "arXiv",
    primaryClass = "hep-ph",
    month = "10",
    year = "2025"
}

@article{Papucci:2011wy,
    author = "Papucci, Michele and Ruderman, Joshua T. and Weiler, Andreas",
    title = "{Natural SUSY Endures}",
    eprint = "1110.6926",
    archivePrefix = "arXiv",
    primaryClass = "hep-ph",
    reportNumber = "DESY-11-193, CERN-PH-TH-265",
    doi = "10.1007/JHEP09(2012)035",
    journal = "JHEP",
    volume = "09",
    pages = "035",
    year = "2012"
}

@article{Larsen:2012rq,
    author = "Larsen, Grant and Nomura, Yasunori and Roberts, Hannes L. L.",
    title = "{Supersymmetry with Light Stops}",
    eprint = "1202.6339",
    archivePrefix = "arXiv",
    primaryClass = "hep-ph",
    reportNumber = "UCB-PTH-12-03",
    doi = "10.1007/JHEP06(2012)032",
    journal = "JHEP",
    volume = "06",
    pages = "032",
    year = "2012"
}

@article{Barbieri:2012tu,
    author = "Barbieri, Riccardo and Buttazzo, Dario and Sala, Filippo and Straub, David M. and Tesi, Andrea",
    title = "{A 125 GeV composite Higgs boson versus flavour and electroweak precision tests}",
    eprint = "1211.5085",
    archivePrefix = "arXiv",
    primaryClass = "hep-ph",
    doi = "10.1007/JHEP05(2013)069",
    journal = "JHEP",
    volume = "05",
    pages = "069",
    year = "2013"
}

@article{Matsedonskyi:2014iha,
    author = "Matsedonskyi, Oleksii",
    title = "{On Flavour and Naturalness of Composite Higgs Models}",
    eprint = "1411.4638",
    archivePrefix = "arXiv",
    primaryClass = "hep-ph",
    doi = "10.1007/JHEP02(2015)154",
    journal = "JHEP",
    volume = "02",
    pages = "154",
    year = "2015"
}

@article{Panico:2016ull,
    author = "Panico, Giuliano and Pomarol, Alex",
    title = "{Flavor hierarchies from dynamical scales}",
    eprint = "1603.06609",
    archivePrefix = "arXiv",
    primaryClass = "hep-ph",
    reportNumber = "CERN-TH-2016-065",
    doi = "10.1007/JHEP07(2016)097",
    journal = "JHEP",
    volume = "07",
    pages = "097",
    year = "2016"
}

@article{Esteban:2024eli,
    author = "Esteban, Ivan and Gonzalez-Garcia, M. C. and Maltoni, Michele and Martinez-Soler, Ivan and Pinheiro, Jo{\~a}o Paulo and Schwetz, Thomas",
    title = "{NuFit-6.0: updated global analysis of three-flavor neutrino oscillations}",
    eprint = "2410.05380",
    archivePrefix = "arXiv",
    primaryClass = "hep-ph",
    reportNumber = "IFT-UAM/CSIC-24-140, YITP-SB-2024-24, IPPP/24/64, IPPP/24/64, IFT-UAM/CSIC-24-140, YITP-SB-2024-24",
    doi = "10.1007/JHEP12(2024)216",
    journal = "JHEP",
    volume = "12",
    pages = "216",
    year = "2024"
}

@article{Wyler:1982dd,
    author = "Wyler, D. and Wolfenstein, L.",
    title = "{Massless Neutrinos in Left-Right Symmetric Models}",
    reportNumber = "CERN-TH-3435",
    doi = "10.1016/0550-3213(83)90482-0",
    journal = "Nucl. Phys. B",
    volume = "218",
    pages = "205--214",
    year = "1983"
}

@article{Magg:1980ut,
    author = "Magg, M. and Wetterich, C.",
    title = "{Neutrino Mass Problem and Gauge Hierarchy}",
    reportNumber = "CERN-TH-2829",
    doi = "10.1016/0370-2693(80)90825-4",
    journal = "Phys. Lett. B",
    volume = "94",
    pages = "61--64",
    year = "1980"
}

@article{Cheng:1980qt,
    author = "Cheng, T. P. and Li, Ling-Fong",
    title = "{Neutrino Masses, Mixings and Oscillations in SU(2) x U(1) Models of Electroweak Interactions}",
    reportNumber = "PRINT-80-0511 (CARNEGIE-MELLON), COO-3066-152",
    doi = "10.1103/PhysRevD.22.2860",
    journal = "Phys. Rev. D",
    volume = "22",
    pages = "2860",
    year = "1980"
}

@proceedings{tHooft:1980xss,
    editor = "'t Hooft, Gerard and Itzykson, C. and Jaffe, A. and Lehmann, H. and Mitter, P. K. and Singer, I. M. and Stora, R.",
    title = "{Recent Developments in Gauge Theories. Proceedings, Nato Advanced Study Institute, Cargese, France, August 26 - September 8, 1979}",
    doi = "10.1007/978-1-4684-7571-5",
    volume = "59",
    pages = "pp.1--438",
    year = "1980"
}
